\documentclass[english,prb, twocolumn,nofootinbib]{revtex4-2}
\usepackage[T1]{fontenc}
\usepackage[latin9]{inputenc}
\usepackage{color}
\usepackage{babel}
\usepackage{amsmath}
\usepackage{amssymb}
\usepackage{graphicx}
\usepackage[unicode=true,pdfusetitle,
 bookmarks=true,bookmarksnumbered=false,bookmarksopen=false,
 breaklinks=false,pdfborder={0 0 0},pdfborderstyle={},backref=false,colorlinks=true]
 {hyperref}

\makeatletter
\hypersetup{colorlinks,allcolors=blue}

\usepackage[caption=false]{subfig}
\newcommand{\phantomsubfloat}[1]{
    {% apply caption setup only temporarily
        \captionsetup[subfigure]{labelformat=empty}
        \subfloat[][]{#1}     }%
} 

\usepackage{MnSymbol}

\makeatother

\begin{document}
\title{Chiral numerical renormalization group}
\author{Matan Lotem}
\email{matanlotem@mail.tau.ac.il}

\author{Eran Sela}
\email{eransx@googlemail.com}

\author{Moshe Goldstein}
\email{mgoldstein@tauex.tau.ac.il}

\affiliation{Raymond and Beverly Sackler School of Physics and Astronomy, Tel Aviv
University, Tel Aviv 6997801, Israel}
\begin{abstract}
The interplay between the Kondo screening of quantum impurities (by
the electronic channels to which they couple) and the interimpurity
RKKY interactions (mediated by the same channels) has been extensively
studied. However, the effect of unidirectional channels (e.g., chiral
or helical edge modes of 2D topological materials) which greatly restrict
the mediated interimpurity interactions, has only more recently come
under scrutiny, and it can drastically alter the physics. Here we
take Wilson's numerical renormalization group (NRG), the most established
numerical method for treating quantum impurity models, and extend
it to systems consisting of two impurities coupled at different locations
to unidirectional channel(s). This is challenging due to the incompatibility
of unidirectionality with one of the main ingredients in NRG---the
mapping of the channel(s) to a Wilson chain---a tight-binding chain
with the impurity at one end and hopping amplitudes which decay exponentially
with the distance. We bridge this gap by introducing a ``Wilson ladder''
consisting of two coupled Wilson chains, and demonstrate that this
construction successfully captures the unidirectionality of the channel(s),
as well as the distance between the two impurities. We use this mapping
in order to study two Kondo impurities coupled to a single chiral
channel, showing that all local properties and thermodynamic quantities
are indifferent to the interimpurity distance, and correspond to two
separate single-impurity models. Extensions to more impurities and/or
helical channels are possible.
\end{abstract}
\maketitle
\renewcommand{\figurename}{FIG.}
\renewcommand{\tablename}{TABLE.}

\section{Introduction}

\begin{figure}
\begin{centering}
\includegraphics[width=1\columnwidth]{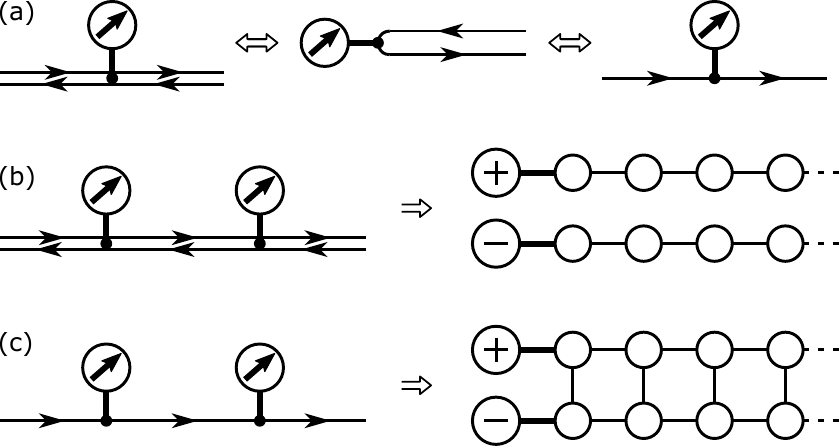}
\par\end{centering}
\begin{centering}
\phantomsubfloat{\label{fig:Intro-Unfolding}}\phantomsubfloat{\label{fig:Intro-Two-Chains}}\phantomsubfloat{\label{fig:Intro-Wilson-Ladder}}\vspace{-2.5em}
\par\end{centering}
\caption{(a) For a single impurity the coupling to a bidirectional channel
can be mapped to a unidirectional channel. (b) For two impurities
coupled to a bidirectional channel, after going to an even-odd basis
each impurity couples to a separate channel which can be mapped onto
a tight-binding (Wilson) chain. (c) However, for two impurities coupled
to a unidirectional channel, this mapping is not possible, and instead,
we map the channel onto two coupled Wilson chains, or a Wilson ladder.}
\end{figure}

Two-dimensional topological materials exhibit the remarkable property
of edge modes in which electrons of a given species can propagate
only in one direction \citep{hasanColloquiumTopologicalInsulators2010,qiTopologicalInsulatorsSuperconductors2011}.
Thus, intrachannel backscattering is forbidden, resulting in channels
which remain ballistic over large distances. Coupling such channels,
or baths, to quantum impurities, i.e., impurities with an internal
degree of freedom (for example, a localized spin), taps onto the exotic
world of Kondo physics \citep{kondoResistanceMinimumDilute1964,hewsonKondoProblemHeavy1993},
in which the bath electrons form a coherent manybody screening cloud
around the impurity. When considering single-impurity physics, the
unidirectionality of the channel(s) does not have significant consequences.
This is illustrated in Fig.\,\ref{fig:Intro-Unfolding}: Generally
one can choose a basis of the bidirectional channel modes such that
the impurity is coupled to the end of the channel, and then ``unfold''
the channel by interpreting the outgoing (backscattered) modes of
the bidirectional channel as the forward-scattered modes of a unidirectional
channel. Indeed, a variety of methods for solving quantum impurity
problems explicitly rely on this mapping \citep{andreiSolutionKondoProblem1983,tsvelickExactResultsTheory1983,affleckCurrentAlgebraApproach1990a,affleckKondoEffectConformal1991,gogolinBosonizationStronglyCorrelated2004}.
However, clearly such a mapping cannot be generalized to multiple
impurities, where the uni- or bidirectionality of the channel(s) becomes
important. The interplay between impurities coupled to bidirectional
channels typically gives rise to effective RKKY \citep{rudermanIndirectExchangeCoupling1954,kasuyaTheoryMetallicFerro1956,yosidaMagneticPropertiesCuMn1957}
interactions, $K\vec{S}_{m}\!\cdot\vec{S}_{m^{\prime}}$, between
impurity spins. This results in the transition of the Kondo lattice
from a Kondo-screened (heavy-fermion) phase to a magnetically ordered
phase \citep{doniachKondoLatticeWeak1977}, and has been extensively
studied by considering two-impurity models \citep{jayaprakashTwoImpurityKondoProblem1981,jonesStudyTwoMagnetic1987,jonesLowTemperaturePropertiesTwoImpurity1988,fyeQuantumMonteCarlo1989,fyeAnomalousFixedPoint1994,affleckConformalfieldtheoryApproachTwoimpurity1995,ganSolutionTwoimpurityKondo1995,silvaParticleHoleAsymmetryTwoImpurity1996,andreiQuantumPhaseTransition1999}.
However, if the impurities are coupled to unidirectional channels,
the picture is more complicated.

Assuming spinful channels, we have two natural scenarios. The first
is \textit{helical} channels, meaning that the two spin species propagate
in opposite directions, as in the quantum spin Hall effect \citep{kaneQuantumSpinHall2005a,bernevigQuantumSpinHall2006},
so that electrons can only backscatter into the opposite spin channel,
flipping the impurity spin in the process. As a result, the $z$ component
of the RKKY interaction is forbidden, but the transverse components
are still allowed \citep{gaoInplaneNoncollinearExchange2009}. Taking
into account Rashba couplings, intrachannel interactions, and bulk
effects further complicates the resulting RKKY structure and its interplay
with the Kondo physics \citep{leeElectricalControlInteraction2015,kurilovichIndirectExchangeInteraction2017,yevtushenkoKondoImpuritiesCoupled2018}.
This leads to dramatic consequences on transport properties, and is
suspected to be responsible for the breaking of quantized conductance
in quantum spin Hall systems \citep{cheianovMesoscopicFluctuationsConductance2013,altshulerLocalizationEdge2D2013,hsuNuclearspininducedLocalizationEdge2017}.
The second scenario is \textit{chiral} channels, meaning that both
spin species propagate in the same direction, as in the integer quantum
Hall effect \citep{halperinQuantizedHallConductance1982}, so that
backscattering is completely forbidden, and RKKY interactions cannot
be generated. This has far reaching implications when combined with
the multichannel Kondo effect, which is known (in the single-impurity
case) to give rise to fractionalized quasiparticles due to frustration.
These quasiparticles come with a fractional residual entropy \citep{tsvelickThermodynamicsMultichannelKondo1985},
reminiscent of a single non-Abelian anyon, the exotic quasiparticles
lying at the heart of topological quantum computing \citep{kitaevFaulttolerantQuantumComputation2003,nayakNonAbelianAnyonsTopological2008}.
In the multi-impurity case, the emergent RKKY interactions lift the
frustration, thus avoiding fractionalization, but recent proposals
try to circumvent this \citep{komijaniIsolatingKondoAnyons2020,lopesAnyonsMultichannelKondo2020}.
As chirality eliminates the RKKY interactions, the decoupled non-Abelian
anyons are expected to survive \citep{lopesAnyonsMultichannelKondo2020,gabayMultiimpurityChiralKondo2022,lotemManipulatingNonAbelianAnyons2022}.
Recently the two- and three-channel Kondo effects have been demonstrated
for a single quantum dot coupled to (multiple) integer quantum Hall
(chiral) edge modes \citep{iftikharTwochannelKondoEffect2015,iftikharTunableQuantumCriticality2018},
with clear signatures of the fractionalization \citep{landauChargeFractionalizationTwoChannel2018,vandalumWiedemannFranzLawNonFermi2020,nguyenThermoelectricTransportThreeChannel2020,hanFractionalEntropyMultichannel2022}.
Likewise, similar devices realizing two impurities have been studied
\citep{pouseQuantumSimulationExotic2023}. Thus, one can expect that
extending this setup to multiple impurities coupled by unidirectional
chiral modes will enable experimental observation of decoupled anyons.
We note that such an extension is more realistic with a \textit{partially
connected} scenario, where in one of the spin species the impurities
are coupled to the same unidirectional channel, while in the other
spin species each impurity is coupled to a separate channel.

It would therefore be useful to have a generic method for analyzing
multiple quantum impurities coupled to the same unidirectional channel(s).
For this we turn to Wilson's numerical renormalization group (NRG)
\citep{wilsonRenormalizationGroupCritical1975,bullaNumericalRenormalizationGroup2008},
one of the most generic and reliable tools for studying quantum impurity
models. A key part in (standard) NRG is mapping the electronic bath
to a so-called Wilson chain---a tight-binding chain with the impurity
at one end and hopping amplitudes decaying exponentially with the
distance. Over the years, NRG has also been applied to two-impurity
systems in different scenarios \citep{jonesStudyTwoMagnetic1987,silvaParticleHoleAsymmetryTwoImpurity1996,ingersentStudyTwoimpurityTwochannel1992,affleckConformalfieldtheoryApproachTwoimpurity1995,campoAlternativeDiscretizationNumerical2005,mitchellTwoChannelKondoPhysics2012,mitchellMultipleMagneticImpurities2015,lechtenbergRealisticQuantumCritical2017,eickhoffEffectiveLowenergyDescription2018,eickhoffStronglyCorrelatedMultiimpurity2020}
but always under the assumption of bidirectional channels, which enables
mapping the bath to two separate Wilson chains, as illustrated in
Fig.\,\ref{fig:Intro-Two-Chains}. Such a mapping is not possible
for unidirectional channels, as a nearest-neighbor tight-binding chain
has no notion of directionality.

Here we overcome this obstacle by mapping the bath to two coupled
Wilson chains, or a \textit{Wilson ladder}, as depicted in Fig.\,\ref{fig:Intro-Wilson-Ladder}.
We note that the resulting structure is formally a particular case
of a channel mixing bath \citep{liuQuantumImpuritiesChannel2016}.
However, due to oscillatory terms which typically arise in two-impurity
systems, the procedure introduced in Ref.\,\citep{liuQuantumImpuritiesChannel2016}
would face technical difficulties for a two-impurity problem. Similar
problems are also expected for a derivation based on open Wilson chains
\citep{bruognoloOpenWilsonChains2017}. In this work, we therefore
introduce an alternative derivation of the Wilson ladder. By exploiting
PT (inversion + time reversal) symmetry, we enforce a real Hamiltonian,
and together with particle-hole symmetry in the bath, we can nullify
both onsite energies and crosslinks in the ladder. We then demonstrate
that the Wilson-ladder structure correctly captures the distance between
the two impurities, with a transition from two weakly coupled chains
at high energies (or short wavelengths), to an effective single chain
at low energies (or long wavelengths). Considering each ladder level
as an enlarged effective site along a Wilson chain, we can proceed
with iterative diagonalization by standard NRG procedure.

We test the mapping on two resonant levels coupled at different locations
to a spinless chiral channel. Such a noninteracting system can be
solved exactly both in the continuum limit (of the chiral channel)
and after discretization, and so serves as an excellent benchmark
for the method. We find that thermodynamic quantities are accurately
captured at all temperatures, and that most features of zero-temperature
spectral properties are also captured. However, some of the (exact)
spectral quantities exhibit oscillations at a frequency corresponding
to the interimpurity distance, which by construction cannot be captured
by a logarithmic discretization procedure. Still, we find that we
do successfully reproduce the envelope of the oscillations. This implies
that static temperature-dependent correlation functions are successfully
captured at low temperatures, while at high temperatures, for which
the correlations should drop to zero exponentially, we get artificial
oscillations around zero.

We then turn to study a single-channel chiral two-impurity Anderson
model (in the local-moment limit). We find that by looking only at
local impurity quantities, e.g., the impurity entropy and magnetic
susceptibly, one cannot discern the difference between the chiral
system and two separate copies of a single-impurity problem. Thus,
at high temperatures, we have a free spin at each impurity, and at
low temperatures both spins are fully screened, with the crossover
(Kondo) temperature $T_{K}$ independent of the distance between the
impurities. This is actually consistent with the Bethe-ansatz solution
of the Kondo problem \citep{andreiSolutionKondoProblem1983,tsvelickExactResultsTheory1983},
which has also been applied to multiple impurities coupled to a chiral
channel. It can also be explained by the following intuitive argument:
Due to the absence of backscattering, the first impurity cannot ``know''
about the second, and thus ``behaves'' as in the single-impurity
case. Applying a PT transformation, the same holds for the second
(last) impurity. We point out that both in our solution, and implicitly
in the Bethe-ansatz solution, one assumes some (possibly small) separation
between the impurities. Thus, neither solution is applicable to two
impurities exactly at the same point, but such a case is trivial,
corresponding to an enlarged single-impurity problem, and is not the
focus of this work. Looking at static impurity-impurity correlations,
we find that they do depend on the interimpurity distance, and are
nonzero at low temperatures. This is, however, expected---even in
a trivial noninteracting chiral bath we have spatial correlations,
and once the impurities are in the strongly-coupled regime, these
correlations are reflected by the impurity-impurity correlations.
Such \textit{static} correlations do not affect the local impurity
physics, and we demonstrate that a local perturbation at one impurity
does not affect the local physics of the other, i.e., response functions
(\textit{retarded} correlations) are chiral.

The remainder of this paper will be ordered as follows: In Sec.\,\ref{sec:Method},
we derive the Wilson ladder and comment about the NRG implementation,
leaving some of the technical details to Appendices \ref{sec:App-Campo-Oliveira}
and \ref{sec:App-Exploiting-Symmetries}. In Sec.\,\ref{sec:Noninteracting-Results},
we test the quality of this mapping on a noninteracting system, demonstrating
its advantages and limitations; some technical details are relegated
to Appendix~\ref{sec:App-Single-Particle-Calculations}. We then
apply the method to a spinful chiral channel coupled to two Kondo
impurities in Sec.\,\ref{sec:Results} and analyze the results. Finally,
we conclude in Sec.\,\ref{sec:Summary} and comment on possible applications
of the presented method.

\section{The Wilson Ladder\label{sec:Method}}

We start this section by formally defining the problem we wish to
address, pointing out where previous solutions break down, and setting
the stage for the derivation of the Wilson ladder, which will then
be outlined in the subsections. The Hamiltonian of a generic quantum
impurity problem can be written as
\begin{equation}
H=H_{\mathrm{imp}}+H_{\mathrm{coupling}}+H_{\mathrm{bath}},
\end{equation}
where $H_{\mathrm{bath}}$ describes a quadratic (noninteracting)
fermionic bath, $H_{\mathrm{imp}}$ describes the (typically interacting)
impurity degrees of freedom, and $H_{\mathrm{coupling}}$ couples
the bath to the impurity(ies). In this section we will assume an Anderson
impurity model, for which $H_{\mathrm{coupling}}$ is also quadratic.
Wilson's NRG can be decomposed into three stages:
\begin{enumerate}
\item[A.] \textit{Logarithmic discretization}, or coarse graining, of the bath
Hamiltonian\textsl{.}
\item[B.] \textit{}\textsl{Tridiagonalization} of the discrete bath Hamiltonian
to a tight-binding (Wilson) chain with the impurity at one end.
\item[C.] \textit{}Numerical \textsl{iterative diagonalization} of the full
Hamiltonian, probing ever shrinking energy scales with each iteration.
\end{enumerate}
The first two steps are indifferent to the interaction $U$ within
the impurity sites {[}see Eq.\,(\ref{eq:H-Imp-Anderson}) below{]}.
Therefore, we set the impurity Hamiltonian to zero, perform the mapping,
and reintroduce it only for the iterative diagonalization. Assuming
the bath and coupling Hamiltonians are diagonal in spin and flavor
indices, we can apply the mapping to a single flavor of spinless fermions,
and then duplicate the resulting structure. Thus, for the derivation
of the Wilson ladder, we will consider a single channel of noninteracting
spinless right-moving fermions coupled to two impurities at $\pm R/2$:
\begin{subequations}
\label{eq:H}
\begin{align}
H_{\mathrm{imp}} & =0,\\
H_{\mathrm{bath}} & =\int_{-\infty}^{\infty}\psi^{\dagger}\left({-}iv_{F}\partial_{x}\right)\psi dx,\label{eq:H-Bath}\\
H_{\mathrm{coupling}} & =\tilde{t}_{0}d_{1}^{\dagger}\psi\left({-}\tfrac{R}{2}\right)+\tilde{t}_{0}d_{2}^{\dagger}\psi\left({+}\tfrac{R}{2}\right)+\mathrm{H.c.},\label{eq:H-Coupling}
\end{align}
\end{subequations}
where $\hbar{=}1$ throughout, $v_{F}$ is the Fermi velocity, and
we have assumed both impurities couple only locally and with equal
real amplitude $\tilde{t}_{0}$ (but this can be generalized to more
complicated setups). The fermionic field and impurity operators satisfy
$\left\{ \psi\left(x\right),\psi^{\dagger}\left(x^{\prime}\right)\right\} =2\pi\delta\left(x-x^{\prime}\right)$
and $\left\{ d_{m},d_{m^{\prime}}^{\dagger}\right\} =\delta_{mm^{\prime}}$,
respectively. Observe that Eq.\,(\ref{eq:H}) conserves total charge,
and is invariant under the following transformations
\begin{subequations}
\begin{align}
\text{particle-hole\ :\ \ \ } & \psi\!\left(x\right)\to\psi^{\dagger}\!\left(x\right),\ d_{m}\to-d_{m}^{\dagger},\label{eq:PH-symmetry}\\
\text{PT\ \ \ \ \ \ :\ \ \  } & x\to-x,\ i\to-i,\ d_{1}\leftrightarrow d_{2},\label{eq:PT-symmetry}
\end{align}
\end{subequations}
but not under inversion or time reversal individually.

In order to point out where previous NRG approaches break down we
will consider the impurities' retarded Green function, which can be
written as a $2\times2$ matrix
\begin{equation}
\mathbf{G}^{R}\left(\omega\right)=\left[\omega\mathbf{1}-\mathbf{h}-\mathbf{\Sigma}^{R}\left(\omega\right)\right]^{-1},\label{eq:Noninteracting-Retarded-Green-Function}
\end{equation}
with $\mathbf{1}$ the identity matrix, $\mathbf{h}$ the (single-particle)
impurity Hamiltonian, which in our case is zero, and
\begin{equation}
\mathbf{\Sigma}^{R}\left(\omega\right)=\frac{\tilde{t}_{0}^{2}}{2iv_{F}}\begin{pmatrix}1 & 2e^{i\omega R/v_{F}}\\
0 & 1
\end{pmatrix},\label{eq:Noninteracting-Self-Energy}
\end{equation}
the retarded self-energy contribution due to hybridization with the
bath. The zero element below the diagonal of $\mathbf{\Sigma}^{R}$
is the formal manifestation of chirality, which implies that a retarded
quantity at $-R/2$ cannot depend on anything that happens at $+R/2$.
This immediately carries on to $\mathbf{G}^{R}$$\left(\omega\right)$,
as the inverse of an upper-triangular matrix is an upper-triangular
matrix, and is not affected by the introduction of local potentials
at the impurities or asymmetric couplings,
\begin{equation}
\mathbf{h}\,{\to}\begin{pmatrix}\mu_{1} & 0\\
0 & \mu_{2}
\end{pmatrix},\ \ \mathbf{\Sigma}^{R}\!\left(\omega\right){\to}\frac{1}{2iv_{F}}\begin{pmatrix}\left|\tilde{t}_{1}\right|^{2} & 2\tilde{t}_{1}\tilde{t}_{2}^{*}e^{i\omega R/v_{F}}\\
0 & \left|\tilde{t}_{2}\right|^{2}
\end{pmatrix}.\label{eq:RLM-asymmetry}
\end{equation}
We define the impurity spectral function 
\begin{equation}
\mathbf{A}\left(\omega\right)=-\frac{1}{2\pi i}\left[\mathbf{G}^{R}\left(\omega\right)-\mathbf{G}^{A}\left(\omega\right)\right],\label{eq:Noninteracting-Spectral-Function}
\end{equation}
with $\mathbf{G}^{A}=\mathbf{G}^{R\dagger}$. In Sec.\,\ref{sec:Noninteracting-Results},
we will test the quality of the discretization scheme by how well
it reproduces $\mathbf{A}\left(\omega\right)$. An important consequence
of the upper-triangular structure of $\mathbf{G}^{R}\left(\omega\right)$
is that the diagonal elements of the spectral function are equal to
those of a single impurity with the corresponding local potential
and coupling. It is convenient to also define the hybridization function
\begin{equation}
\mathbf{\Gamma}\!\left(\omega\right)\!=\!-\frac{1}{2i}\!\left[\mathbf{\Sigma}^{R}\!\left(\omega\right)\!-\!\mathbf{\Sigma}^{A}\!\left(\omega\right)\right]\!=\!\Gamma\begin{pmatrix}1 & \!\!e^{i\omega R/v_{F}}\\
e^{-i\omega R/v_{F}}\!\! & 1
\end{pmatrix},
\end{equation}
where in the last form we returned to the case of symmetrically-coupled
impurities, and defined $\Gamma\equiv\frac{\tilde{t}_{0}^{2}}{2v_{F}}$.
As $\boldsymbol{\Gamma}$$\left(\omega\right)$ encodes all the information
about the bath, any bath which reproduces this hybridization function
will result in the same impurity physics. NRG relies precisely on
this property in order to replace the bath by a Wilson chain, to which
we can apply iterative diagonalization.

\begin{figure*}[t]
\includegraphics[width=1\textwidth]{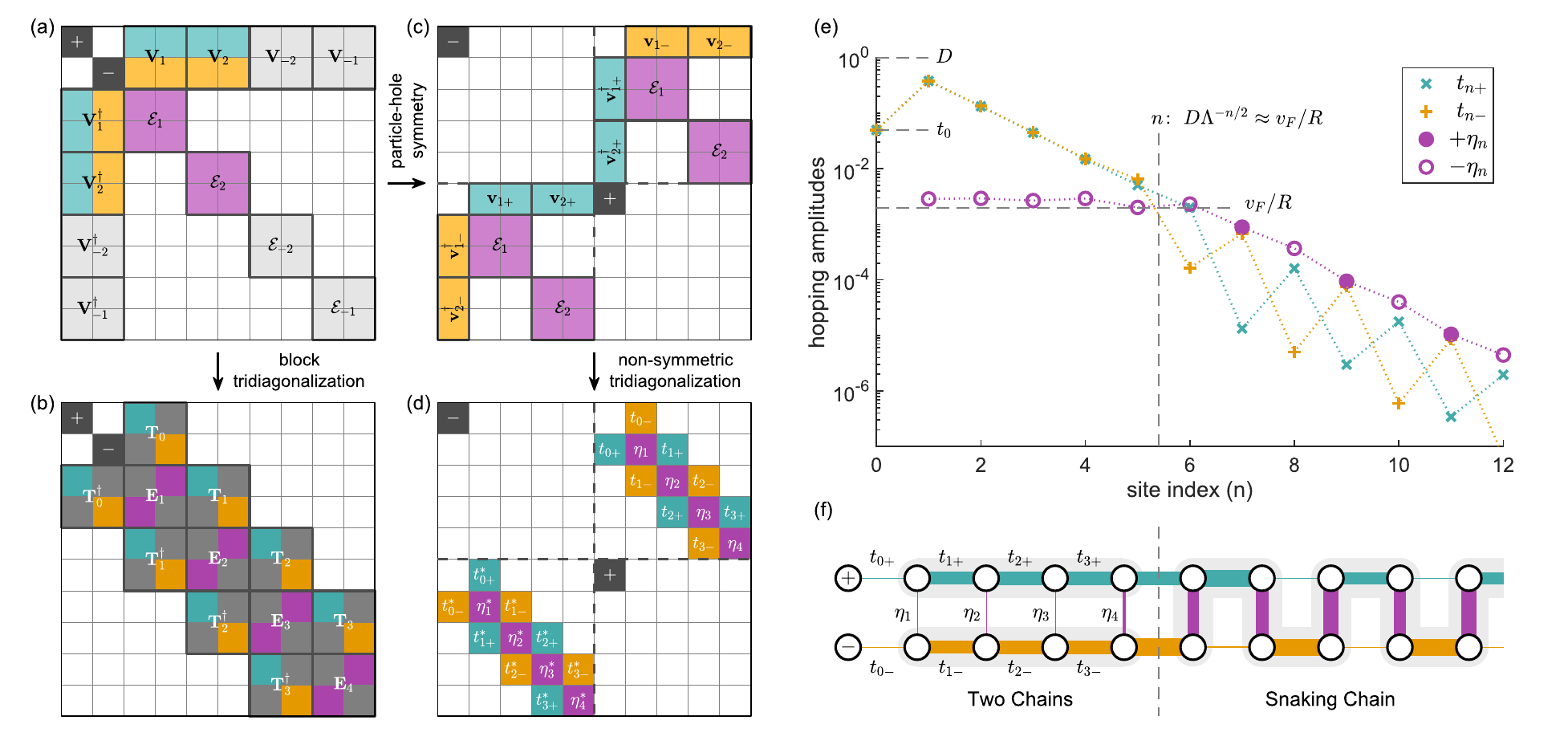}

\phantomsubfloat{\label{fig:Tridiag-Block-Star}}\phantomsubfloat{\label{fig:Tridiag-Block-Chain}}\phantomsubfloat{\label{fig:Tridiag-Symmetric}}\phantomsubfloat{\label{fig:Tridiag-Symmetric-Chain}}\phantomsubfloat{\label{fig:Tridiag-Hopping}}\phantomsubfloat{\label{fig:Tridiag-Ladder}}
\begin{centering}
\vspace{-2.5em}
\par\end{centering}
\caption{(a) Block structure of the discretized Hamiltonian, corresponding
to Eq.\,(\ref{eq:H-Star}), with two positive and two negative energy
intervals. White tiles are zero, and so are the black tiles, which
indicate the location of the even ($+$) and odd ($-$) impurity modes.
(b) Applying block tridiagonalization, one arrives at the generalized
Wilson chain structure in Eqs.\,(\ref{eq:H-Chain-B}) {[}or Eqs.\,(\ref{eq:H-Chain}){]}.
White and black tiles are zero, and in the presence of particle-hole
symmetry, the gray tiles can also be nullified, resulting in a Wilson
ladder structure. (c) Transforming to a particle-hole symmetric basis,
we get the structure in Eq.\,(\ref{eq:H-PHsym-Mat}). Color-coding
relates tiles of equal value between (a) and (c). Note that the indices
of the even and odd impurity modes have changed. (d) Applying nonsymmetric
tridiagonalization to the upper-right and lower-left quarters of (c)
brings them to tridiagonal form. Color-coding relates tiles of equal
value between (b) and (d). Note the alternating relation to even and
odd hopping-amplitudes. (e) Hopping amplitudes along the Wilson ladder
for an example case with $\Lambda=9$, $z=1$, $R/v_{F}=2\pi\Lambda^{2}/D$,
$t_{0}=0.05D$ ($2v_{F}/R\approx\Gamma\approx0.004D$), with the different
scales indicated on the graph by horizontal dashed lines. The vertical
dashed line indicates the change in the qualitative behavior, which
occurs when the hopping amplitudes cross the energy scale $v_{F}/R$.
(f) Wilson ladder with link widths proportional to the value of the
rescaled hopping amplitudes $t_{n\pm}\Lambda^{n/2},\,\eta_{n}\Lambda^{n/2+1/4}$.
The dashed line indicates the change from two weakly-coupled chains
to a single snaking chain. \label{fig:Tridiagonalization}}
\end{figure*}

Observe that $\mathbf{\Gamma}\left(\omega\right)$ cannot be diagonalized
by a frequency-independent transformation. This is the key difference
between the unidirectional channel two-impurity models studied in
this paper, and most bidirectional channel two-impurity models previously
studied with NRG \citep{jonesStudyTwoMagnetic1987,silvaParticleHoleAsymmetryTwoImpurity1996,ingersentStudyTwoimpurityTwochannel1992,affleckConformalfieldtheoryApproachTwoimpurity1995,campoAlternativeDiscretizationNumerical2005,mitchellTwoChannelKondoPhysics2012,mitchellMultipleMagneticImpurities2015,lechtenbergRealisticQuantumCritical2017,eickhoffEffectiveLowenergyDescription2018,eickhoffStronglyCorrelatedMultiimpurity2020}.
For the latter, choosing an even-odd impurity basis diagonalizes the
hybridization function, resulting in the impurities coupling to two
separate channels, each of which can be mapped to a separate Wilson
chain.\footnote{The mapping of the bath onto two separate Wilson chains is possible
as long as the impurities couple to two bath modes which are invariant
under inversion $x\to-x$. This holds for typical bidirectional channels,
but inherently breaks down in the unidirectional case.} In our case, transforming to an even-odd (PT-symmetric) impurity
basis
\begin{equation}
d_{+}=\frac{1}{\sqrt{2}}\left(d_{1}+d_{2}\right),\quad d_{-}=\frac{i}{\sqrt{2}}\left(d_{1}-d_{2}\right),\label{eq:even-odd-transformation}
\end{equation}
results in a real hybridization function
\begin{equation}
\mathbf{\Gamma}\left(\omega\right)\to\Gamma\begin{pmatrix}1+\cos\left(\frac{\omega R}{v_{F}}\right) & -\sin\left(\frac{\omega R}{v_{F}}\right)\\
-\sin\left(\frac{\omega R}{v_{F}}\right) & 1-\cos\left(\frac{\omega R}{v_{F}}\right)
\end{pmatrix}.\label{eq:Hybridization-Even-Odd}
\end{equation}
This guarantees a (numerically more stable and efficient) real representation
for the discretized bath Hamiltonian, but clearly does not diagonalize
$\mathbf{\Gamma}\left(\omega\right)$. When the hybridization function
cannot be diagonalized (by a frequency independent transformation),
one can still take the approach by \citet{liuQuantumImpuritiesChannel2016}
and arrive at a generalized Wilson chain structure consisting of coupled
chains. Introducing a 2-vector notation, $\mathbf{d}\thinspace{\equiv}\left(\!\begin{smallmatrix}d_{+}\\
d_{-}
\end{smallmatrix}\!\right),\,\mathbf{f}_{n}{\equiv}\left(\!\begin{smallmatrix}f_{n+}\\
f_{n-}
\end{smallmatrix}\!\right)$, with $f_{n\pm}$ a discrete set of fermionic bath operators, our
goal will be to write the coupling and bath Hamiltonians as 
\begin{subequations}
\label{eq:H-Chain}
\begin{align}
H_{\mathrm{coupling}} & =\mathbf{d}^{\dagger}\mathbf{T}_{0}\mathbf{f}_{0}+\mathrm{H.c.},\label{eq:H-Coupling-Chain}\\
H_{\mathrm{bath}} & =\sum_{n=0}^{2N}\mathbf{f}_{n}^{\dagger}\mathbf{E}_{n}\mathbf{f}_{n}+\sum_{n=0}^{2N-1}\mathbf{f}_{n}^{\dagger}\mathbf{T}_{n+1}\mathbf{f}_{n+1}+\mathrm{H.c.},\label{eq:H-Bath-Chain}
\end{align}
\end{subequations}
where $\mathbf{E}_{n}$ and $\mathbf{T}_{n}$ are $2\times2$ coefficient
matrices. Eqs.\,(\ref{eq:H-Chain}) resemble the expressions obtained
in the standard Wilson chain scheme for a single-impurity \citep{bullaNumericalRenormalizationGroup2008}.
The difference is that fermionic operators are promoted to 2-vectors
of operators, and scalar coefficients are promoted to $2\times2$
matrices. We will demonstrate that due to particle-hole symmetry in
the bath, we can nullify the off-diagonal elements of $\mathbf{T}_{n}$
and the diagonal elements of $\mathbf{E}_{n}$, resulting in a ladder
structure, as depicted in Fig.\,\ref{fig:Tridiag-Ladder}.

In order to arrive at these expressions, we will first need to logarithmically
discretize the Hamiltonian, bringing it into the so-called star geometry:
\begin{subequations}
\label{eq:H-Star}
\begin{align}
H_{\mathrm{coupling}} & =\sum_{n=\pm1}^{\pm N}\mathbf{d}^{\dagger}\mathbf{V}_{n}\mathbf{c}_{n}+\mathrm{H.c.},\label{eq:H-Coupling-Star}\\
H_{\mathrm{bath}} & =\sum_{n=\pm1}^{\pm N}\mathbf{c}_{n}^{\dagger}\boldsymbol{\mathcal{E}}_{n}\mathbf{c}_{n},\label{eq:H-Bath-Star}
\end{align}
\end{subequations}
with $\mathbf{c}_{n}$ being 2-vectors of fermionic operators, $\mathbf{V}_{n}$
and $\boldsymbol{\mathcal{E}}_{n}$ coefficient matrices, and $4N$
the number of discrete modes. Note that apart from the introduction
of a high-energy cutoff, Eq.\,(\ref{eq:H-Coupling-Star}) will be
exact, while writing the bath Hamiltonian as in Eq.\,(\ref{eq:H-Bath-Star})
is the main approximation in NRG. The discretization scheme used in
Ref.\,\citep{liuQuantumImpuritiesChannel2016} in order to obtain
$\boldsymbol{\mathcal{E}}_{n}$ relies on separately diagonalizing
the hybridization function $\mathbf{\Gamma}\left(\omega\right)$ for
each frequency and then proceeding as in Refs.\,\citep{zitkoAdaptiveLogarithmicDiscretization2009,zitkoEnergyResolutionDiscretization2009}
to numerically solve a differential equation containing an integral
over the diagonalized hybridization function. Due to the highly oscillatory
structure of $\mathbf{\Gamma}\left(\omega\right)$, we found this
scheme to be ill suited for our case, and in Sec.\,\ref{subsec:Discretization},
we will present an alternative (more traditional) discretization scheme
which circumvents this. In Sec.\,\ref{subsec:Tridiagonalization},
we will apply a generalized tridiagonalization procedure and arrive
at the Wilson ladder of Eqs.\,(\ref{eq:H-Chain}), explicitly exploiting
particle-hole symmetry. Finally, in Sec.\,\ref{subsec:Iterative-Diagonalization},
we will comment on several implementation details for the iterative
diagonalization.

\subsection{Logarithmic discretization\label{subsec:Discretization}}

We will now outline a discretization procedure in terms of a general
matrix hybridization function $\mathbf{\Gamma}\left(\omega\right)$,
with the goal of plugging Eq.\,(\ref{eq:Hybridization-Even-Odd})
into the obtained expressions. However, one can equivalently discretize
at the level of a specific continuous Hamiltonian, which for the chiral
Hamiltonian in Eqs.\,(\ref{eq:H}) is very convenient. In Appendix
\ref{sec:App-Campo-Oliveira}, we take this approach, as it allows
for a clearer comparison with traditional discretization schemes,
and specifically the derivation in Ref.\,\citep{campoAlternativeDiscretizationNumerical2005},
which we here generalize to a matrix hybridization.

In order to discretize the bath, we must first introduce a UV (high
energy / short wavelength) cutoff $D$, typically related to the half-bandwidth.
For a chiral model, e.g., an integer quantum Hall edge state, $D$
is related to the bulk gap, and so $D/v_{F}\equiv k_{\max}$ should
not be associated with the Fermi wavevector $k_{F}$ or an underlying
lattice spacing. We point out that introducing this sharp cutoff,
which is required for the numerics, breaks chirality, e.g., leading
to a loss of orthogonality {[}which decays as $\mathrm{sinc}\left(DR/v_{F}\right)\equiv\frac{\sin\left(DR/v_{F}\right)}{DR/v_{F}}$---see
Appendix \ref{sec:App-Campo-Oliveira}{]} between the two bath modes
coupling to the real-space impurities $d_{1}$ and $d_{2}$. We thus
wish to stay in the large-bandwidth limit, i.e., $D$ much larger
than any relevant energy scale, so that the breaking of chirality
is negligible. We can further negate these effects, improving numerical
stability, by taking the cutoff to be commensurate with $R$, i.e.,
\begin{equation}
DR/v_{F}=k_{\max}R=\pi l\ ;\ l\in\mathbb{N}.\label{eq:Commensurate-Cutoff}
\end{equation}
We then introduce a logarithmic discretization grid 
\begin{equation}
\epsilon_{1}^{z}=D,\quad\epsilon_{n>1}^{z}=D\Lambda^{2-n-z},\quad\epsilon_{n<0}^{z}=-\epsilon_{-n}^{z},
\end{equation}
where $\Lambda{>}1$ controls the logarithmic level spacing and $z{\in}\nolinebreak\left(0,1\right]$
shifts the levels, upon which we define intervals
\begin{equation}
I_{n>0}^{z}=\left[\epsilon_{n+1}^{z},\epsilon_{n}^{z}\right],\quad I_{n<0}^{z}=\left[-\epsilon_{-n}^{z},-\epsilon_{-n-1}^{z}\right].
\end{equation}
In order to keep the notation compact, in what follows we will drop
the $z$ index, but remember it is implied whenever $n$ appears.

Each impurity couples to a single mode in each interval. These two
modes are generally not orthogonal with respect to each other, and
in the continuum limit $\Lambda\to1$ they actually coincide. However,
for a finite $\Lambda$ we can always choose linear combinations of
the two which are orthonormal, to which we will refer as $\mathbf{c}_{n}{=}\left(\!\begin{smallmatrix}c_{n+}\\
c_{n-}
\end{smallmatrix}\!\right)$. The coupling to these modes can be obtained by integrating over
the hybridization function
\begin{equation}
\left[\mathbf{V}_{n}\right]^{2}=\frac{1}{\pi}\int_{I_{n}}\boldsymbol{\Gamma}\left(\omega\right)d\omega.\label{eq:Vn2}
\end{equation}
As $\boldsymbol{\Gamma}\left(\omega\right)$ is a Hermitian positive
matrix for each $\omega$, so is its integral, which can thus be expressed
as the square of some (Hermitian) matrix. Hence, its matrix square
root $\mathbf{V}_{n}$ and its inverse $\mathbf{V}_{n}^{-1}$ are
both well defined (and can be chosen to be Hermitian). Projecting
the bath Hamiltonian onto the modes $\mathbf{c}_{n}$, we obtain the
coefficient matrix 
\begin{equation}
\boldsymbol{\mathcal{E}}_{n}^{\mathrm{naive}}=\pi\int_{I_{n}}d\omega\,\omega\mathbf{V}_{n}^{-1}\boldsymbol{\Gamma}\left(\omega\right)\mathbf{V}_{n}^{-1}.\label{eq:Naive-En}
\end{equation}
As $\mathbf{c}_{n}$ are not bath eigenmodes, this is where the main
approximation in NRG enters.\footnote{This procedure can be understood as the expansion of the hybridization
function in each interval $I_{n}$ as a continued fraction, which
is then truncated after one step, with the remaining continuum of
states in the interval discarded following standard Wilsonian philosophy.
See Ref.\,\citep{bruognoloOpenWilsonChains2017} for an outline of
how to write down such an expansion for a hybridization function supported
on some arbitrary interval, and specifically Appendix B therein, which
demonstrates how to do so for a matrix hybridization function. Note
a factor of $\pi$ difference in the definition of our $\boldsymbol{\Gamma}\left(\omega\right)$
with respect to Ref.\,\citep{bruognoloOpenWilsonChains2017}.} Eqs.\,(\ref{eq:Vn2}) and (\ref{eq:Naive-En}) generalize the standard
(naive) single-impurity discretization used in NRG \citep{bullaNumericalRenormalizationGroup2008}
to a matrix hybridization. Following the notation of Ref.\,\citep{zitkoEnergyResolutionDiscretization2009},\footnote{See end of Appendix \ref{sec:App-Campo-Oliveira} for an explicit
comparison with the notation of Ref.\,\citep{zitkoEnergyResolutionDiscretization2009}.} one can immediately write the matrix generalization of the scheme
by \citet{campoAlternativeDiscretizationNumerical2005} as
\begin{equation}
\boldsymbol{\mathcal{E}}_{n}^{\mathrm{CO}}=\left[\pi\int_{I_{n}}\frac{d\omega}{\omega}\mathbf{V}_{n}^{-1}\boldsymbol{\Gamma}\left(\omega\right)\mathbf{V}_{n}^{-1}\right]^{-1}.\label{eq:CampoOliveira-En}
\end{equation}
In Appendix \ref{sec:App-Campo-Oliveira}, we formulate the derivation
of $\mathbf{V}_{n},\boldsymbol{\mathcal{E}}_{n}^{\mathrm{naive}}$
and $\boldsymbol{\mathcal{E}}_{n}^{\mathrm{CO}}$ (up to a unitary
rotation) for our specific model in the language used in Ref.\,\citep{campoAlternativeDiscretizationNumerical2005},
explicitly demonstrating the source of the obtained expression. 

We point out that for the PT-symmetric hybridization function of Eq.\,(\ref{eq:Hybridization-Even-Odd}),
the coefficient matrices are all real, and for positive and negative
$n$ (regardless of the discretization scheme) are related by
\begin{equation}
\mathbf{V}_{-n}=\sigma_{z}\mathbf{V}_{+n}\sigma_{z}\ ;\ \mathbf{\boldsymbol{\mathcal{E}}}_{-n}=-\sigma_{z}\mathbf{\boldsymbol{\mathcal{E}}}_{+n}\sigma_{z},\label{eq:Symmetry-Coef}
\end{equation}
with $\sigma_{z}=\left(\begin{smallmatrix}1 & 0\\
0 & -1
\end{smallmatrix}\right)$. This is a manifestation of particle-hole symmetry, and will be exploited
in the next section.

\subsection{Tridiagonalization\label{subsec:Tridiagonalization}}

In the single-impurity case, one has $N$ positive-energy modes, $N$
negative-energy modes, and one impurity mode, so that the single-particle
Hamiltonian can be written as a $\left(2N+1\right)\times\left(2N+1\right)$
matrix. One can then numerically find a unitary transformation which
brings it into tridiagonal form, e.g., by the Lanczos or Householder
algorithms. This scheme can be readily generalized to the two-impurity
case \citep{liuQuantumImpuritiesChannel2016}: We now have $2N$ positive-energy
modes, $2N$ negative-energy modes, and 2 impurity modes, so that
the single-particle Hamiltonian can be written as a $\left(4N+2\right)\times\left(4N+2\right)$
matrix, or as a $\left(2N+1\right)\times\left(2N+1\right)$ block
matrix with $2\times2$ elements $\mathbf{V}_{n}$ and $\boldsymbol{\mathcal{E}}_{n}$,
as shown in Fig.\,\ref{fig:Tridiag-Block-Star}. One can then apply
a generalized Lanczos or Householder procedure in order to bring it
to a block-tridiagonal form, as shown in Fig.\,\ref{fig:Tridiag-Block-Chain}.
The Hamiltonian is thus given by Eqs.\,(\ref{eq:H-Chain}), which
are rewritten here for the sake of clarity,
\begin{subequations}
\label{eq:H-Chain-B}
\begin{align}
H_{\mathrm{coupling}} & =\mathbf{d}^{\dagger}\mathbf{T}_{0}\mathbf{f}_{0}+\mathrm{H.c.},\\
H_{\mathrm{bath}} & =\sum_{n=0}^{2N}\mathbf{f}_{n}^{\dagger}\mathbf{E}_{n}\mathbf{f}_{n}+\sum_{n=0}^{2N-1}\mathbf{f}_{n}^{\dagger}\mathbf{T}_{n+1}\mathbf{f}_{n+1}+\mathrm{H.c.},
\end{align}
\end{subequations}
with general matrices $\mathbf{T}_{n}$ and Hermitian matrices $\mathbf{E}_{n}$.
Since the coefficient matrices $\mathbf{V}_{n}$ and $\boldsymbol{\mathcal{E}}_{n}$
obtained in the previous section (for a PT-symmetric hybridization)
were all real, $\mathbf{T}_{n}$ ($\mathbf{E}_{n}$) can be written
as real (and symmetric) matrices. One still has the freedom to apply
arbitrary unitary transformations to the $2\times2$ blocks. In the
presence of particle-hole symmetry, these transformations can be used
to nullify the diagonal terms in $\mathbf{E}_{n}$ and the off-diagonal
terms in $\mathbf{T}_{n}$, bringing us to the Wilson ladder structure
shown in Fig.\,\ref{fig:Tridiag-Ladder}. This is a generalization
to the zero onsite energies in the ordinary (single-impurity) Wilson
chain in the presence of particle-hole symmetry. Here we present an
alternative tridiagonalization procedure which exploits particle-hole
symmetry, and thus by construction enforces the ladder structure.

The modes $\mathbf{c}_{n}{=}\left(\!\begin{smallmatrix}c_{n+}\\
c_{n-}
\end{smallmatrix}\!\right)$ do not respect particle-hole symmetry, meaning a particle-hole transformation
mixes different such modes $\mathbf{c}_{n}\to\sigma_{z}\left[\mathbf{c}_{-n}^{\dagger}\right]^{T}$.
However, we can take linear combinations $\mathbf{a}_{n+}$ and $\mathbf{a}_{n-}$
(introducing 4-vector notation),
\begin{equation}
\begin{pmatrix}\mathbf{a}_{n+}\\
\mathbf{a}_{n-}
\end{pmatrix}=\frac{1}{\sqrt{2}}\begin{pmatrix}\mathbf{1} & \sigma_{z}\\
\mathbf{1} & -\sigma_{z}
\end{pmatrix}\begin{pmatrix}\mathbf{c}_{n}\\
\mathbf{c}_{-n}
\end{pmatrix},
\end{equation}
which transform as $\mathbf{a}_{n\pm}\to\pm\left[\mathbf{a}_{n\pm}^{\dagger}\right]^{T}$,
and thus respect this symmetry. Together with the symmetry condition
of Eq.\,(\ref{eq:Symmetry-Coef}), the bath and coupling Hamiltonians
obtain the form
\begin{subequations}
\label{eq:H-PHsym}
\begin{align}
H_{\mathrm{coupling}} & =\!\sum_{n=1}^{N}\!\!\begin{pmatrix}d_{+}^{\dagger} & \!d_{-}^{\dagger}\end{pmatrix}\!\sqrt{2}\!\begin{pmatrix}\mathbf{v}_{n+} & \begin{smallmatrix}\!\!0\, & 0\end{smallmatrix}\\
\begin{smallmatrix}0 & \,0\!\!\end{smallmatrix} & \mathbf{v}_{n-}
\end{pmatrix}\!\begin{pmatrix}\mathbf{a}_{n+}\\
\mathbf{a}_{n-}
\end{pmatrix}+\mathrm{H.c.},\label{eq:H-Coupling-PHsym}\\
H_{\mathrm{bath}} & =\!\sum_{n=1}^{N}\!\!\begin{array}{c}
\begin{pmatrix}\mathbf{a}_{n+}^{\dagger} & \mathbf{a}_{n-}^{\dagger}\end{pmatrix}\\
\\
\end{array}\!\!\!\begin{pmatrix}0 & \mathbf{\boldsymbol{\mathcal{E}}}_{n}\\
\boldsymbol{\mathbf{\mathcal{E}}}_{n} & 0
\end{pmatrix}\!\begin{pmatrix}\mathbf{a}_{n+}\\
\mathbf{a}_{n-}
\end{pmatrix},\label{eq:H-Bath-PHsym}
\end{align}
\end{subequations}
where $\mathbf{v}_{n+}$ and $\mathbf{v}_{n-}$ are, respectively,
the top and bottom rows of $\mathbf{V}_{n}=\left(\!\begin{smallmatrix}\mathbf{v}_{n+}\\
\mathbf{v}_{n-}
\end{smallmatrix}\!\right)$.

We can now embed Eq.\,(\ref{eq:H-PHsym}) into a single-particle
Hamiltonian. We choose a seemingly peculiar order of rows (and columns):
The first row corresponds to the odd impurity mode\textbf{ $d_{-}$},
rows $2\to2N+1$ correspond to the particle-hole even bath modes $\mathbf{a}_{n+}$,
row $2N+2$ corresponds to the even impurity mode $d_{+}$, and rows
$2N+3\to4N+2$ correspond to the particle-hole odd bath modes $\mathbf{a}_{n-}$.
Thus, the single-particle Hamiltonian has the following structure
\begin{equation}
H=\begin{pmatrix}O & M\\
M^{\dagger} & O
\end{pmatrix},\quad M=\begin{pmatrix}0 & \mathbf{v}_{1-} & \cdots & \mathbf{v}_{N-}\\
\mathbf{v}_{1+}^{\dagger} & \boldsymbol{\mathcal{E}}_{1} &  & \vspace{-0.8em}\\
 &  &  & \hspace{-2em}\begin{smallmatrix}0 & \cdots & 0\\
 & \ddots & \vdots\\
 &  & 0
\end{smallmatrix}\vspace{-0.8em}\\
\vdots & \vspace{-0.8em} & \ddots\\
 &  & \hspace{-2em}\begin{smallmatrix}0\\
\vdots & \ddots\\
0 & \cdots & 0
\end{smallmatrix}\vspace{-0.8em}\\
\mathbf{v}_{N+}^{\dagger} &  &  & \boldsymbol{\mathcal{E}}_{N}
\end{pmatrix},\label{eq:H-PHsym-Mat}
\end{equation}
as shown in Fig.\,\ref{fig:Tridiag-Symmetric}, with $M$ a $\left(2N+1\right)\times\left(2N+1\right)$
matrix and $O$ a zeros matrix of the same dimension. $M$ is not
a Hermitian (symmetric, assuming real coefficients) matrix, but can
still be brought to tridiagonal form by applying different unitary
(orthogonal) transformations from the left and right 
\begin{equation}
\mathcal{M}=U_{+}MU_{-}^{\dagger}=\begin{pmatrix}0 & t_{0-} & \begin{smallmatrix}0\end{smallmatrix} & \begin{smallmatrix}\cdots\end{smallmatrix} & \begin{smallmatrix}0\end{smallmatrix}\\
t_{0+} & \eta_{1} & t_{1+} &  & \begin{smallmatrix}\vdots\end{smallmatrix}\\
\begin{smallmatrix}0\end{smallmatrix} & t_{1-} & \eta_{2} & t_{2-} & \begin{smallmatrix}0\end{smallmatrix}\\
\begin{smallmatrix}\vdots\end{smallmatrix} &  & t_{2+} & \eta_{3} & \ddots\\
\begin{smallmatrix}0\end{smallmatrix} & \begin{smallmatrix}\cdots\end{smallmatrix} & \begin{smallmatrix}0\end{smallmatrix} & \ddots & \mathbf{\ddots}
\end{pmatrix},\label{eq:Tridiag-m}
\end{equation}
with the labels of the matrix elements explained below. Then, specifying
the unitary transformation $\mathcal{U}=\left(\begin{smallmatrix}U_{+} & 0\\
0 & U_{-}
\end{smallmatrix}\right)$ we can transform $H$ into 
\begin{equation}
\mathcal{U}H\mathcal{U}^{\dagger}=\begin{pmatrix}O & \mathcal{M}\\
\mathcal{M}^{\dagger} & O
\end{pmatrix},
\end{equation}
as shown in Fig.\,\ref{fig:Tridiag-Symmetric-Chain}. In the transformed
Hamiltonian, rows 1 and $2N+2$ still correspond to $d_{-}$ and $d_{+}$,
respectively. If we identify rows $\footnotesize{\begin{cases}
n+2 & \text{even }n\\
n+3+2N & \text{odd }n
\end{cases}}$ with $f_{n+}$ and rows $\footnotesize{\begin{cases}
n+3+2N & \text{even }n\\
n+2 & \text{odd }n
\end{cases}}$ with $f_{n-}$, we find that the elements of $\mathcal{M}$ in Eq.\,(\ref{eq:Tridiag-m})
give us $\mathbf{T}_{n}=\left(\begin{smallmatrix}t_{n+} & 0\\
0 & t_{n-}
\end{smallmatrix}\right)$ and $\mathbf{E}_{n}=\left(\begin{smallmatrix}0 & \eta_{n}\\
\eta_{n}^{*} & 0
\end{smallmatrix}\right)$. Thus, we indeed get the ladder structure as advertised.

Let us study the obtained hopping amplitudes, which are plotted for
an example case in Fig.\,\ref{fig:Tridiag-Hopping}. Without loss
of generality we can choose real and positive ladder-rail amplitudes
$t_{n\pm}$, and let any required phase fall on the rung amplitudes
$\eta_{n}$. The latter are also real {[}due to the real coefficients
in Eq.\,(\ref{eq:H-Star}){]}, and so this phase amounts at most
to a sign. Observe that the hopping amplitudes fall into two distinct
regimes, separated by the energy scale corresponding to the interimpurity
distance, $v_{F}/R$. This distinction is further emphasized in Fig.\,\ref{fig:Tridiag-Ladder},
in which the width of the different links corresponds to the hopping
amplitudes when rescaled by $\Lambda^{-n/2}$.

Small values of $n$ correspond to wavelengths shorter than the distance
between the impurities, or conversely, impurities which are very far
apart with respect to the wavelengths considered, so that we expect
them to hardly affect each other. We indeed observe that the rail
amplitudes decay exponentially $t_{n\pm}\sim\Lambda^{-n/2}$, with
$t_{n+}$ and $t_{n-}$ of the same order for a given $n$, leading
to two well-defined Wilson chains, while the rung amplitudes $\eta_{n}\approx v_{F}/R$
are constant and small, resulting in weak coupling between the chains,
as depicted in the left part of Fig.\,\ref{fig:Tridiag-Ladder}.
Thus, in the limit of extremely distant impurities $R\to\infty$,
the chains completely decouple, as expected. 

Large $n$ correspond to wavelengths larger than the distance between
the impurities, or conversely, impurities which are very close to
each other. In the limit $R\to0$, we expect both impurities to couple
to the same bath mode, resulting in a single Wilson chain.\footnote{One has to be careful with the order of limits with respect to the
high-energy cutoff $D$. We assume $R\gg v_{F}/D$, which even in
the limit of small $R$ (and $D\to\infty$) retains a finite short
chiral wire segment between the two impurities, such that inversion
and time reversal symmetry remain independently broken (but PT is
conserved). This assumption is also implied, although not explicitly
stated, in the multi-impurity Bethe-ansatz solution.} Indeed, we observe that both the rail and the rung amplitudes decay
exponentially $\sim\Lambda^{-n/2}$, but we have alternating weak
rail amplitudes, and rung amplitudes which are of order of the strong
rail, as depicted in the right part of Fig.\,\ref{fig:Tridiag-Ladder}.
This corresponds to a single Wilson chain snaking through the ladder,
with small next-next-nearest-neighbor corrections. Note that this
snaking chain has two sites for each $n$. Thus, renumbering the sites
along the snake by index $m$, we see that the hopping amplitudes
decay along the effective chain as $\sim\sqrt{\Lambda}^{\,-m/2}$,
which can be interpreted as an effective smaller logarithmic discretization
parameter $\sqrt{\Lambda}$ in this regime. Note also that although
this has no consequences, one can get a slightly cleaner picture by
transferring the alternating signs of the rung amplitudes to the weak
links, so that all weak links are negative, and the snaking chain
has positive amplitudes.

Before moving on, we return to shortly discuss the freedom of applying
arbitrary $2\times2$ unitary rotations $u_{n}$ between the two modes
at each level of the ladder (defining $\mathbf{f}_{-1}\equiv\mathbf{d}$
as the impurity modes)
\begin{equation}
\mathbf{f}_{n}\to u_{n}^{\dagger}\mathbf{f}_{n},\ \mathbf{E}_{n}\to u_{n}^{\dagger}\mathbf{E}_{n}u_{n},\ \mathbf{T}_{n}\to u_{n-1}^{\dagger}\mathbf{T}_{n}u_{n}.
\end{equation}
Setting $\mathbf{E}_{n}=\eta_{n}\sigma_{x}$ and $\mathbf{T}_{n}=\frac{t_{n+}+t_{n-}}{2}\mathbf{1}+\frac{t_{n+}-t_{n-}}{2}\sigma_{z}$
with real coefficients, we have fully utilized this freedom (up to
signs), and this comes naturally in the construction of Eq.\,(\ref{eq:H-Bath-PHsym})
{[}or Figs.\,\ref{fig:Tridiag-Symmetric}\, and \ref{fig:Tridiag-Symmetric-Chain}{]}.
However, we can return to the block-tridiagonal structure in Fig.\,\ref{fig:Tridiag-Block-Chain}
and play with the $u_{n}$ rotations, investigating their effect on
the Wilson ladder. An immediate observation is that if we wish to
retain zero onsite energies, we can modify the phases of the ladder
rungs, but not their amplitudes. We can also take the impurities back
to the real-space basis by inverting Eq.\,(\ref{eq:even-odd-transformation}),
i.e., choosing $u_{-1}=u\equiv\frac{1}{\sqrt{2}}\left(\begin{smallmatrix}1 & 1\\
i & -i
\end{smallmatrix}\right)$:
\begin{equation}
\mathbf{d}=\begin{pmatrix}d_{+}\\
d_{-}
\end{pmatrix}\to u^{\dagger}\mathbf{d}=\frac{1}{\sqrt{2}}\begin{pmatrix}1 & -i\\
1 & i
\end{pmatrix}\begin{pmatrix}d_{+}\\
d_{-}
\end{pmatrix}=\begin{pmatrix}d_{1}\\
d_{2}
\end{pmatrix}.
\end{equation}
This comes at the cost of a nondiagonal and complex $\mathbf{T}_{0}$,
compromising the ladder structure. However, we can proceed to choose
$u_{0}=u$, transforming the first ladder sites $\mathbf{f}_{0}$
to the real-space basis, resulting in purely imaginary rung hopping
amplitudes (of magnitude $\eta_{0}$), and, assuming equal rails $t_{0+}=t_{0-}$,
regaining the diagonal structure of $\mathbf{T}_{0}$. Now the next
level has cross terms, but as long as we have equal rails,\textbf{
}$t_{n+}=t_{n-}$, i.e., at the earlier levels of the ladder, as demonstrated
in Fig.\,\ref{fig:Tridiag-Hopping}, sequentially applying $u_{n}=u$
will preserve the ladder structure (with purely imaginary rungs).
Once we arrive at the snaking regime, where $t_{n+}\neq t_{n-}$,
such transformations will start compromising the ladder structure.
However, in that regime, it is indeed more natural to remain in the
even-odd basis, with both impurities coupling to the even mode, and
the odd mode (approximately) decoupling, resulting in a single chain.

As a final note, we mention that the entire procedure is quite sensitive
to the oscillations in the hybridization function, and in order to
retain double-precision (${\sim}10^{-16}$) accuracy in the Wilson
ladder couplings, we had to resort to higher precision arithmetics
(with ${\sim}10^{-32}$ accuracy), both in the discretization, i.e.,
in the evaluation of Eqs.\,(\ref{eq:Vn2})--(\ref{eq:CampoOliveira-En}),
as well as in the tridiagonalization. Once the Wilson ladder is obtained,
the couplings can be cast back to double precision, with the numerical
iterative diagonalization performed at that level.

\subsection{Iterative diagonalization\label{subsec:Iterative-Diagonalization}}

In Sec.\,\ref{sec:Results}, we will reintroduce interactions at
the impurities, so that the system can no longer be solved in the
single-particle basis. Thus, we will need to proceed with iterative
diagonalization. As this is standard NRG procedure, we only give a
quick overview for completeness, and refer the reader to Refs.\,\citep{bullaNumericalRenormalizationGroup2008,weichselbaumTensorNetworksNumerical2012}
for further details. We then comment on several points which proved
important in the implementation for our specific problem.

The challenge which the iterative diagonalization procedure comes
to address is the exponential scaling of the Hilbert space size with
the number of bath (and impurity) modes, so that full exact diagonalization
is feasible only for very small systems. However, thanks to the logarithmic
discretization, we can use the low-energy spectrum of a finite Wilson
chain (or ladder) of length $n^{\prime}$ in order to calculate the
low-energy spectrum of a chain (or ladder) of length $n^{\prime}+1$.
Thus, we start with a short chain, which we can fully diagonalize,
(i) keep only a fixed number $N_{K}$ of low-energy states, (ii) add
a new site, enlarging the considered Hilbert space, (iii) diagonalize
the new Hamiltonian, and return to step (i), proceeding iteratively
to construct a chain of any desired length, while for each chain length
we have only a fixed number, $N_{K}$, of low-energy states. The reason
this works is the exponential decay of the hopping amplitudes along
the chain, which induces energy scale separation, i.e., the coupling
of site $n^{\prime}+1$ to site $n^{\prime}$ serves only as a small
perturbation to a chain of length $n^{\prime}$. Thus, it can only
mix states of similar energy, implying that the $N_{K}$ low-energy
states of a chain of length $n^{\prime}+1$ indeed only depend on
states already contained in the $N_{K}$ low-energy states of the
chain of length $n^{\prime}$. We therefore obtain an effective low-energy
Hamiltonian for any desired energy scale (or chain length), and the
iterative diagonalization is really a numerical implementation of
a renormalization group flow from high to low energies (or temperatures).
We can then track the changes in the effective Hamiltonians in order
to identify fixed points, and also extract thermodynamic and (zero
or finite temperature) static and dynamical quantities.

When dealing with two-impurity problems (also in the bidirectional
case), each channel is mapped to two effective channels (odd and even).
As the computational cost of NRG, i.e., the required number of kept
states, $N_{K}$, for a desired precision, scales exponentially with
the number of channels, this proves a major challenge. When the different
channels are decoupled, as in single-impurity-multichannel or two-impurity-bidirectional-channel
calculations, one can interleave the different channels \citep{mitchellGeneralizedWilsonChain2014,stadlerInterleavedNumericalRenormalization2016},
introducing each new site channel by channel, and reducing the number
of states back to $N_{K}$ after each channel. In order to preserve
energy scale separation after the introduction of each channel, the
different channels must be shifted with respect to each other by a
channel-dependent $z$ shift. However, once the channels are coupled,
as in our case, it is not clear how to introduce this shift. Thus,
in each iteration we must introduce the even and odd channels, i.e.,
the two ladder sites at the new level, together.

The computational cost can also be reduced significantly by exploiting
different symmetries. The interacting model considered in this paper
(in Sec.\,\ref{sec:Results}) exhibits $\mathrm{SU(2)_{charge}\otimes SU(2)_{spin}}$
symmetry, with the former due to charge conservation together with
particle-hole symmetry. As we use the contemporary formulation of
NRG as a matrix-product-state algorithm \citep{weichselbaumTensorNetworksNumerical2012},
exploiting these symmetries can be delegated to the underlying tensor-network
library, in our case \textsc{QSpace} \citep{weichselbaumTensorNetworksNumerical2012,weichselbaumNonabelianSymmetriesTensor2012,weichselbaumXsymbolsNonAbelianSymmetries2020},
which treats Abelian and non-Abelian symmetries on equal footing.
This requires formulating the problem (e.g., the Hamiltonian) in terms
of operators which respect the symmetry. It is usually straightforward,
but in our case incurs several technical issues, which are addressed
in Appendix~\ref{sec:App-Exploiting-Symmetries}. The considered
model is also PT symmetric, and so when written in terms of PT-symmetric
fermionic operators (as discussed in the Sec.\,\ref{subsec:Tridiagonalization}
for the Wilson ladder and in Appendix~\ref{sec:App-Exploiting-Symmetries}
for the interaction terms), must have real coefficients. As the fermionic
operators themselves can also be written as real tensors, we can resort
to real (double-precision) arithmetics, which result in a factor of
${\sim}4$ speedup with respect to complex (double-precision) arithmetics.

\section{Noninteracting Benchmark\label{sec:Noninteracting-Results}}

\begin{figure*}
\includegraphics[width=1\textwidth]{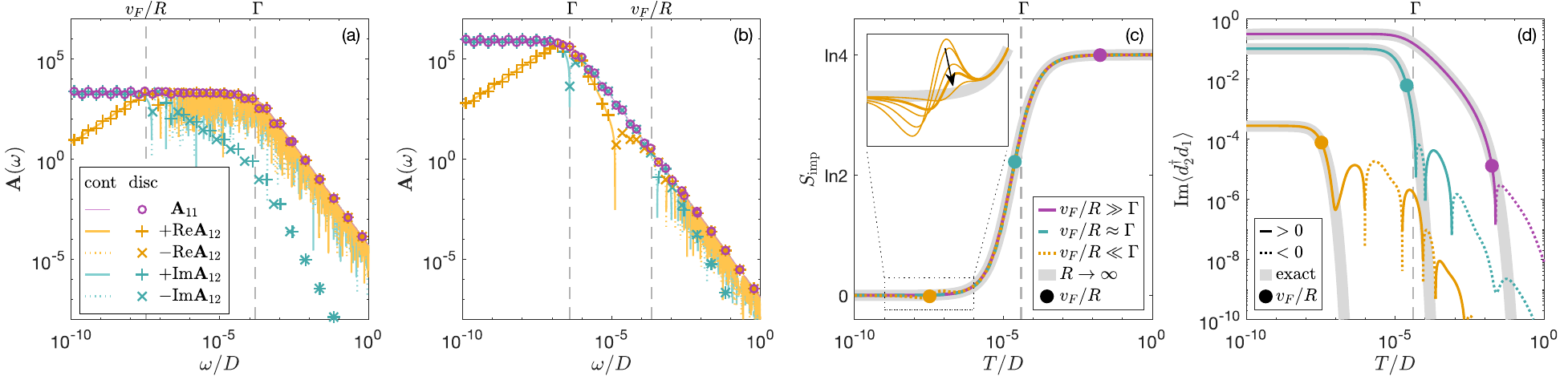}

\phantomsubfloat{\label{fig:NI-Spec-Large-Gamma}}\phantomsubfloat{\label{fig:NI-Spec-Small-Gamma}}\phantomsubfloat{\label{fig:NI-Entropy}}\phantomsubfloat{\label{fig:NI-CorrT}}
\begin{centering}
\vspace{-2.5em}
\par\end{centering}
\caption{Comparison of the Wilson-ladder numerical discretization (with $\Lambda=3$)
and the analytical continuum limit for a noninteracting two-impurity
model. (a,b) The elements of the impurity spectral function matrix
for $3.3\times10^{-8}D=v_{F}/R\ll\Gamma=1.5\times10^{-4}D$ and $2.2\times10^{-4}D=v_{F}/R\gg\Gamma=3.9\times10^{-7}D$,
respectively, taking $z=1$. The continuum limit result according
to Eq.\,(\ref{eq:NI-Spectral-Function}) is indicated by solid (dotted)
lines for positive (negative) values, and discrete system results
are indicated by the different markers. (c) The impurity entropy is
plotted as a function of temperature for fixed $\Gamma\approx4\times10^{-5}D$
($t_{0}=0.005D$) and various finite distances $R$ (with $v_{F}/R$
indicated by circles) as well as $R\to\infty$ (shaded gray), averaged
over four $z$ shifts. The inset demonstrates the elimination of a
numerical artifact at $v_{F}/R\ll\Gamma$ upon reducing the logarithmic
discretization parameter $\Lambda=9,6,4,3,2$. (d) Imaginary part
of the temperature-dependent static impurity-impurity correlator for
the same parameters as in (c), calculated after discretization. The
continuum limit results (shaded gray) for the same distances are plotted
as a reference.\label{fig:Noninteracting-Results}}
\end{figure*}

We can now test the quality of the presented mapping by looking at
how well it reproduces the chiral behavior in the exactly-solvable
noninteracting case. Note that the tridiagonalization procedure in
Sec.\,\ref{subsec:Tridiagonalization} is exact, and the iterative
diagonalization in Sec.\,\ref{subsec:Iterative-Diagonalization}
can in principle be brought to any desired accuracy. Thus, we are
only testing the quality of the discretization presented in Sec.\,\ref{subsec:Discretization},
and calculate all quantities from the single-particle Hamiltonian
(see Appendix~\ref{sec:App-Single-Particle-Calculations}). However,
the results have a clearer interpretation when considering the structure
of the Wilson ladder. Assuming the large bandwidth limit, i.e., the
high-energy cutoff $D$ much larger than all other scales, we are
left with two characteristic energy scales in the problem: the inverse
interimpurity distance $v_{F}/R$, and $\Gamma=\frac{\tilde{t}_{0}^{2}}{2v_{F}}=\frac{\pi t_{0}^{2}}{2D}$,
which quantifies the hybridization strength, and plays the role of
the Kondo temperature in the noninteracting limit.

We start by comparing the continuum-limit expression for the impurity
spectral function with its discrete version. In the limit $D\to\infty$
(while keeping $\Gamma$ and $v_{F}/R$ finite), the spectral function
is given by Eq.\,(\ref{eq:Noninteracting-Spectral-Function}), with
its components rewritten here explicitly
\begin{subequations}
\label{eq:NI-Spectral-Function}
\begin{align}
\mathbf{A}_{11}\left(\omega\right)=\mathbf{A}_{22}\left(\omega\right) & =\frac{\Gamma/\pi}{\omega^{2}+\Gamma^{2}},\\
\mathbf{A}_{12}\left(\omega\right)=\mathbf{A}_{21}^{*}\left(\omega\right) & =\frac{\Gamma/\pi}{\omega^{2}+\Gamma^{2}}\frac{\omega-i\Gamma}{\omega+i\Gamma}e^{i\omega R/v_{F}}.
\end{align}
\end{subequations}
Observe that the diagonal elements (which are real by construction)
are equal, $\mathbf{A}_{11}\left(\omega\right)=\mathbf{A}_{22}\left(\omega\right)$,
independent of $R$, and equal to the spectral function of a single-impurity
resonant-level model (i.e., the $R\to\infty$ limit). This is a key
signature of chirality, and any local perturbation, e.g., an onsite
potential or different coupling $t_{0}$, at either of the impurities
will not affect the other impurity {[}i.e., its spectral function---see
discussion around Eq.\,(\ref{eq:RLM-asymmetry}){]}. The discrete
version of the spectral function can be evaluated at the single-particle
eigenenergies, with its different unique components plotted in Figs.\,\ref{fig:NI-Spec-Large-Gamma}
and \ref{fig:NI-Spec-Small-Gamma} for $v_{F}/R\ll\Gamma$ and $v_{F}/R\gg\Gamma$,
respectively. The diagonal (local) elements are correctly captured
at all energy scales, as indicated by the circles. The off-diagonal
elements are trickier: Their magnitude follows the same Lorentzian
as the diagonal terms and is correctly captured at all scales, but
they also have a phase, which for $\omega<v_{F}/R$ is correctly captured,
but for $\omega\gg v_{F}/R$ is highly oscillatory, and thus cannot
be captured by the logarithmic discretization.

We then turn to consider the impurity contribution to the entropy
$\mathcal{S}_{\mathrm{imp}}\equiv\mathcal{S}-\mathcal{S}_{0}$, defined
as the difference between $\mathcal{S}$, the entropy of the full
system, and $\mathcal{S}_{0}$, the entropy of the bath when decoupled
from the impurity. In Fig.\,\ref{fig:NI-Entropy}, we plot $\mathcal{S}_{\mathrm{imp}}$
as a function of temperature (taking $k_{B}=1$) for different interimpurity
distances $R$. Let us first examine the $R\to\infty$ limit (gray),
which corresponds to two copies of single-impurity models, so that
the impurity entropy is additive. At high temperatures the two impurities
are effectively decoupled from the system, so that each can be either
full or empty, yielding four possible impurity configurations, resulting
in $\mathcal{S}_{\mathrm{imp}}=\ln4$. At low temperatures each impurity
is strongly coupled to the bath, so that there are no free impurity
degrees of freedom and we have $\mathcal{S}_{\mathrm{imp}}=0$. The
transition between the two regimes naturally occurs at the scale $\Gamma$.
Going to finite $R$, with $v_{F}/R$ larger (solid purple), smaller
(dotted orange), and of order (dashed green) of $\Gamma$, we observe
that as long as $v_{F}/R$ is far below the cutoff, $\mathcal{S}_{\mathrm{imp}}$
is completely independent of $R$. Thus, the $R\to\infty$ behavior
in fact holds for arbitrary distances between the impurities, and
we expect it to apply for any (global) impurity thermodynamic property.
Note that for $v_{F}/R<\Gamma$, a kink appears in the entropy at
$T\sim v_{F}/R$, as shown in the inset. However, this kink, which
will also appear in the interacting case when $v_{F}/R<T_{K}$, is
a numerical artifact, and disappears upon shrinking $\Lambda$.

Lastly we turn to consider the thermal correlation functions between
the impurities' fermionic operators:
\begin{align}
\left\langle d_{m^{\prime}}^{\dagger}d_{m}\right\rangle _{T} & =\frac{1}{\beta}\sum_{i\nu_{l}}\mathbf{G}_{mm^{\prime}}\left(i\nu_{l}\right)e^{i\nu_{l}0^{+}}\label{eq:NI-CorrT}\\
 & =\int_{-\infty}^{\infty}\mathbf{A}_{mm^{\prime}}^{\!\left(T=0\right)}\left(\omega\right)f_{\mathrm{FD}}\left(\omega;T\right)d\omega,\nonumber 
\end{align}
where $\mathbf{G}\left(i\nu_{l}\right)=\footnotesize{\begin{cases}
\mathbf{G}^{R}\left(i\nu_{l}\right) & \nu_{l}>0\\
\mathbf{G}^{A}\left(i\nu_{l}\right) & \nu_{l}<0
\end{cases}}$ is the Matsubara Green function, with $\nu_{l}=\left(2l+1\right)\pi/\beta$
the fermionic Matsubara frequencies and $\mathbf{G}^{R}=\mathbf{G}^{A\dagger}$
given by Eq.\,(\ref{eq:Noninteracting-Retarded-Green-Function}).
In the second line, $\mathbf{A}$ is the zero temperature spectral
function, and the temperature dependence enters only through the Fermi-Dirac
distribution, $f_{\mathrm{FD}}$. From symmetry considerations we
expect both impurities to be at half-filling, i.e., $\left\langle d_{m}^{\dagger}d_{m}\right\rangle _{T}=\frac{1}{2}$
for all temperatures, and this is indeed reproduced (not shown). Thus,
the only relevant information appears in the off-diagonal terms, which
are evaluated numerically both in the continuum limit and for the
discrete system. We find that at temperatures $T<v_{F}/R$ the correlations
are captured correctly by the discrete system. However, as is evident
from the second line of Eq.\,(\ref{eq:NI-CorrT}), the oscillations
in the spectral function impair the high-temperature results, which
should fall exponentially, but instead oscillate around zero, decaying
at a slower rate.

Before moving to the interacting case, we draw the main conclusions
from this section. Our scheme reliably reproduces the thermodynamics,
as well as local dynamics, but its validity regarding interimpurity
properties depends on the interimpurity distance. As one would have
expected, the discretized NRG scheme does not reproduce the fast oscillations
at energies exceeding $v_{F}/R$. However, these high-energy discrepancies
are washed out as one looks at low energies, where the spectral functions
are accurately reproduced. Most importantly, these discrepancies do
not compromise chirality, which is preserved on all energy scales,
as demonstrated by the distance independence of the thermodynamic
and local spectral quantities at all energy scales.

\section{Chiral Two-Impurity Kondo Model\label{sec:Results}}

We are now ready to proceed to a spinful system and introduce local
Coulomb interactions at each impurity. We start with a two-impurity
Anderson model
\begin{equation}
H_{\mathrm{imp}}=\sum_{m=1,2}\left[\varepsilon_{d}\left(n_{m\uparrow}+n_{m\downarrow}\right)+Un_{m\uparrow}n_{m\downarrow}\right],\label{eq:H-Imp-Anderson}
\end{equation}
where $n_{m\sigma}=d_{m\sigma}^{\dagger}d_{m\sigma}$ is the occupation
operators of impurity $m$, and $H_{\mathrm{coupling}}$ and $H_{\mathrm{bath}}$
are as in Eq.\,(\ref{eq:H}) with the introduction of spin indices.
In order to restrict the parameter space, in this work, we focus on
the particle-hole symmetric regime, $\varepsilon_{d}=-U/2$, and take
the local-moment limit $U\gg\Gamma$. Thus, each impurity retains
a spin-half degree of freedom $\vec{S}_{m}$, and, employing a Schrieffer-Wolff
transformation \citep{schriefferRelationAndersonKondo1966a}, we arrive
at the two-impurity chiral Kondo model, with $H_{\mathrm{bath}}$
as before and
\begin{equation}
H_{\mathrm{imp}}+H_{\mathrm{coupling}}=J\vec{S}_{1}\cdot\vec{s}\left({-}\tfrac{R}{2}\right)+J\vec{S}_{2}\cdot\vec{s}\left({+}\tfrac{R}{2}\right),\label{eq:H-Kondo}
\end{equation}
where $\vec{s}\left(x\right)=\frac{1}{2}\sum_{\sigma\sigma^{\prime}}\psi_{\sigma}^{\dagger}\!\left(x\right)\vec{\sigma}_{\sigma\sigma^{\prime}}\psi_{\sigma^{\prime}}\!\left(x\right)$
is the bath spin operator at position $x$, $\vec{\sigma}=\left(\sigma_{x},\sigma_{y},\sigma_{z}\right)$
are the Pauli matrices, and $J=\frac{16}{\pi}D\Gamma/U=8t_{0}^{2}/U$
is the Kondo coupling.\footnote{In practice we employ a useful NRG trick: Instead of implementing
the (two-impurity) Kondo Hamiltonian, we use the (two-impurity) Anderson
Hamiltonian, setting $U\gg\Gamma\gg D$, which numerically implements
the Schrieffer-Wolf transformation.} $H_{\mathrm{bath}}$ is unaffected by the Schrieffer-Wolff transformation,
and can still be mapped to the Wilson ladder in Eq.\,(\ref{eq:H-Bath-Chain}).
Note that the interaction terms of both models, as defined above,
are invariant under both inversion and time reversal, and so the full
Hamiltonian is still PT symmetric.

As a baseline we consider infinitely spaced impurities, i.e., $R\to\infty$,
in which case the system corresponds to two separate copies of a single-impurity
Kondo (or Anderson) model. Each copy undergoes a regular Kondo effect,
with each spin fully screened at temperatures below the Kondo temperature
\begin{equation}
T_{K}=D\sqrt{\rho J}\exp\left(-\frac{1}{\rho J}\right),
\end{equation}
where $\rho=\frac{1}{2D}$ is the local bath density of states (at
the Fermi energy). Thus, whenever we refer to the $R\to\infty$ limit,
we actually run a standard NRG calculation for the single-impurity
problem, and multiply results by 2 (due to the two impurities) when
necessary. In what follows, we then hold the single-impurity Kondo
temperature $T_{K}$ fixed, and introduce a finite interimpurity separation
$R$.

\begin{figure*}
\begin{centering}
\includegraphics[width=1\textwidth]{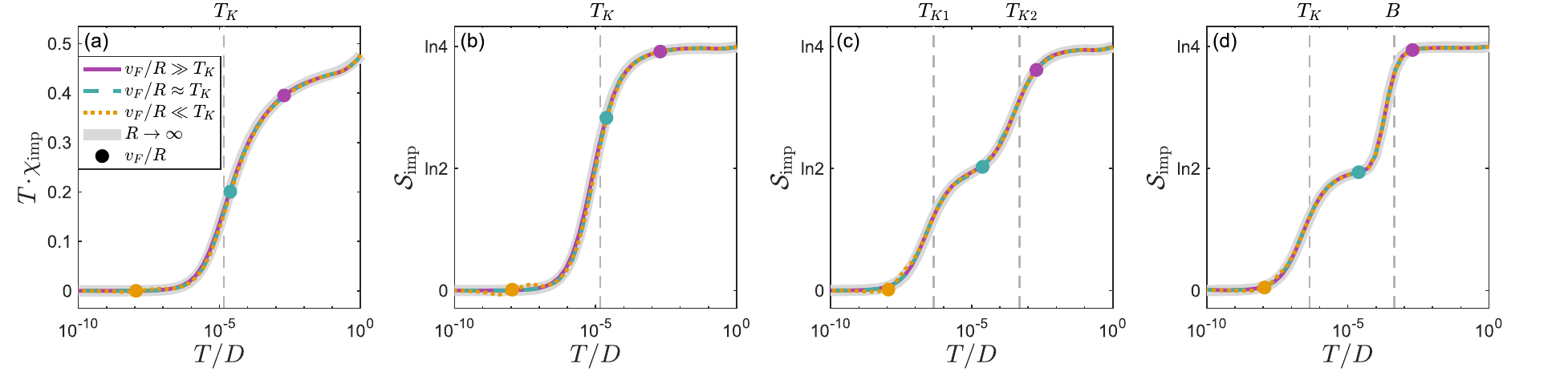}
\par\end{centering}
\begin{centering}
\phantomsubfloat{\label{fig:Int-Tchi}}\phantomsubfloat{\label{fig:Int-Sent}}\phantomsubfloat{\label{fig:Int-SentJ}}\phantomsubfloat{\label{fig:Int-SentB}}\phantomsubfloat{\label{fig:Int-Kondo}}
\par\end{centering}
\begin{centering}
\vspace{-2.5em}
\par\end{centering}
\caption{Distance independence of thermodynamic properties of the chiral two-impurity
Kondo model. The Kondo coupling $J$ is fixed, with the corresponding
$T_{K}$ indicated by a dashed line. The interimpurity distance is
taken at three typical values $DR/v_{F}=2\pi\Lambda^{4}$ (purple)
$2\pi\Lambda^{8}$ (green), 2$\pi\Lambda^{15}\text{ (orange)}$ and
in the $R\to\infty$ separate-impurities limit (shaded gray), with
$\Lambda=3$, up to 5000 kept multiplets, and a single $z$ shift
$z=1$. (a) Impurity magnetic susceptibility multiplied by temperature
and (b) Impurity entropy with equal couplings $J/D=0.2$; (c) Impurity
entropy with impurity-dependent couplings $J_{1}/D=0.15$ and $J_{2}/D=0.3$;
(d) Impurity entropy with equal couplings $J/D=0.15$ and a magnetic
field $B=1000T_{K}$ only at the second impurity. \label{fig:Interacting-Local}}
\end{figure*}

We start with global thermodynamic quantities: The magnetic susceptibility
$\chi_{\mathrm{imp}}\equiv\chi-\chi_{0}$ (multiplied by temperature)
in Fig.\,\ref{fig:Int-Tchi} and the impurity entropy $\mathcal{S}_{\mathrm{imp}}\equiv\mathcal{S}-\mathcal{S}_{0}$
in Fig.\,\ref{fig:Int-Sent}, both defined as the difference between
the thermodynamic quantity of the full system ($\mathcal{S}$ or $\chi$)
and that of a decoupled bath ($\mathcal{S}_{0}$ or $\chi_{0}$).
Observe that for all $R$ they follow the universal curve of the single-impurity
spin-$\frac{1}{2}$ Kondo effect (for each impurity, so multiplied
by 2 for both, in shaded gray): At high temperatures we have a free
spin at each impurity, resulting in $\mathcal{S}_{\mathrm{imp}}=\ln4$
entropy and $T\chi\to\frac{1}{2}$ ($\mathcal{S}_{\mathrm{imp}}=\ln2$
and $T\chi\to\frac{1}{4}$ per impurity), while at low temperatures
we go to the strong coupling fixed point, where the spins are fully
screened, with no residual impurity degrees of freedom, resulting
in zero entropy and $T\chi\to0$. We can further probe the additivity
of global quantitates by breaking the symmetry between the impurities,
so that each is expected to contribute differently to the global quantities.
In Fig.\,\ref{fig:Int-SentJ}, we introduce impurity-dependent Kondo
couplings $J_{1}$ and $J_{2}$, breaking only PT symmetry but retaining
all other symmetries. In Fig.\,\ref{fig:Int-SentB}, we retain equal
couplings but introduce a magnetic field $B$ at one impurity (results
are indifferent to which impurity), thus breaking both $\mathrm{SU(2)_{spin}}$
and PT symmetry. In both cases the resulting impurity entropy is simply
the sum of two single-impurity models (shaded gray), either with different
couplings, or with different magnetic fields (0 and $B$). 

We thus arrive at the conclusion that the local physics of each impurity
is indifferent to the distance between the impurities (also for $T\ll v_{F}/R$),
and each undergoes a single-impurity Kondo effect. Although initially
surprising, this is consistent with the Bethe ansatz \citep{andreiSolutionKondoProblem1983},\footnote{The interimpurity distance does not appear explicitly in the Bethe-ansatz
solution, but finite separation between the impurities is assumed
implicitly by taking the large bandwidth (or continuous channel) limit.} which also accounts for multiple impurities connected to a chiral
channel, and actually has an intuitive explanation: Due to chirality,
electrons cannot backscatter, and thus can only forward scatter first
off the first impurity and then off the second, at most acquiring
a phase shift at each impurity. However, nothing in this description
depends on the distance between the impurities, and so each impurity
contributes a $\frac{\pi}{2}$ phase shift just as in the $R\to\infty$
limit, resulting in a global $\pi$ phase shift, regardless of the
distance. In a more colloquial language, the first impurity does not
``know'' about the second impurity, so that from its perspective
this is a single-impurity problem, and from PT symmetry the same holds
for the second (last) impurity.

\begin{figure*}
\begin{centering}
\includegraphics[width=1\textwidth]{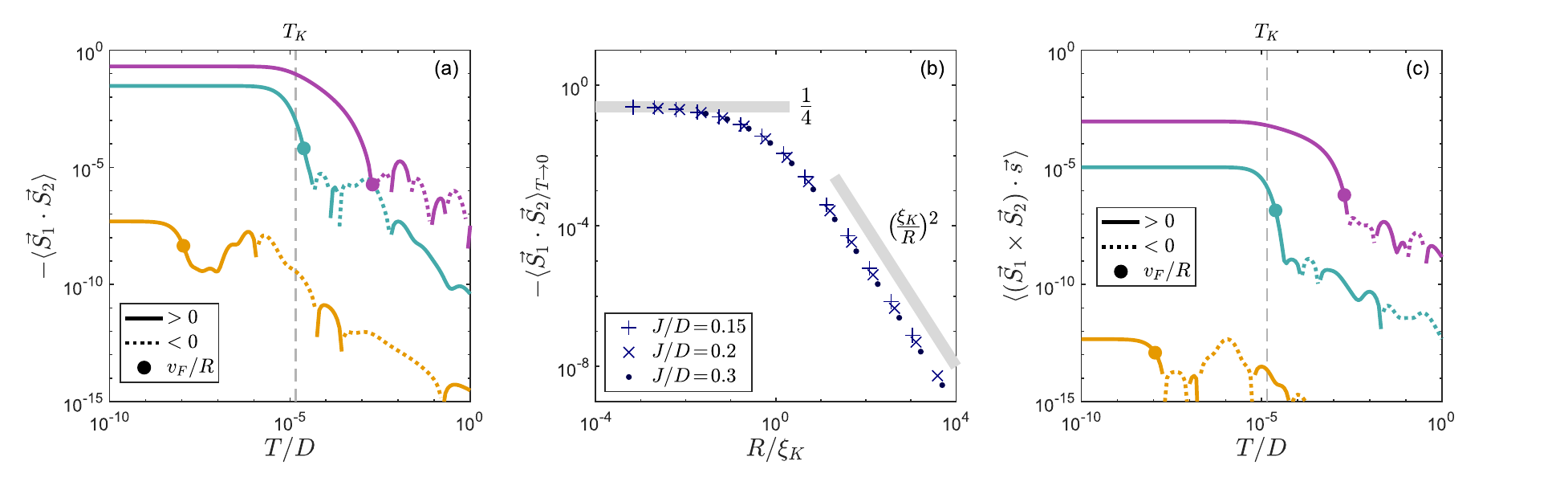}
\par\end{centering}
\begin{centering}
\phantomsubfloat{\label{fig:Int-Corr-Imps}}\phantomsubfloat{\label{fig:Int-Corr-Imps-T0}}\phantomsubfloat{\label{fig:Int-Corr-Chiral}}
\par\end{centering}
\begin{centering}
\vspace{-4em}
\par\end{centering}
\caption{Static temperature-dependent correlations for the chiral two-impurity
Kondo model with the same parameters as in Fig.\,\ref{fig:Interacting-Local}
(unless stated otherwise), and averaged over four $z$ shifts: (a)
Temperature-dependent impurity-impurity correlations; (b) $T\to0$
impurity-impurity correlations for different Kondo-couplings $J$
as a function of the interimpurity distance $R$ rescaled by the Kondo
length scale $\xi_{K}\equiv v_{F}/T_{K}$; (c) Temperature-dependent
chiral correlations which violate both inversion and time reversal
symmetry, but not their product PT symmetry. \label{fig:Interacting-Correlations}}
\end{figure*}

In order to appreciate these results, we first recall the paradigmatic
\citet{doniachKondoLatticeWeak1977} scenario for the bidirectional
case. There, for finite separation between the impurities, the bath
can mediate effective RKKY \citep{rudermanIndirectExchangeCoupling1954,kasuyaTheoryMetallicFerro1956,yosidaMagneticPropertiesCuMn1957}
interactions between the two impurities, i.e., $K\!\!\left(\!R\right)\vec{S}_{1}\!\cdot\vec{S}_{2}$,
which can be either ferromagnetic ($K{<}0$) or antiferromagnetic
($K{>}0$), driving the two impurities towards an impurity-impurity
triplet or singlet, respectively. This competes with the Kondo effect,
which drives each impurity towards a singlet with a collective bath
mode in its vicinity. Due to the finite distance between the impurities,
a single bath generically provides two effective Kondo screening channels,
e.g., even and odd. Thus, in the case of ferromagnetic interaction
(or weak enough antiferromagnetic interaction), one gets a fully Kondo-screened
triplet, while for strong antiferromagnetic interaction one gets a
self-screened interimpurity singlet, which requires no further Kondo
screening \citep{jonesStudyTwoMagnetic1987,silvaParticleHoleAsymmetryTwoImpurity1996}.\footnote{There are exceptions to this rule, e.g., if one of the screening channels
decouples (at low or all temperatures), so that for ferromagnetic
RKKY we are left with an under-screened triplet. One such example
is in the limit of vanishing interimpurity distance $R\ll v_{F}/D=1/k_{\max}$
(for an arbitrary dispersion), so that the two impurities are effectively
at the same site, and thus couple to the same single (even) screening
channel. Another example is a 1D bidirectional channel with interimpurity
distances $R$ that are commensurate with the high-energy cutoff,
i.e., integer $k_{\max}R/\pi$ \citep{lechtenbergRealisticQuantumCritical2017},
which for a linearized dispersion at half-filling (so that $k_{\max}=k_{F}$)
corresponds to impurities that are separated by an even number of
lattice spacings. Note that this condition differs from the one in
Eq.\,(\ref{eq:Commensurate-Cutoff}), since for a chiral channel
$k_{\max}$ is not associated with $k_{F}$ and the underlying lattice
spacing.} The two-impurity chiral model, however, clearly does not adhere to
this picture, as demonstrated, e.g., by the impurity entropy in Fig.\,\ref{fig:Interacting-Local},
which shows that:
\begin{enumerate}
\item[(i)\,] \textit{}The two impurities are fully screened, as $\mathcal{S}_{\mathrm{imp}}^{T\to0}\to0$.
\item[(ii)] \textit{}The screening mechanism is of Kondo nature, as it occurs
at $T_{K}$ for all interimpurity distances.
\end{enumerate}
Point (i) rules out an interimpurity triplet, as, in contrast to a
bidirectional channel, a chiral channel cannot provide two independent
screening channels; hence, if a triplet were to form, it would remain
under-screened. The antiferromagnetic interimpurity correlations (to
be discussed below) in Fig.\,\ref{fig:Int-Corr-Imps} also contradict
such a triplet. Point (ii) rules out a self-screened RKKY singlet,
which would be expected to form at the RKKY energy scale $K\!\!\left(\!R\right)$.
The absence of any distance-related scale in all local quantities
suggests that such effective interactions indeed do not emerge. However,
this is to be expected, as chirality prevents the emergence of RKKY
interactions.

We are thus left with the question: What are the consequences of both
impurities being coupled to the same (chiral) bath? In order to address
it, we need to consider nonlocal quantities, and thus turn to static
temperature-dependent correlations between the impurities, starting
with $\left\langle \vec{S}_{1}\!\cdot\vec{S}_{2}\right\rangle _{T}$
in Fig.\,\ref{fig:Int-Corr-Imps}. At high temperatures we observe
only numerical noise, which is both positive and negative. However,
once we reach temperatures $T<v_{F}/R$, antiferromagnetic (negative)
correlations set in. Observe that the magnitude of the correlations
becomes substantial (with respect to its $T\to0$ value) once $T<T_{K}$,
i.e., in the strong coupling regime. The $T\to0$ value of these correlations
is plotted in Fig.\,\ref{fig:Int-Corr-Imps-T0} as a function of
the interimpurity distance $R$ and for different Kondo couplings
$J$. We find it to be a function of a single parameter $R/\xi_{K}$,
with $\xi_{K}\equiv v_{F}/T_{K}$ the single-impurity Kondo length
scale (which is inevitable, as this is the only remaining scale in
the regime $T\to0,\ T_{K},v_{F}/R\ll J,D$). For large separation,
the correlations decay as $R^{-2}$, while for interimpurity distance
smaller than $\xi_{K}$ they saturate to a constant, which, curiously,
appears to be $-\frac{1}{4}$, the strongest possible correlation
in the absence of interimpurity entanglement (meaning the two-impurity
density matrix remains separable \citep{horodeckiQuantumEntanglement2009}).
In Fig.\,\ref{fig:Int-Corr-Chiral}, we plot a chiral correlation
$\left\langle \left(\vec{S}_{1}\!\times\vec{S}_{2}\right)\cdot\vec{s}\right\rangle _{T}$
with $\vec{s}=\vec{s}\left({-}R/2\right)+\vec{s}\left({+}R/2\right)$,
which follows a similar trend to $\left\langle \vec{S}_{1}\!\cdot\vec{S}_{2}\right\rangle _{T}$.
The interesting property of this quantity is that its average usually
must vanish, as it breaks both inversion and time reversal symmetry
separately. However, it does not break the product of the two symmetries,
i.e., PT symmetry, and is thus allowed to be finite in the chiral
case.

\begin{figure*}
\begin{centering}
\includegraphics[width=1\textwidth]{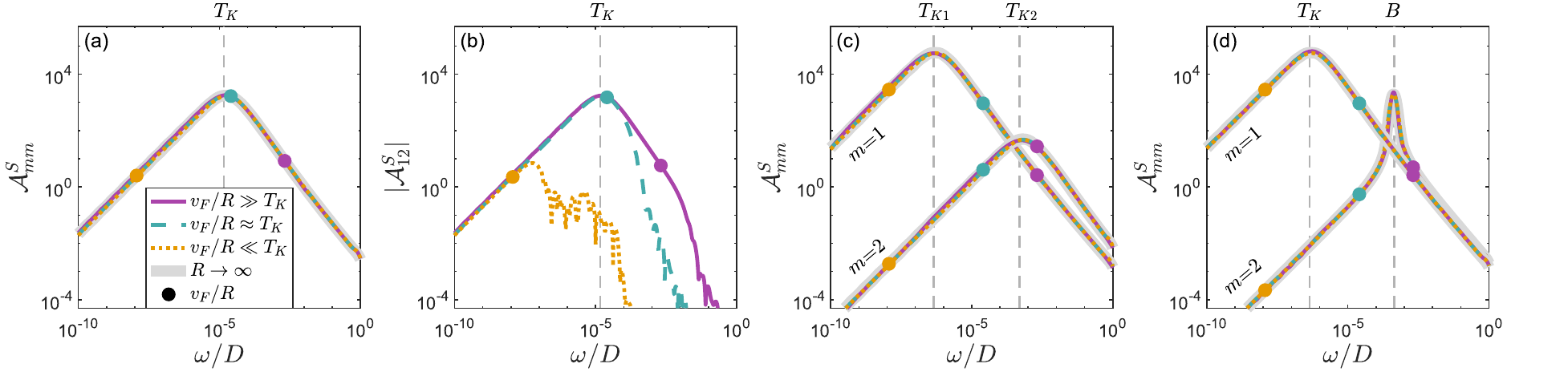}
\par\end{centering}
\begin{centering}
\phantomsubfloat{\label{fig:Int-Spec}}\phantomsubfloat{\label{fig:Int-Spec12}}\phantomsubfloat{\label{fig:Int-SpecJ}}\phantomsubfloat{\label{fig:Int-SpecB}}
\par\end{centering}
\begin{centering}
\vspace{-2.5em}
\par\end{centering}
\caption{Dynamic zero-temperature spin-spin correlations, for the same parameters
as in Fig.\,\ref{fig:Interacting-Local} and averaged over four $z$
shifts\protect\textsuperscript{\ref{fn:Lambda-Extrap}}: (a) Diagonal
(local) $\mathcal{A}_{mm}^{S}\left(\omega\right)=-\frac{1}{\pi}\mathrm{Im}\llangle S_{m}^{+};S_{m}^{-}\rrangle_{\omega}$,
and (b) the envelope of the off-diagonal $\left|\mathcal{A}_{12}^{S}\left(\omega\right)\right|=\left|\mathcal{A}_{21}^{S*}\left(\omega\right)\right|$,
spin spectral function, both calculated for equal couplings $J/D=0.2$;
{[}(c) and (d){]} Response of the local spin spectral function to
the breaking of impurity symmetry, either (c) by impurity-dependent
couplings $J_{1}$ and $J_{2}$, or (d) by taking equal coupling and
introducing a magnetic field at the second impurity. The spectral
function belonging to each impurity is labeled accordingly, and flipping
between the couplings or magnetic fields returns the same result with
flipped labels. \label{fig:Interacting-Spectral}}
\end{figure*}

It should be stressed that the observed correlations do not imply
effective interactions, nor bidirectional causality between the impurities.
Indeed, for temperatures lower than $v_{F}/R$ a (noninteracting)
chiral channel exhibits static correlations between $\psi\left({-}R/2\right)$
and $\psi\left({+}R/2\right)$, even though an event at $+R/2$ cannot
affect $\psi\left({-}R/2\right)$ {[}see the chiral structure of $\boldsymbol{\Sigma}^{R}\left(\omega\right)$
in Eq.\,(\ref{eq:Noninteracting-Self-Energy}){]}. These correlations
decay as $\sim\!R^{-1}$, resulting in $\sim\!R^{-2}$ decay of the
bath spin-spin correlations. Once the Kondo effect sets in, the impurity
spins fuse into the bath, so that the bath correlations are reflected
onto the impurity-impurity correlations. This onset of the correlations
at $T_{K}$ is clearly demonstrated in Fig.\,\ref{fig:Int-Corr-Imps},
and the $\sim\!R^{-2}$ decay of their $T\to0$ value in Fig.\,\ref{fig:Int-Corr-Imps-T0}
matches that of the bath. The saturation to the constant $-\frac{1}{4}$
at short distances can also be understood in this picture: The collective
bath spins to which the impurities fuse have a finite size, $\xi_{K}$,
and so for distances $R<\xi_{K}$, they are highly correlated, resulting
in the two impurity spins also becoming maximally correlated. However,
had the correlations exceed $-\frac{1}{4}$, this would have implied
entanglement between the impurities, so that a measurement of each
would affect the other, in violation of chirality.

Finally, in order to establish that despite the observed correlations,
the local dynamics of the impurities are independent of the distance,
we turn to the (retarded) spin-spin correlator and its corresponding
spectral function $\boldsymbol{\mathcal{A}}^{S}\!\left(\omega\right)$
{[}defined analogously to Eq.\,(\ref{eq:Noninteracting-Spectral-Function}){]}
\begin{subequations}
\begin{align}
\!\!\!\!\llangle S_{m}^{+};S_{m^{\prime}}^{-}\rrangle_{\omega} & \!=\!-i\!\!\int_{-\infty}^{\infty}\!\!\!\!\!\!\Theta\!\left(t\right)\!\left\langle \left[S_{m}^{+}\!\left(t\right)\!,S_{m^{\prime}}^{-}\!\left(0\right)\right]\right\rangle _{T\to0}\!e^{i\omega t}dt,\\
\mathcal{A}_{mm^{\prime}}^{S}\left(\omega\right) & \!=\!-\frac{1}{2\pi i}\left(\llangle S_{m}^{+};S_{m^{\prime}}^{-}\rrangle_{\omega}-\llangle S_{m^{\prime}}^{+};S_{m}^{-}\rrangle_{\omega}^{*}\right).
\end{align}
\end{subequations}
In NRG we naturally obtain $\boldsymbol{\mathcal{A}}^{S}\!\left(\omega\right)$
from the Lehman representation, and plot its diagonal elements $-\frac{1}{\pi}\mathrm{Im\llangle S_{m}^{+};S_{m}^{-}\rrangle_{\omega}}$
and (the envelope of its) off-diagonal elements in Figs.\,\ref{fig:Int-Spec}
and \ref{fig:Int-Spec12}, respectively. Observe that for both impurities
the local correlations {[}in Fig.\,\ref{fig:Int-Spec}{]} coincide
with the single-impurity (or $R\to\infty$, in shaded gray) results
at all frequencies, displaying no signatures of the finite distance.
The envelope of the off-diagonal elements {[}in Fig.\,\ref{fig:Int-Spec12}{]}
coincides with the diagonal elements for $\omega<v_{F}/R$, as in
the noninteracting case, while for higher frequencies it is noisy,
which is unavoidable for a logarithmic discretization, and decays
rapidly.

Due to chirality, we expect the retarded correlator to be upper-triangular,\footnote{This chirality of the retarded correlators, i.e., upper-triangular
structure, also in the presence of interactions, could be shown order-by-order
in Keldysh perturbation theory, in a similar way to showing that the
retardedness is preserved \citep{kamenevFieldTheoryNonEquilibrium2011}.} i.e., $\llangle S_{2}^{+};S_{1}^{-}\rrangle_{\omega}=0$, implying
no response at the first impurity to an event at the second impurity.
Explicitly showing this requires numerically applying a Hilbert transform
to $\boldsymbol{\mathcal{A}}^{S}\!\left(\omega\right)$, which is
infeasible due to the noise in the off-diagonal terms. However, we
can still observe signatures of this chiral property directly in the
spectral function by breaking the symmetry between the two impurities,
i.e., by introducing impurity dependent couplings or magnetic fields,
as shown in Figs.\,\ref{fig:Int-SpecJ} and \ref{fig:Int-SpecB},
respectively. For a chiral (i.e., upper-triangular) retarded correlator,
the local (i.e., diagonal) elements of the corresponding spectral
function should only depend on the local coupling and magnetic field
{[}in analogy to hopping and chemical potential in the noninteracting
case---see Eq.\,(\ref{eq:RLM-asymmetry}) and the following discussion{]}.
This is indeed what we observe, conclusively demonstrating the effectiveness
of our chiral NRG scheme in capturing chiral behavior, and showing
that the local physics of each impurity are simply the single-impurity
Kondo physics of separate impurities.\footnotetext{The discretization
can impair chirality, leaving signatures on the spectral function
for individual $z$ shifts. These are typically washed out after $z$-averaging
or by taking smaller $\Lambda$, as these are better approximations
for the (chiral) continuum limit. The smallest $R$ (in purple) turns
out to be most sensitive to this, and thus require cranking up the
NRG parameters: In Fig.\,\ref{fig:Int-SpecB}, taking $\Lambda=2.5$
(with 7000 kept multiplets) suffices, while in Fig.\,\ref{fig:Int-SpecJ},
we also linearly extrapolate to the continuum limit in $\Lambda=3,2\to1$\label{fn:Lambda-Extrap}
(with 8000 kept multiplets for $\Lambda=2$).}

\section{Summary and Outlook\label{sec:Summary}}

In this paper, we have introduced an extension of Wilson's NRG for
two impurities coupled to unidirectional channel(s) at a finite separation
$R$. It is based on mapping each bath (channel) onto two coupled
tight-binding Wilson chains, or a Wilson ladder, with the impurities
coupled to one end of the ladder. As in a Wilson chain, the energy
scales along the Wilson ladder decay exponentially with the distance
from the impurities. We find that in the vicinity of the impurities,
corresponding to high energies or short wavelengths with respect to
$R$, the two chains are only weakly coupled. On the other hand, at
low energies or long wavelengths with respect to the interimpurity
separation, we effectively get a single Wilson chain (with small next-next
nearest neighbor corrections), which corresponds to the impurities
being coupled to the bath at \textit{almost} the same point. However,
it should be stressed that the impurities are never exactly at the
same point, and we always retain the notion of some finite separation,
which ensures the unidirectionality of the channel. In this work,
we restricted ourselves to two impurities, but in principle the same
procedure should apply to an arbitrary number of impurities, resulting
in a Wilson chain for each impurity, which is then coupled to the
other chains. We have also restricted ourselves to a featureless bath
density of states, corresponding to a linear chiral dispersion, but
the presented method can be applied to an arbitrary chiral or helical
dispersion.

We demonstrated the power of the mapping to the Wilson ladder on a
spinless noninteracting (analytically solvable) system consisting
of two resonant levels coupled to a single unidirectional channel.
We showed that local impurity properties and thermodynamic quantities
are indifferent to the distance between the impurities, and are thus
equal to those of two infinitely spaced, i.e., independent, impurity
models. This is a key signature of chirality. We also successfully
captured interimpurity correlations at low energies, while at high
energies, for which the correlations are highly oscillatory (and decay),
we only captured noise (albeit around zero). Introducing spin and
interactions, we applied the method to two impurities coupled antiferromagnetically
to a single spinful chiral channel. Similarly to the noninteracting
test case, we found that all considered local and thermodynamic quantities
are independent of the interimpurity distance, and thus follow the
universal curves (of two copies) of the single-impurity Kondo effect.
Again, interimpurity correlations were successfully captured at low
energies while oscillatory around zero at high energies. The key observation
is that although such \textit{static} correlations exist, they are
not due to effective (e.g., RKKY) interactions between the impurities,
which are forbidden due to chirality. Hence, interimpurity response
functions (\textit{retarded} correlations) remain chiral, and the
system locally behaves like two separate single-impurity problems,
regardless of the \textit{finite} distance between the two impurities.

Our goal in this paper was to demonstrate how NRG can be applied to
unidirectional channels. Once we have established this, there are
many prospective applications. One interesting direction is to look
at helical systems, even in the single channel case. Such systems
were explored analytically in different limits \citep{gaoInplaneNoncollinearExchange2009,leeElectricalControlInteraction2015,kurilovichIndirectExchangeInteraction2017,yevtushenkoKondoImpuritiesCoupled2018},
but the presented method enables a quantitative study of the transition
between an RKKY phase and a Kondo-screened phase. Another direction
relates to our initial motivation for this work---studying multi-impurity
chiral multichannel Kondo systems. In Ref.\,\citep{lotemManipulatingNonAbelianAnyons2022},
we studied such a system, demonstrating the emergence of decoupled
non-Abelian anyons, for which the fusion channel could be measured
by looking at interimpurity spin correlations. There we relied on
arguments for interimpurity distance independence in order to simplify
the problem by taking the limit $R\to0^{+}$. In this work, we support
these arguments with numerical results for a single-channel model,
but also demonstrate that the distance does enter in nonuniversal
quantities such as zero-temperature value of the interimpurity correlations.
Thus, it would be interesting to study the correlations in the multichannel
case with the help of the Wilson ladder, in order to discern between
the nonuniversal distant-dependent properties, and the universal properties
related to the measurement of the fusion channel.

\textit{Note added}: Recently, we were made aware of a related work
by \citet{ferrerRKKYKondoCrossover2022}, dealing with multiple impurities
on a helical edge, and agreed on a synchronized preprint submission.

\textit{Acknowledgments:} We would like to thank J. von Delft, A.
Weichselbaum, S.-S. Lee, and N. Andrei for fruitful discussions. Numerical
simulations were performed using the \textsc{QSpace} tensor-network
library and accompanying code \citep{weichselbaumTensorNetworksNumerical2012,weichselbaumNonabelianSymmetriesTensor2012,leeAdaptiveBroadeningImprove2016,weichselbaumXsymbolsNonAbelianSymmetries2020}.
Support from the Rutgers Tel-Aviv university joined fund is gratefully
acknowledged. E.S. was further supported by the Synergy funding for
Project No. 941541, ARO (W911NF-20-1-0013), the US-Israel Binational
Science Foundation (BSF) Grant No. 2016255, and the Israel Science
Foundation (ISF) Grant No. 154/19. M.G. was supported by the ISF and
the Directorate for Defense Research and Development (DDR\&D) Grant
No. 3427/21 and by the BSF Grant No. 2020072.

\appendix

\section{Discretization Following Ref.\,\citep{campoAlternativeDiscretizationNumerical2005}\label{sec:App-Campo-Oliveira}}

In this appendix, we will derive explicit expressions for the coefficient
matrices $\mathbf{V}_{n}$ and $\boldsymbol{\mathcal{E}}_{n}$ in
the spirit of the discretization procedure by \citet{campoAlternativeDiscretizationNumerical2005}.
The obtained expressions will differ from those in Sec.\,\ref{subsec:Discretization}
by a unitary rotation of each set of modes $\mathbf{c}_{n}$ {[}see
Eq.\,(\ref{eq:Coeff-Rotation}){]}. At the end of the appendix, we
also compare with the notation of Ref.\,\citep{zitkoEnergyResolutionDiscretization2009}.

We start this derivation by Fourier transforming to $k$-space, in
which the bath Hamiltonian is diagonal, and there we introduce the
high-energy cutoff $D=v_{F}k_{\max}$, where, as mentioned, for a
chiral system $k_{\max}$ should not be associated with the Fermi
wavevector $k_{F}$ or an underlying lattice spacing:
\begin{subequations}
\begin{align}
H_{\mathrm{bath}} & =\int_{-k_{\max}}^{k_{\max}}\frac{dk}{2\pi}v_{F}kc_{k}^{\dagger}c_{k},\label{eq:H-Bath-k}\\
\psi\left(x\right) & =\int_{-k_{\max}}^{k_{\max}}\frac{dk}{2\pi}e^{ikx}c_{k}\equiv\sqrt{\frac{k_{\max}}{\pi}}\tilde{\psi}\left(x\right).
\end{align}
\end{subequations}
The fermionic operators $c_{k}$ and $\tilde{\psi}\left(x\right)$
satisfy $\left\{ c_{k},c_{k^{\prime}}^{\dagger}\right\} {=}\nolinebreak2\pi\delta\left(k{-}k^{\prime}\right)$
and $\left\{ \tilde{\psi}\left(x\right),\tilde{\psi}^{\dagger}\left(x\right)\right\} {=}1$,
respectively. Note that the cutoff impairs the orthogonality relations
of the field operators at the impurities $\left\{ \tilde{\psi}\left({-}R/2\right),\tilde{\psi}^{\dagger}\left({+}R/2\right)\right\} =\mathrm{sinc}\left(k_{\max}R\right)$,
but this can be amended by choosing the cutoff to be commensurate
with $R$ as in Eq.\,(\ref{eq:Commensurate-Cutoff}).

We then write the normalized combinations of bath modes to which the
impurity modes $d_{+}$ and $d_{-}$ {[}defined in Eq.\,(\ref{eq:even-odd-transformation}){]}
couple:
\begin{subequations}
\label{eq:even-odd-f0}
\begin{align}
f_{0+} & =\int_{-k_{\max}}^{k_{\max}}\frac{dk}{2\pi}\sqrt{\frac{2\pi}{k_{\max}}}\cos\left(kR/2\right)c_{k},\\
f_{0-} & =\int_{-k_{\max}}^{k_{\max}}\frac{dk}{2\pi}\sqrt{\frac{2\pi}{k_{\max}}}\sin\left(kR/2\right)c_{k}.
\end{align}
\end{subequations}
Switching to 2-vector notation, we can write the coupling Hamiltonian
as
\begin{subequations}
\begin{align}
 & H_{\mathrm{coupling}}=\mathbf{d}^{\dagger}\mathbf{T}_{0}\mathbf{f}_{0}+\mathrm{H.c.},\\
\mathbf{T}_{0}= & \begin{pmatrix}t_{0} & 0\\
0 & t_{0}
\end{pmatrix},\quad t_{0}=\sqrt{\frac{k_{\max}}{\pi}}\tilde{t}_{0}=\sqrt{\frac{2D\Gamma}{\pi}}.
\end{align}
\end{subequations}
Note that taking a cutoff which is not commensurate with $R$ only
affects the normalization of the even-odd modes $\mathbf{f}_{0}$,
but by construction they are still orthogonal. Thus, introducing normalization
constants $\mathcal{N}_{\pm}=1\pm\mathrm{sinc}\left(k_{\max}R\right)$,
we get 
\begin{equation}
\mathbf{f}_{0}\to\begin{pmatrix}\sqrt{\mathcal{N}_{+}} & 0\\
0 & \sqrt{\mathcal{N}_{-}}
\end{pmatrix}^{-1}\!\!\!\mathbf{f}_{0},\quad\mathbf{T}_{0}\to\begin{pmatrix}\sqrt{\mathcal{N}_{+}} & 0\\
0 & \sqrt{\mathcal{N}_{-}}
\end{pmatrix}\mathbf{T}_{0},
\end{equation}
which leaves the product $\mathbf{T}_{0}\mathbf{f}_{0}$ unchanged.
As this does not affect the rest of the discretization procedure in
any way, i.e., the expressions for $\mathbf{V}_{n}$ and $\boldsymbol{\mathcal{E}}_{n}$
remain unchanged, and the modified $\mathbf{T}_{0}$ is automatically
obtained from the numerical tridiagonalization, we can forget about
it for the rest of the derivation.

Exploiting particle-hole symmetry {[}see Eq.\,(\ref{eq:PH-symmetry}),
which translates to invariance under $c_{k}\to c_{-k}^{\dagger},d_{m}\to-d_{m}^{\dagger}]$,
we can restrict the discretization procedure to positive $k$, and
obtain the negative $k$ coefficients by symmetry. We redefine the
discretization gridpoints to $k$-space (focusing on positive $k$
and $n$)
\begin{equation}
k_{1}^{z}=k_{\max},\quad k_{n>1}^{z}=k_{\max}\Lambda^{2-n-z},
\end{equation}
and the intervals $I_{n}^{z}=\left[k_{n+1}^{z},k_{n}^{z}\right]$,
denoting their width and midpoints
\begin{equation}
\Delta k_{n}^{z}\equiv k_{n}^{z}-k_{n+1}^{z},\quad\bar{k}_{n}^{z}\equiv\frac{k_{n}^{z}+k_{n+1}^{z}}{2}.
\end{equation}
For conciseness, we now drop the $z$ index, but it is implied whenever
$n$ appears.

In each interval, each impurity is coupled only to a single mode,
defined by limiting the support of the integrals in Eq.\,(\ref{eq:even-odd-f0})
to $I_{n}$. These modes are generally not orthogonal with respect
to each other, but we can always choose linear combinations of the
two which are orthonormal, for example:
\begin{subequations}
\label{eq:Discrete-Orthogonal-Modes}
\begin{align}
c_{n\pm} & =\int_{I_{n}}\frac{dk}{2\pi}\varphi_{n\pm}\left(k\right)c_{k},\\
\varphi_{n+}\!\left(k\right) & =\sqrt{\frac{4\pi}{\Delta k_{n}}}\frac{\cos\left[\left(k-\bar{k}_{n}\right)R/2\right]}{\sqrt{1+\mathrm{sinc}\left(\Delta k_{n}R/2\right)}},\\
\varphi_{n-}\!\left(k\right) & =\sqrt{\frac{4\pi}{\Delta k_{n}}}\frac{\sin\left[\left(k-\bar{k}_{n}\right)R/2\right]}{\sqrt{1-\mathrm{sinc}\left(\Delta k_{n}R/2\right)}},
\end{align}
\end{subequations}
with their coupling to the impurities given by the product of a rotation
matrix and a rescaling matrix
\begin{align}
\mathbf{V}_{n} & =t_{0}\sqrt{\frac{\Delta k_{n}}{2k_{\max}}}\mathbf{R}_{n}\mathbf{S}_{n},\label{eq:Vn-App}\\
\mathbf{R}_{n} & =\begin{pmatrix}\cos\left(\bar{k}_{n}R/2\right) & -\sin\left(\bar{k}_{n}R/2\right)\\
+\sin\left(\bar{k}_{n}R/2\right) & \cos\left(\bar{k}_{n}R/2\right)
\end{pmatrix},\nonumber \\
\mathbf{S}_{n} & =\begin{pmatrix}\sqrt{1+\mathrm{sinc}\left(\frac{\Delta k_{n}R}{2}\right)} & 0\\
0 & \sqrt{1-\mathrm{sinc}\left(\frac{\Delta k_{n}R}{2}\right)}
\end{pmatrix}.\nonumber 
\end{align}
Observe that the chosen $\mathbf{c}_{n}$ are PT symmetric, which,
as expected, results in real $\mathbf{V}_{n}$. The hybridization
function of Eq.\,(\ref{eq:Hybridization-Even-Odd}) is given by
\begin{equation}
\boldsymbol{\Gamma}\left(\omega\right)=\Gamma\frac{\Delta k_{n}}{2\pi}\mathbf{R}_{n}\mathbf{S}_{n}\left(\begin{smallmatrix}\varphi_{n+}^{2}\left(k\right) & \varphi_{n+}\left(k\right)\varphi_{n-}\left(k\right)\\
\varphi_{n+}\left(k\right)\varphi_{n-}\left(k\right) & \varphi_{n-}^{2}\left(k\right)
\end{smallmatrix}\right)\mathbf{S}_{n}\mathbf{R}_{n}^{\dagger},
\end{equation}
where $n$ and $k$ are chosen such that $k=\omega/v_{F}\in I_{n}$.
It is then evident that the coefficient matrices and modes $\mathbf{c}_{n}{\equiv}\left(\!\begin{smallmatrix}c_{n+}\\
c_{n-}
\end{smallmatrix}\!\right)$ defined in this appendix differ from those in Sec.\,\ref{subsec:Discretization}
by the rotation $\mathbf{R}_{n}$
\begin{equation}
\mathbf{c}_{n}^{\text{Sec.\ref{sec:Method}}}{=}\mathbf{R}_{n}\mathbf{c}_{n},\ \mathbf{V}_{n}^{\text{Sec.\ref{sec:Method}}}{=}\mathbf{V}_{n}\mathbf{R}_{n}^{\dagger},\ \ \mathbf{\boldsymbol{\mathcal{E}}}_{n}^{\text{Sec.\ref{sec:Method}}}{=}\mathbf{R}_{n}\mathbf{\boldsymbol{\mathcal{E}}}_{n}\mathbf{R}_{n}^{\dagger}.\label{eq:Coeff-Rotation}
\end{equation}
$\mathbf{V}_{n}\mathbf{c}_{n}$ and $\mathbf{c}_{n}^{\dagger}\mathbf{\boldsymbol{\mathcal{E}}}_{n}\mathbf{c}_{n}$,
however, are unaffected, and Eq.\,(\ref{eq:Symmetry-Coef}) still
relates the negative $n$ coefficient matrices to those with positive
$n$. With this in hand, one can show that the expressions below in
Eqs.\,(\ref{eq:Naive-En-App}) and (\ref{eq:CampoOliveira-En-App})
lead to Eqs.\,(\ref{eq:Naive-En}) and (\ref{eq:CampoOliveira-En}),
respectively.

Following in the steps of Ref.\,\citep{campoAlternativeDiscretizationNumerical2005},
we start by extending the traditional \textit{naive} discretization
scheme (dating back to Wilson) to the chiral two-impurity scenario.
In each interval $I_{n}$, we define a real orthonormal basis of functions
$\varphi_{n\xi}\left(k\right)$ (to be specified later on), which
satisfy the orthogonality condition
\begin{equation}
\int_{I_{n}}\frac{dk}{2\pi}\varphi_{n\xi}\left(k\right)\varphi_{n\xi^{\prime}}\left(k\right)=\delta_{\xi\xi^{\prime}}.\label{eq:Naive-OC}
\end{equation}
We then use these functions to define a canonical basis of fermionic
modes
\begin{subequations}
\begin{align}
c_{n\xi} & =\int_{I_{n}}\frac{dk}{2\pi}\varphi_{n\xi}\left(k\right)c_{k}\\
c_{k} & =\sum_{\xi}\varphi_{n\xi}\left(k\right)c_{n\xi}\ ;\ k\in I_{n}
\end{align}
\end{subequations}
Substituting this into the bath Hamiltonian in Eq.\,(\ref{eq:H-Bath-k}),
we obtain 
\begin{align}
H_{\mathrm{bath}} & =\sum_{n=\pm1}^{\pm N}\int_{I_{n}}\frac{dk}{2\pi}v_{F}kc_{k}^{\dagger}c_{k}\\
 & =\sum_{n=\pm1}^{\pm N}\sum_{\xi\xi^{\prime}}\left[\int_{I_{n}}\frac{dk}{2\pi}v_{F}k\varphi_{n\xi}\left(k\right)\varphi_{n\xi^{\prime}}\left(k\right)\right]c_{n\xi}^{\dagger}c_{n\xi^{\prime}}.\nonumber 
\end{align}
We now wish to choose two specific modes (in each interval), $c_{n1}$
and $c_{n2}$, which span the space of modes coupled to the impurities,
i.e., that spanned by $c_{n+}$ and $c_{n-}$. We find that $\varphi_{n1}\left(k\right)\equiv\varphi_{n+}\left(k\right)$
and $\varphi_{n2}\left(k\right)\equiv\varphi_{n-}\left(k\right)$
 {[}specified in Eq.\,(\ref{eq:Discrete-Orthogonal-Modes}){]} already
satisfy Eq.\,(\ref{eq:Naive-OC}), and so $\mathbf{c}_{n}{=}\left(\!\begin{smallmatrix}c_{n+}\\
c_{n-}
\end{smallmatrix}\!\right){=}\left(\!\begin{smallmatrix}c_{n1}\\
c_{n2}
\end{smallmatrix}\!\right)$. Thus, projecting out all other modes (which are never constructed
explicitly) leaves us with the coefficient matrix
\begin{equation}
\mathbf{\boldsymbol{\mathcal{E}}}_{n}^{\mathrm{naive}}=\int_{I_{n}}\frac{dk}{2\pi}v_{F}k\left(\begin{smallmatrix}\varphi_{n+}^{2}\left(k\right) & \varphi_{n+}\left(k\right)\varphi_{n-}\left(k\right)\\
\varphi_{n+}\left(k\right)\varphi_{n-}\left(k\right) & \varphi_{n-}^{2}\left(k\right)
\end{smallmatrix}\right).\label{eq:Naive-En-App}
\end{equation}

We can then solve the integrals explicitly and obtain 
\begin{subequations}
\begin{align}
\mathbf{\boldsymbol{\mathcal{E}}}_{n}^{\mathrm{naive}} & =v_{F}\bar{k}_{n}\begin{pmatrix}1 & \alpha_{n}\\
\alpha_{n} & 1
\end{pmatrix},\\
\alpha_{n} & =\frac{\mathrm{sinc}\left(\Delta k_{n}R/2\right)-\cos\left(\Delta k_{n}R/2\right)}{\bar{k}_{n}R\sqrt{1-\mathrm{sinc}^{2}\left(\Delta k_{n}R/2\right)}}.
\end{align}
\end{subequations}
Observe that the diagonal terms are simply the interval midpoints,
as one would indeed expect from the naive scheme. At short wavelengths
(high energies) the off diagonal terms are negligible (of order $v_{F}/R$),
corresponding to the picture of two decoupled chains, while for long
wavelengths they are of order of (but smaller than) the diagonal terms
(for any finite $\Lambda$), and one can no longer consider the chiral
channel as two separate baths.

We now turn to extend the \textit{improved} discretization scheme
of Ref.\,\citep{campoAlternativeDiscretizationNumerical2005} to
the chiral two-impurity scenario. We start by defining a real set
of functions $\tilde{\varphi}_{n\xi}\left(k\right)$ (to be specified
later on), which satisfy a weighted orthogonality condition, but are
still normalized with respect to the unweighted inner product
\begin{subequations}
\label{eq:CampoOliveira-Conditions}
\begin{align}
 & \int_{I_{n}}\frac{dk}{2\pi}\frac{1}{k}\tilde{\varphi}_{n\xi}\left(k\right)\tilde{\varphi}_{n\xi^{\prime}}\left(k\right)=\frac{\delta_{\xi\xi^{\prime}}}{\boldsymbol{\mathcal{K}}_{n\xi}},\label{eq:CampoOliveira-OC}\\
 & \int_{I_{n}}\frac{dk}{2\pi}\tilde{\varphi}_{n\xi}\left(k\right)\tilde{\varphi}_{n\xi^{\prime}}\left(k\right)=1,\label{eq:CampoOliveira-Norm}
\end{align}
\end{subequations}
where $\boldsymbol{\mathcal{K}}_{n}$ is a diagonal matrix (with elements
$\boldsymbol{\mathcal{K}}_{n\xi})$ chosen such that Eq.\,(\ref{eq:CampoOliveira-Norm})
is satisfied. We use these to define a nonorthogonal (but normalized)
basis of modes
\begin{subequations}
\begin{align}
\tilde{c}_{n\xi} & =\int_{I_{n}}\frac{dk}{2\pi}\tilde{\varphi}_{n\xi}\left(k\right)c_{k},\\
c_{k} & =\sum_{\xi}\frac{\boldsymbol{\mathcal{K}}_{n\xi}}{k}\tilde{\varphi}_{n\xi}\left(k\right)\tilde{c}_{n\xi}\ ;\ k\in I_{n}.
\end{align}
\end{subequations}
As before, we substitute $c_{k}$ into the bath Hamiltonian, and find
that in our nonorthogonal basis, $H_{\mathrm{bath}}$ is diagonal
\begin{equation}
H_{\mathrm{bath}}=\sum_{n=\pm1}^{\pm N}\sum_{\xi}v_{F}\boldsymbol{\mathcal{K}}_{n\xi}\tilde{c}_{n\xi}^{\dagger}\tilde{c}_{n\xi}.\label{eq:CampoOliveria-Hbath}
\end{equation}
Again we wish to keep only two specific modes, $\tilde{c}_{n1}$ and
$\tilde{c}_{n2}$, which span the space of modes coupled to the impurities.
We are thus looking for modes related to our original $\mathbf{c}_{n}{=}\left(\!\begin{smallmatrix}c_{n1}\\
c_{n2}
\end{smallmatrix}\!\right)$ by an orthogonal transformation \textbf{$\mathbf{O}_{n}$}:
\begin{equation}
\begin{pmatrix}\tilde{c}_{n1}\\
\tilde{c}_{n2}
\end{pmatrix}=\mathbf{O}_{n}\begin{pmatrix}c_{n1}\\
c_{n2}
\end{pmatrix},\quad\begin{pmatrix}\tilde{\varphi}_{n1}\left(k\right)\\
\tilde{\varphi}_{n2}\left(k\right)
\end{pmatrix}=\mathbf{O}_{n}\begin{pmatrix}\varphi_{n1}\left(k\right)\\
\varphi_{n2}\left(k\right)
\end{pmatrix},
\end{equation}
such that $\tilde{\varphi}_{n1}\left(k\right)$ and $\tilde{\varphi}_{n2}\left(k\right)$
satisfy Eq.\,(\ref{eq:CampoOliveira-Conditions}), with $\varphi_{n1}\left(k\right)=\varphi_{n+}\left(k\right)$
and $\varphi_{n2}\left(k\right)=\varphi_{n-}\left(k\right)$ as specified
above. We find $\mathbf{O}_{n}$ by diagonalizing the symmetric matrix
\begin{subequations}
\begin{align}
\mathbf{K}_{n}^{-1} & =\int_{I_{n}}\frac{dk}{2\pi}\frac{1}{k}\left(\begin{smallmatrix}\varphi_{n1}^{2}\left(k\right) & \varphi_{n1}\left(k\right)\varphi_{n2}\left(k\right)\\
\varphi_{n1}\left(k\right)\varphi_{n2}\left(k\right) & \varphi_{n2}^{2}\left(k\right)
\end{smallmatrix}\right),\\
\boldsymbol{\mathcal{K}}_{n}^{-1} & =\mathbf{O}_{n}\mathbf{K}_{n}^{-1}\mathbf{O}_{n}^{\dagger}=\begin{pmatrix}\boldsymbol{\mathcal{K}}_{n1} & 0\\
0 & \boldsymbol{\mathcal{K}}_{n2}
\end{pmatrix}^{-1}.
\end{align}
\end{subequations}
We can now discard all modes in Eq.\,(\ref{eq:CampoOliveria-Hbath})
except $\tilde{c}_{n1},\tilde{c}_{n2}$, and use $\mathbf{O}_{n}$
to rewrite $H_{\mathrm{bath}}$ in terms of $c_{n1},c_{n2}$, to get
$\mathbf{\boldsymbol{\mathcal{E}}}_{n}^{\mathrm{CO}}=v_{F}\mathbf{K}_{n}$
(switching notation back to $\left\{ +,-\right\} =\left\{ 1,2\right\} $):
\begin{equation}
\boldsymbol{\mathcal{E}}_{n}^{\mathrm{CO}}=\left[\int_{I_{n}}\frac{dk}{2\pi}\frac{1}{v_{F}k}\left(\begin{smallmatrix}\varphi_{n+}^{2}\left(k\right) & \varphi_{n+}\left(k\right)\varphi_{n-}\left(k\right)\\
\varphi_{n+}\left(k\right)\varphi_{n-}\left(k\right) & \varphi_{n-}^{2}\left(k\right)
\end{smallmatrix}\right)\right]^{-1}\!\!\!\!\!\!.\label{eq:CampoOliveira-En-App}
\end{equation}
Note that in practice we only calculate $\mathbf{K}_{n}$, and never
construct $\mathbf{O}_{n}$ and $\boldsymbol{\mathcal{K}}_{n}$.

Solving the integral for $\mathbf{K}_{n}^{-1}$, we obtain
\begin{subequations}
\begin{align}
 & \left[\mathbf{K}_{n}^{-1}\right]_{\pm\pm}=\frac{1}{\mathcal{N}_{n\pm}}\left[\log r_{n}\pm C_{n}\Delta\mathcal{C}_{n}\pm S_{n}\Delta\mathcal{S}_{n}\right],\\
 & \left[\mathbf{K}_{n}^{-1}\right]_{\pm\mp}=\frac{1}{\sqrt{\mathcal{N}_{n+}\mathcal{N}_{n-}}}\left[C_{n}\Delta\mathcal{S}_{n}-S_{n}\Delta\mathcal{C}_{n}\right],\\
 & \mathcal{N}_{n\pm}=\Delta k_{n}\left(1\pm\mathrm{sinc}\left(\Delta k_{n}R/2\right)\right),\\
 & r_{n}=k_{n}/k_{n+1},\ C_{n}=\cos\left(\bar{k}_{n}R\right),\ S_{n}=\sin\left(\bar{k}_{n}R\right),\\
 & \Delta\mathcal{C}_{n}{=}\!\int_{k_{n+1}R}^{k_{n}R}\frac{\cos\left(x\right)}{x}dx,\ \Delta\mathcal{S}_{n}{=}\!\int_{k_{n+1}R}^{k_{n}R}\frac{\sin\left(x\right)}{x}dx.
\end{align}
\end{subequations}
Observe that for short wavelengths (high energies) the off diagonal
terms $\left[\mathbf{K}_{n}^{-1}\right]_{\pm\mp}$ vanish, corresponding
to two decoupled baths, while the diagonal terms approach $\left[\mathbf{K}_{n}^{-1}\right]_{\pm\pm}\to\frac{\log k_{n}-\log k_{n+1}}{k_{n}-k_{n+1}}$,
so that inverting we get the discrete energies $\boldsymbol{\mathcal{E}}_{n\pm\pm}\to\frac{1-\Lambda^{-1}}{\log\Lambda}D\Lambda^{2-n-z}$
expected from the scheme of Ref.\,\citep{campoAlternativeDiscretizationNumerical2005}
{[}see Eq.\,(46) therein{]}. In the opposite limit of long wavelengths,
the chiral channel can no longer be considered as two separate baths,
and indeed all elements of $\mathbf{K}_{n}^{-1}$ survive.

As a final step, we would like to compare with notation used in Ref.\,\citep{zitkoEnergyResolutionDiscretization2009}
to describe the discretization scheme of Ref.\,\citep{campoAlternativeDiscretizationNumerical2005}.
We thus first write the local density of states for the modes $\mathbf{f}_{0}$
coupled to the impurities
\begin{equation}
\boldsymbol{\rho}\!\left(\omega\right)=\pi\mathbf{T}_{0}^{-1}\boldsymbol{\Gamma}\!\left(\omega\right)\mathbf{T}_{0}^{-1},\ \ \left[\mathbf{T}_{0}\right]^{2}=\frac{1}{\pi}\int_{-D}^{D}\boldsymbol{\Gamma}\!\left(\omega\right)d\omega,
\end{equation}
so that $\int_{-D}^{D}\boldsymbol{\rho}\left(\omega\right)=\mathbf{1}$.
Then Eq.\,(30) in Ref.\,\citep{zitkoEnergyResolutionDiscretization2009}
takes the form
\begin{equation}
\mathbf{f}_{0}=\sum_{n}\left[\int_{I_{n}}\boldsymbol{\rho}\left(\omega\right)d\omega\right]^{\frac{1}{2}}\mathbf{c}_{n},
\end{equation}
where taking the square root is well defined because $\boldsymbol{\rho}\left(\omega\right)$
is positive. We then rewrite Eqs.\,(\ref{eq:Naive-En}) and (\ref{eq:CampoOliveira-En})
as
\begin{align}
\!\boldsymbol{\mathcal{E}}_{n}^{\mathrm{naive}} & \!=\!\left[\!\int_{\!I_{n}}\!\!\!\boldsymbol{\rho}\!\left(\omega\right)\!d\omega\right]^{-\!\frac{1}{2}}\!\left[\!\int_{\!I_{n}}\!\!\!\boldsymbol{\rho}\!\left(\omega\right)\!\omega d\omega\right]\!\left[\!\int_{\!I_{n}}\!\!\!\boldsymbol{\rho}\!\left(\omega\right)\!d\omega\right]^{-\!\frac{1}{2}}\!\!\!\!\!\!,\\
\boldsymbol{\mathcal{E}}_{n}^{\mathrm{CO}} & \!=\!\left[\!\int_{\!I_{n}}\!\!\!\boldsymbol{\rho}\!\left(\omega\right)\!d\omega\right]^{\frac{1}{2}}\!\left[\!\int_{\!I_{n}}\!\!\!\boldsymbol{\rho}\!\left(\omega\right)\!\frac{d\omega}{\omega}\right]^{-1}\!\left[\!\int_{\!I_{n}}\!\!\!\boldsymbol{\rho}\!\left(\omega\right)\!d\omega\right]^{\frac{1}{2}}\!\!\!\!,
\end{align}
which is the matrix generalization of Eqs.\,(33) and (32) in Ref.\,\citep{zitkoEnergyResolutionDiscretization2009},
respectively.

\section{Exploiting Symmetries\label{sec:App-Exploiting-Symmetries}}

In this appendix we discuss technical aspects in the exploitation
of symmetries as part of our chiral NRG scheme. We focus on the $\mathrm{SU(2)_{charge}\otimes SU(2)_{spin}}$
symmetric case, but the discussion below also applies (with some simplifications)
to the $\mathrm{U(1)_{charge}\otimes SU(2)_{spin}}$ symmetric case,
while for the $\mathrm{U(1)_{charge}\otimes U(1)_{spin}}$ symmetric
case it becomes trivial (and so irrelevant).

We start with a brief overview of the way symmetry-respecting tensors
are formulated in the \textsc{QSpace} library, with the full details
given in Ref.\,\citep{weichselbaumNonabelianSymmetriesTensor2012}.
Similarly to the decomposition in the Wigner-Eckart theorem \citep{sakuraiModernQuantumMechanics2017b},
for a general non-Abelian symmetry, each tensor can be decomposed
into an outer product of (generalized) reduced matrix elements, and
(generalized) Clebsch-Gordan coefficients. We write (and track) only
the tensors describing the reduced matrix elements, which limits the
operators we can access (to those respecting the symmetry, which should
be all we need). In the case of $\mathrm{SU(2)_{charge}\otimes SU(2)_{spin}}$,
these are given by (see Appendix~A.9 of Ref.\,\citep{weichselbaumNonabelianSymmetriesTensor2012})
\begin{equation}
\mathbf{F}_{m}{=}\left(\!\begin{smallmatrix}\left(-1\right)^{m}\!\!\! & f_{m\uparrow}^{\dagger}\\
 & f_{m\downarrow}\\
\left(-1\right)^{m}\!\!\! & f_{m\downarrow}^{\dagger}\\
\quad\ -\!\!\! & f_{m\uparrow}
\end{smallmatrix}\!\right),\ \mathbf{S}_{m}{=}\left(\!\begin{smallmatrix}-\tfrac{1}{\sqrt{2}}S_{m}^{+}\\
\quad\ \ \,S_{m}^{z}\\
+\tfrac{1}{\sqrt{2}}S_{m}^{-}
\end{smallmatrix}\!\!\right),\ \mathbf{C}_{m}{=}\left(\!\begin{smallmatrix}-\tfrac{1}{\sqrt{2}}C_{m}^{+}\\
\quad\ \ \,C_{m}^{z}\\
+\tfrac{1}{\sqrt{2}}C_{m}^{-}
\end{smallmatrix}\!\!\right),\label{eq:Sym-Operators}
\end{equation}
where $f_{m\sigma}$ are fermionic operators at site $m$ with spin
$\sigma$, and $S_{i}$ ($C_{i}$) are the spin (charge) $\mathrm{SU(2)}$
operators. We then use these operators in order to construct operators
which are scalars with respect to the symmetry, with the Hamiltonian
being a sum of such scalars.

The site index $m$ bipartitions the sites according to its parity
(and can, but does not have to, coincide with the fermionic order
for the Jordan-Wigner string). The structure of Eq.\,(\ref{eq:Sym-Operators})
then allows only purely real (imaginary) hopping terms if $m$ and
$m^{\prime}$ have different (same) parity
\begin{subequations}
\label{eq:Sym-Hopping}
\begin{align}
m{+}m^{\prime}\text{ odd}\ \!:\ \ \  & \mathbf{F}_{m}^{\dagger}\mathbf{F}_{m^{\prime}}\!=\ \sum_{\sigma}\!\left(f_{m\sigma}^{\dagger}f_{m^{\prime}\sigma}{+}f_{m^{\prime}\sigma}^{\dagger}f_{m\sigma}\right)\!,\!\label{eq:Sym-Real-Hopping}\\
m{+}m^{\prime}\text{ even}\!:\ i & \mathbf{F}_{m}^{\dagger}\mathbf{F}_{m^{\prime}}\!=i\sum_{\sigma}\!\left(f_{m\sigma}^{\dagger}f_{m^{\prime}\sigma}{-}f_{m^{\prime}\sigma}^{\dagger}f_{m\sigma}\right)\!.\!\label{eq:Sym-Imag-Hopping}
\end{align}
\end{subequations}
We recall that the Wilson ladder (in the even-odd PT-symmetric basis)
has only real hopping amplitudes, and indeed, by choosing $m$ to
snake along the ladder ($d_{+}{\to}d_{-}{\to}f_{0-}{\to}f_{0+}{\to}f_{1+}{\to}f_{1-}{\to}f_{2-}{\to}...$)
all hopping terms are between sites of opposite parity, as dictated
by symmetry. If we return to the real-space basis, as discussed at
the end of Sec.\,\ref{subsec:Tridiagonalization}, we will have both
purely imaginary and purely real terms (but no complex terms), and
can choose a different ordering for $m$ such that all terms adhere
to Eq.\,(\ref{eq:Sym-Hopping}).

We now turn to the interaction terms, which can indeed be constructed
from the operators in Eq.\,(\ref{eq:Sym-Operators}), both for the
Kondo model {[}Eq.\,(\ref{eq:H-Kondo}){]}: $\mathbf{S}_{m}^{\dagger}\mathbf{S}_{m^{\prime}}=\vec{S}_{m}\cdot\vec{S}_{m^{\prime}}$,
and for the Anderson model {[}Eq.\,(\ref{eq:H-Imp-Anderson}), after
taking $\varepsilon_{d}=-U/2${]}
\begin{align}
\tfrac{4}{3}\mathbf{C}_{m}^{\dagger}\mathbf{C}_{m} & =\left(n_{m\uparrow}+n_{m\downarrow}-1\right)^{2}\\
 & =2n_{m\uparrow}n_{m\downarrow}-\left(n_{m\uparrow}+n_{m\downarrow}\right)+1\nonumber \\
 & =1-\left(n_{m\uparrow}-n_{m\downarrow}\right)^{2}=1-\tfrac{4}{3}\mathbf{S}_{m}^{\dagger}\mathbf{S}_{m}.\nonumber 
\end{align}
These (real) terms are defined in the real-space basis, whereas the
Wilson ladder is real in the even-odd PT-symmetric basis. If we transform
the entire (or part of the) Wilson ladder back to real-space, we will
generate imaginary terms. However, as the interaction terms are invariant
under a PT transformation, if we write them in terms of PT-symmetric
fermionic (or spin) operators, we are guaranteed to have real coefficients.
Thus, the full Hamiltonian is purely real, and we can restrict the
numerical iterative diagonalization to real (double precision) arithmetic,
which significantly reduces the computational cost. The transforming
of the interaction terms to the PT-symmetric basis is derived in the
remainder of this appendix.

Consider two fermionic sites $m\in\left\{ 1,2\right\} $ with spinfull
operators $f_{m\sigma}$, which correspond e.g., to the two impurity
sites $\mathbf{d}$ or the first two ladder sites $\mathbf{f}_{0}$.
We define a basis transformation with respect to the site index
\begin{equation}
\tilde{f}_{m\sigma}=\sum_{m^{\prime}}u_{mm^{\prime}}f_{m^{\prime}\sigma},
\end{equation}
where $u$ is a $2\times2$ unitary matrix, e.g., $u=\frac{1}{\sqrt{2}}\left(\begin{smallmatrix}1 & 1\\
i & -i
\end{smallmatrix}\right)$ for the transformation to PT-symmetric modes in Eqs.\,(\ref{eq:even-odd-transformation})
and (\ref{eq:even-odd-f0}). The corresponding operators $\mathbf{F}_{1},\mathbf{F}_{2}$
trivially transform accordingly, so that we can immediately write
them in terms of the operators $\tilde{\mathbf{F}}_{1},\tilde{\mathbf{F}}_{2}$,
and vice versa. However, it is not \textit{a priory} clear how the
$\mathrm{SU}(2)$ operators, e.g., $\mathbf{S}_{1},\mathbf{S}_{2}$,
transform, and how to write them if we only have access to $\tilde{\mathbf{S}}_{1},\tilde{\mathbf{S}}_{2}$.
Generically we would like to relate operators which are quadratic
in $f_{m\sigma}$ and diagonal with respect to $m$ in the original
basis, i.e., can be written as
\begin{equation}
O_{m}=\sum_{\sigma\sigma^{\prime}}O_{\sigma\sigma^{\prime}}f_{m\sigma}^{\dagger}f_{m\sigma^{\prime}},
\end{equation}
to those which are quadratic in $\tilde{f}_{m\sigma}$ and diagonal
with respect to $m$ in the new basis,
\begin{equation}
\tilde{O}_{m}=\sum_{\sigma\sigma^{\prime}}\!O_{\sigma\sigma^{\prime}}\tilde{f}_{m\sigma}^{\dagger}\tilde{f}_{m\sigma^{\prime}}=\sum_{\sigma\sigma^{\prime}}\!\sum_{ll^{\prime}}\!O_{\sigma\sigma^{\prime}}f_{l\sigma}^{\dagger}u_{lm}^{\dagger}u_{ml^{\prime}}f_{l^{\prime}\sigma^{\prime}}.\label{eq:Sym-Trans-Otilde}
\end{equation}
If we exploit only Abelian symmetries, we have access to the individual
fermionic operators, and can write the individual quadratic $f_{l\sigma}^{\dagger}f_{l^{\prime}\sigma^{\prime}}$
terms, but for non-Abelian symmetries we only have access to specific
combinations, e.g., only $\mathbf{F}_{1}^{\dagger}\mathbf{F}_{2}$
(and $\tilde{\mathbf{F}}_{1}^{\dagger}\tilde{\mathbf{F}}_{2}$) in
the $\mathrm{SU(2)_{charge}\otimes SU(2)_{spin}}$ case. We would
thus like to directly express $O_{m}$ in terms of $\tilde{O}_{m}$,
using only the allowed combinations.

In order to do so, we first observe that if we write each operator
$f_{m\sigma}$ as a matrix acting on the two-site Hilbert space, then
we have a unitary matrix $U$ acting on the same space such that
\begin{equation}
\tilde{f}_{m\sigma}{=}\thinspace U^{\dagger}f_{m\sigma}U,\ \ U{\equiv}\exp\!\left(\sum_{mm^{\prime}\sigma}\!\!\!\left[\log u\right]_{mm^{\prime}}f_{m\sigma}^{\dagger}f_{m^{\prime}\sigma}\!\right),\label{eq:Sym-Trans-UfU}
\end{equation}
where $\log u$ is the matrix logarithm of the transformation matrix
$u$. Substituting this into into Eq.\,(\ref{eq:Sym-Trans-Otilde})
we immediately get
\begin{equation}
\tilde{O}_{m}=\sum_{\sigma\sigma^{\prime}}O_{\sigma\sigma^{\prime}}U^{\dagger}f_{m\sigma}^{\dagger}UU^{\dagger}f_{m\sigma^{\prime}}U=U^{\dagger}O_{m}U.\label{eq:Sym-Trans-UOU}
\end{equation}
For completeness we prove Eq.\,(\ref{eq:Sym-Trans-UfU}). It is convenient
to first diagonalize $u$, and write $U$ in the basis defined by
this diagonalization
\begin{subequations}
\begin{align}
 & u_{mm^{\prime}}=\sum_{l}s_{ml}^{\dagger}e^{i\phi_{l}}s_{lm^{\prime}},\quad\bar{f}_{m\sigma}=\sum_{m^{\prime}}s_{mm^{\prime}}f_{m^{\prime}\sigma},\\
\!\!\!U & {=}\exp\biggl(\!i\!\!\!\!\!\!\!\!\!\!\sum_{\ \ \ \ lmm^{\prime}\sigma}\!\!\!\!\!\!\!\!\!\!s_{ml}^{\dagger}\phi_{l}s_{lm^{\prime}}f_{m\sigma}^{\dagger}f_{m^{\prime}\sigma}\!\biggr){=}\exp\biggl(\!i\!\sum_{l\sigma}\!\phi_{l}\bar{f}_{l\sigma}^{\dagger}\bar{f}_{l\sigma}\!\biggr),
\end{align}
\end{subequations}
with $s$ a unitary matrix and $\phi_{l}$ a phase. Substituting this
into the r.h.s. of Eq.\,(\ref{eq:Sym-Trans-UfU}), we arrive at its
l.h.s.,
\begin{align}
U^{\dagger}f_{m\alpha}U & =\exp\left(-i\sum_{a\alpha}\phi_{a}\bar{f}_{a\alpha}^{\dagger}\bar{f}_{a\alpha}\right)\\
 & \qquad\cdot\left(\sum_{lm^{\prime}}s_{ml}^{\dagger}s_{lm^{\prime}}f_{m^{\prime}\sigma}\right)\cdot\exp\left(\sum_{b\beta}i\phi_{b}\bar{f}_{b\beta}^{\dagger}\bar{f}_{b\beta}\right)\nonumber \\
 & =\sum_{l}s_{ml}^{\dagger}e^{-i\phi_{l}\bar{f}_{l\sigma}^{\dagger}\bar{f}_{l\sigma}}\bar{f}_{l\sigma}e^{i\phi_{l}\bar{f}_{l\sigma}^{\dagger}\bar{f}_{l\sigma}}\nonumber \\
 & =\sum_{lm^{\prime}}s_{ml}^{\dagger}e^{i\phi_{l}}s_{lm^{\prime}}f_{m^{\prime}\sigma}=\sum_{m^{\prime}}u_{mm^{\prime}}f_{m^{\prime}\sigma}=\tilde{f}_{m\sigma}.\nonumber 
\end{align}

We have thus reduced the question to whether for a given $u$, we
can write $U$ as in Eq.\,(\ref{eq:Sym-Trans-UfU}) using only operators
respected by the symmetry, i.e., can we write $\log u=i\alpha\sigma_{x}$,
and then $U{=}\exp\left(i\alpha\mathbf{F}_{1}^{\dagger}\mathbf{F}_{2}\right)$.
This is generically not possible, and in particular not for $u=\frac{1}{\sqrt{2}}\left(\begin{smallmatrix}1 & 1\\
i & -i
\end{smallmatrix}\right)$. However, we are only interested in finding a $U$ which transforms
quadratic operators, and do not require it to correctly transform
the individual fermionic operators. As $O_{m}$ and $\tilde{O}_{m}$
are indifferent to phases $f_{m\sigma}{\to}e^{i\theta_{m}}f_{m\sigma}$
and $\tilde{f}_{m\sigma}{\to}e^{i\tilde{\theta}_{m}}\tilde{f}_{m\sigma}$,
if we can find 
\begin{equation}
v=\begin{pmatrix}e^{i\tilde{\theta}_{1}} & 0\\
0 & e^{i\tilde{\theta}_{2}}
\end{pmatrix}u\begin{pmatrix}e^{-i\theta_{1}} & 0\\
0 & e^{-i\theta_{2}}
\end{pmatrix}\text{ s.t. }\log v=i\alpha\sigma_{x},
\end{equation}
then $U{=}\exp\left(i\alpha\mathbf{F}_{1}^{\dagger}\mathbf{F}_{2}\right)$
will satisfy Eq.\,(\ref{eq:Sym-Trans-UOU}) {[}although not Eq.\,(\ref{eq:Sym-Trans-UfU}){]}.
 Indeed, in our specific case, we find such a $v$:
\begin{equation}
v=\frac{1}{\sqrt{2}}\begin{pmatrix}1 & i\\
i & 1
\end{pmatrix}=\frac{1}{\sqrt{2}}\begin{pmatrix}1 & 1\\
i & -i
\end{pmatrix}\begin{pmatrix}1 & 0\\
0 & i
\end{pmatrix},\ \ \log v=i\frac{\pi}{4}\sigma_{x}.
\end{equation}

For the Anderson model, we then transform the operators $\mathbf{S}_{+},\mathbf{S}_{-}$
of the even-odd impurities to those in real-space $\mathbf{S}_{1},\mathbf{S}_{2}$
and use them to write the interaction term. For the Kondo model, we
similarly use the even-odd basis spin operators at the first ladder
sites $\mathbf{s}_{0\pm}$ in order to write those in real space $\mathbf{s}\left(\pm R/2\right)$.
We also define $\mathbf{S}_{\pm}$ as the even-odd basis impurity
operators, and these transform trivially to real-space $\left(\!\begin{smallmatrix}\mathbf{S}_{1}\\
\mathbf{S}_{2}
\end{smallmatrix}\!\right)=u^{\dagger}\left(\!\begin{smallmatrix}\mathbf{S}_{+}\\
\mathbf{S}_{-}
\end{smallmatrix}\!\right),$ so that we can write the Kondo term. Assuming equal Coulomb or Kondo
interactions at both impurities, we thus arrive at a real interaction
Hamiltonian.

\section{Single-Particle Calculations\label{sec:App-Single-Particle-Calculations}}

For the noninteracting calculations of the discrete model, we write
the full single-particle Hamiltonian as a matrix $H$ and the bath
single-particle Hamiltonian as a matrix $H_{0}$ (which is the same
as $H$, excluding the impurity rows and columns). We then diagonalize
$H=UEU^{\dagger}$ to get single-particle energies $E_{a}$ (and $E_{\alpha}^{0}$
for $H_{0}$). The impurity entropy as a function of temperature $T$
is then given by

\begin{subequations}
\begin{align}
\mathcal{S}_{\mathrm{imp}}\!\left(T\right) & =\mathcal{S}\!\left(T\right)-\mathcal{S}_{0}\!\left(T\right),\\
\mathcal{S}_{\left(0\right)}\!\left(T\right) & =-\!\sum_{\alpha}f_{\alpha}^{\left(0\right)}\!\log f_{\alpha}^{\left(0\right)}{+}\!\left(\!1{-}f_{\alpha}^{\left(0\right)}\!\right)\!\log\!\left(\!1{-}f_{\alpha}^{\left(0\right)}\!\right)\!,
\end{align}
\end{subequations}
where $f_{\alpha}^{\left(0\right)}{\equiv}f_{\mathrm{FD}}\left(E_{\alpha}^{\left(0\right)};T\right)$
is the occupation of the single-particle eigenmodes. In order to negate
even-odd oscillations, we also need to average over two $z$ shifts
($z=\frac{1}{2},1$).

The impurity spectral function (in the discrete model) is evaluated
at $E_{\alpha}$, with proper broadening
\begin{equation}
\left[\mathbf{A}_{\mathrm{disc}}^{T=0}\left(E_{\alpha}\right)\right]_{mm^{\prime}}=\frac{U_{m\alpha}^{*}U_{m^{\prime}\alpha}}{\left|E_{\alpha}\right|\log\Lambda},
\end{equation}
where the impurity indices $m,m^{\prime}$ on the l.h.s. take the
values $1,2$, and on the r.h.s. select the corresponding rows/columns
in $H$. The thermal impurity correlations are then evaluated according
to the second row of Eq.\,(\ref{eq:NI-CorrT}), replacing the integral
by a discrete sum over the (single-particle) energies 
\begin{equation}
\!\left\langle d_{m^{\prime}}^{\dagger}d_{m}\right\rangle _{T}\!{=}\!\!\int_{-\infty}^{\infty}\!\!\!\!\!\mathbf{A}_{mm^{\prime}}^{\!T=0}\!\left(\omega\right)\!f_{\mathrm{FD}}\!\left(\omega\right)\!d\omega\to\!\sum_{\alpha}\!\frac{U_{m^{\prime}\alpha}U_{m\alpha}^{*}}{e^{E_{\alpha}/T}{+}1}.\label{eq:App-CorrT}
\end{equation}
In the continuum limit, we already start from an expression for the
impurity spectral function, i.e., Eq.\,(\ref{eq:NI-Spectral-Function}),
and then obtain the thermal correlations by numerically solving the
integral in Eq.\,(\ref{eq:NI-CorrT}) {[}Eq.\,(\ref{eq:App-CorrT})
here{]}.

\bibliography{../../../bib}

\begin{thebibliography}{70}%
\makeatletter
\providecommand \@ifxundefined [1]{%
 \@ifx{#1\undefined}
}%
\providecommand \@ifnum [1]{%
 \ifnum #1\expandafter \@firstoftwo
 \else \expandafter \@secondoftwo
 \fi
}%
\providecommand \@ifx [1]{%
 \ifx #1\expandafter \@firstoftwo
 \else \expandafter \@secondoftwo
 \fi
}%
\providecommand \natexlab [1]{#1}%
\providecommand \enquote  [1]{``#1''}%
\providecommand \bibnamefont  [1]{#1}%
\providecommand \bibfnamefont [1]{#1}%
\providecommand \citenamefont [1]{#1}%
\providecommand \href@noop [0]{\@secondoftwo}%
\providecommand \href [0]{\begingroup \@sanitize@url \@href}%
\providecommand \@href[1]{\@@startlink{#1}\@@href}%
\providecommand \@@href[1]{\endgroup#1\@@endlink}%
\providecommand \@sanitize@url [0]{\catcode `\\12\catcode `\$12\catcode
  `\&12\catcode `\#12\catcode `\^12\catcode `\_12\catcode `\%12\relax}%
\providecommand \@@startlink[1]{}%
\providecommand \@@endlink[0]{}%
\providecommand \url  [0]{\begingroup\@sanitize@url \@url }%
\providecommand \@url [1]{\endgroup\@href {#1}{\urlprefix }}%
\providecommand \urlprefix  [0]{URL }%
\providecommand \Eprint [0]{\href }%
\providecommand \doibase [0]{https://doi.org/}%
\providecommand \selectlanguage [0]{\@gobble}%
\providecommand \bibinfo  [0]{\@secondoftwo}%
\providecommand \bibfield  [0]{\@secondoftwo}%
\providecommand \translation [1]{[#1]}%
\providecommand \BibitemOpen [0]{}%
\providecommand \bibitemStop [0]{}%
\providecommand \bibitemNoStop [0]{.\EOS\space}%
\providecommand \EOS [0]{\spacefactor3000\relax}%
\providecommand \BibitemShut  [1]{\csname bibitem#1\endcsname}%
\let\auto@bib@innerbib\@empty
%</preamble>
\bibitem [{\citenamefont {Hasan}\ and\ \citenamefont
  {Kane}(2010)}]{hasanColloquiumTopologicalInsulators2010}%
  \BibitemOpen
  \bibfield  {author} {\bibinfo {author} {\bibfnamefont {M.~Z.}\ \bibnamefont
  {Hasan}}\ and\ \bibinfo {author} {\bibfnamefont {C.~L.}\ \bibnamefont
  {Kane}},\ }\bibfield  {title} {\bibinfo {title} {Colloquium: {{Topological}}
  insulators},\ }\href {https://doi.org/10.1103/RevModPhys.82.3045} {\bibfield
  {journal} {\bibinfo  {journal} {Rev. Mod. Phys.}\ }\textbf {\bibinfo {volume}
  {82}},\ \bibinfo {pages} {3045} (\bibinfo {year} {2010})}\BibitemShut
  {NoStop}%
\bibitem [{\citenamefont {Qi}\ and\ \citenamefont
  {Zhang}(2011)}]{qiTopologicalInsulatorsSuperconductors2011}%
  \BibitemOpen
  \bibfield  {author} {\bibinfo {author} {\bibfnamefont {X.-L.}\ \bibnamefont
  {Qi}}\ and\ \bibinfo {author} {\bibfnamefont {S.-C.}\ \bibnamefont {Zhang}},\
  }\bibfield  {title} {\bibinfo {title} {Topological insulators and
  superconductors},\ }\href {https://doi.org/10.1103/RevModPhys.83.1057}
  {\bibfield  {journal} {\bibinfo  {journal} {Rev. Mod. Phys.}\ }\textbf
  {\bibinfo {volume} {83}},\ \bibinfo {pages} {1057} (\bibinfo {year}
  {2011})}\BibitemShut {NoStop}%
\bibitem [{\citenamefont {Kondo}(1964)}]{kondoResistanceMinimumDilute1964}%
  \BibitemOpen
  \bibfield  {author} {\bibinfo {author} {\bibfnamefont {J.}~\bibnamefont
  {Kondo}},\ }\bibfield  {title} {\bibinfo {title} {Resistance {{Minimum}} in
  {{Dilute Magnetic Alloys}}},\ }\href {https://doi.org/10.1143/PTP.32.37}
  {\bibfield  {journal} {\bibinfo  {journal} {Progress of Theoretical Physics}\
  }\textbf {\bibinfo {volume} {32}},\ \bibinfo {pages} {37} (\bibinfo {year}
  {1964})}\BibitemShut {NoStop}%
\bibitem [{\citenamefont {Hewson}(1993)}]{hewsonKondoProblemHeavy1993}%
  \BibitemOpen
  \bibfield  {author} {\bibinfo {author} {\bibfnamefont {A.~C.}\ \bibnamefont
  {Hewson}},\ }\href {https://doi.org/10.1017/CBO9780511470752} {\emph
  {\bibinfo {title} {The {{Kondo Problem}} to {{Heavy Fermions}}}}},\ Cambridge
  {{Studies}} in {{Magnetism}}\ (\bibinfo  {publisher} {{Cambridge University
  Press}},\ \bibinfo {address} {{Cambridge}},\ \bibinfo {year}
  {1993})\BibitemShut {NoStop}%
\bibitem [{\citenamefont {Andrei}\ \emph {et~al.}(1983)\citenamefont {Andrei},
  \citenamefont {Furuya},\ and\ \citenamefont
  {Lowenstein}}]{andreiSolutionKondoProblem1983}%
  \BibitemOpen
  \bibfield  {author} {\bibinfo {author} {\bibfnamefont {N.}~\bibnamefont
  {Andrei}}, \bibinfo {author} {\bibfnamefont {K.}~\bibnamefont {Furuya}},\
  and\ \bibinfo {author} {\bibfnamefont {J.~H.}\ \bibnamefont {Lowenstein}},\
  }\bibfield  {title} {\bibinfo {title} {Solution of the {{Kondo}} problem},\
  }\href {https://doi.org/10.1103/RevModPhys.55.331} {\bibfield  {journal}
  {\bibinfo  {journal} {Rev. Mod. Phys.}\ }\textbf {\bibinfo {volume} {55}},\
  \bibinfo {pages} {331} (\bibinfo {year} {1983})}\BibitemShut {NoStop}%
\bibitem [{\citenamefont {Tsvelick}\ and\ \citenamefont
  {Wiegmann}(1983)}]{tsvelickExactResultsTheory1983}%
  \BibitemOpen
  \bibfield  {author} {\bibinfo {author} {\bibfnamefont {A.}~\bibnamefont
  {Tsvelick}}\ and\ \bibinfo {author} {\bibfnamefont {P.}~\bibnamefont
  {Wiegmann}},\ }\bibfield  {title} {\bibinfo {title} {Exact results in the
  theory of magnetic alloys},\ }\href
  {https://doi.org/10.1080/00018738300101581} {\bibfield  {journal} {\bibinfo
  {journal} {Advances in Physics}\ }\textbf {\bibinfo {volume} {32}},\ \bibinfo
  {pages} {453} (\bibinfo {year} {1983})}\BibitemShut {NoStop}%
\bibitem [{\citenamefont {Affleck}(1990)}]{affleckCurrentAlgebraApproach1990a}%
  \BibitemOpen
  \bibfield  {author} {\bibinfo {author} {\bibfnamefont {I.}~\bibnamefont
  {Affleck}},\ }\bibfield  {title} {\bibinfo {title} {A current algebra
  approach to the {{Kondo}} effect},\ }\href
  {https://doi.org/10.1016/0550-3213(90)90440-O} {\bibfield  {journal}
  {\bibinfo  {journal} {Nuclear Physics B}\ }\textbf {\bibinfo {volume}
  {336}},\ \bibinfo {pages} {517} (\bibinfo {year} {1990})}\BibitemShut
  {NoStop}%
\bibitem [{\citenamefont {Affleck}\ and\ \citenamefont
  {Ludwig}(1991)}]{affleckKondoEffectConformal1991}%
  \BibitemOpen
  \bibfield  {author} {\bibinfo {author} {\bibfnamefont {I.}~\bibnamefont
  {Affleck}}\ and\ \bibinfo {author} {\bibfnamefont {A.~W.~W.}\ \bibnamefont
  {Ludwig}},\ }\bibfield  {title} {\bibinfo {title} {The {{Kondo}} effect,
  conformal field theory and fusion rules},\ }\href
  {https://doi.org/10.1016/0550-3213(91)90109-B} {\bibfield  {journal}
  {\bibinfo  {journal} {Nuclear Physics B}\ }\textbf {\bibinfo {volume}
  {352}},\ \bibinfo {pages} {849} (\bibinfo {year} {1991})}\BibitemShut
  {NoStop}%
\bibitem [{\citenamefont {Gogolin}\ \emph {et~al.}(2004)\citenamefont
  {Gogolin}, \citenamefont {Nersesyan},\ and\ \citenamefont
  {Tsvelik}}]{gogolinBosonizationStronglyCorrelated2004}%
  \BibitemOpen
  \bibfield  {author} {\bibinfo {author} {\bibfnamefont {A.~O.}\ \bibnamefont
  {Gogolin}}, \bibinfo {author} {\bibfnamefont {A.~A.}\ \bibnamefont
  {Nersesyan}},\ and\ \bibinfo {author} {\bibfnamefont {A.~M.}\ \bibnamefont
  {Tsvelik}},\ }\href@noop {} {\emph {\bibinfo {title} {Bosonization and
  Strongly Correlated Systems}}}\ (\bibinfo  {publisher} {{Cambridge university
  press}},\ \bibinfo {year} {2004})\BibitemShut {NoStop}%
\bibitem [{\citenamefont {Ruderman}\ and\ \citenamefont
  {Kittel}(1954)}]{rudermanIndirectExchangeCoupling1954}%
  \BibitemOpen
  \bibfield  {author} {\bibinfo {author} {\bibfnamefont {M.~A.}\ \bibnamefont
  {Ruderman}}\ and\ \bibinfo {author} {\bibfnamefont {C.}~\bibnamefont
  {Kittel}},\ }\bibfield  {title} {\bibinfo {title} {Indirect {{Exchange
  Coupling}} of {{Nuclear Magnetic Moments}} by {{Conduction Electrons}}},\
  }\href {https://doi.org/10.1103/PhysRev.96.99} {\bibfield  {journal}
  {\bibinfo  {journal} {Phys. Rev.}\ }\textbf {\bibinfo {volume} {96}},\
  \bibinfo {pages} {99} (\bibinfo {year} {1954})}\BibitemShut {NoStop}%
\bibitem [{\citenamefont {Kasuya}(1956)}]{kasuyaTheoryMetallicFerro1956}%
  \BibitemOpen
  \bibfield  {author} {\bibinfo {author} {\bibfnamefont {T.}~\bibnamefont
  {Kasuya}},\ }\bibfield  {title} {\bibinfo {title} {A {{Theory}} of {{Metallic
  Ferro-}} and {{Antiferromagnetism}} on {{Zener}}'s {{Model}}},\ }\href
  {https://doi.org/10.1143/PTP.16.45} {\bibfield  {journal} {\bibinfo
  {journal} {Progress of Theoretical Physics}\ }\textbf {\bibinfo {volume}
  {16}},\ \bibinfo {pages} {45} (\bibinfo {year} {1956})}\BibitemShut {NoStop}%
\bibitem [{\citenamefont {Yosida}(1957)}]{yosidaMagneticPropertiesCuMn1957}%
  \BibitemOpen
  \bibfield  {author} {\bibinfo {author} {\bibfnamefont {K.}~\bibnamefont
  {Yosida}},\ }\bibfield  {title} {\bibinfo {title} {Magnetic {{Properties}} of
  {{Cu-Mn Alloys}}},\ }\href {https://doi.org/10.1103/PhysRev.106.893}
  {\bibfield  {journal} {\bibinfo  {journal} {Phys. Rev.}\ }\textbf {\bibinfo
  {volume} {106}},\ \bibinfo {pages} {893} (\bibinfo {year}
  {1957})}\BibitemShut {NoStop}%
\bibitem [{\citenamefont {Doniach}(1977)}]{doniachKondoLatticeWeak1977}%
  \BibitemOpen
  \bibfield  {author} {\bibinfo {author} {\bibfnamefont {S.}~\bibnamefont
  {Doniach}},\ }\bibfield  {title} {\bibinfo {title} {The {{Kondo}} lattice and
  weak antiferromagnetism},\ }\href
  {https://doi.org/10.1016/0378-4363(77)90190-5} {\bibfield  {journal}
  {\bibinfo  {journal} {Physica B+C}\ }\textbf {\bibinfo {volume} {91}},\
  \bibinfo {pages} {231} (\bibinfo {year} {1977})}\BibitemShut {NoStop}%
\bibitem [{\citenamefont {Jayaprakash}\ \emph {et~al.}(1981)\citenamefont
  {Jayaprakash}, \citenamefont {{Krishna-murthy}},\ and\ \citenamefont
  {Wilkins}}]{jayaprakashTwoImpurityKondoProblem1981}%
  \BibitemOpen
  \bibfield  {author} {\bibinfo {author} {\bibfnamefont {C.}~\bibnamefont
  {Jayaprakash}}, \bibinfo {author} {\bibfnamefont {H.~R.}\ \bibnamefont
  {{Krishna-murthy}}},\ and\ \bibinfo {author} {\bibfnamefont {J.~W.}\
  \bibnamefont {Wilkins}},\ }\bibfield  {title} {\bibinfo {title}
  {Two-{{Impurity Kondo Problem}}},\ }\href
  {https://doi.org/10.1103/PhysRevLett.47.737} {\bibfield  {journal} {\bibinfo
  {journal} {Phys. Rev. Lett.}\ }\textbf {\bibinfo {volume} {47}},\ \bibinfo
  {pages} {737} (\bibinfo {year} {1981})}\BibitemShut {NoStop}%
\bibitem [{\citenamefont {Jones}\ and\ \citenamefont
  {Varma}(1987)}]{jonesStudyTwoMagnetic1987}%
  \BibitemOpen
  \bibfield  {author} {\bibinfo {author} {\bibfnamefont {B.~A.}\ \bibnamefont
  {Jones}}\ and\ \bibinfo {author} {\bibfnamefont {C.~M.}\ \bibnamefont
  {Varma}},\ }\bibfield  {title} {\bibinfo {title} {Study of two magnetic
  impurities in a {{Fermi}} gas},\ }\href
  {https://doi.org/10.1103/PhysRevLett.58.843} {\bibfield  {journal} {\bibinfo
  {journal} {Phys. Rev. Lett.}\ }\textbf {\bibinfo {volume} {58}},\ \bibinfo
  {pages} {843} (\bibinfo {year} {1987})}\BibitemShut {NoStop}%
\bibitem [{\citenamefont {Jones}\ \emph {et~al.}(1988)\citenamefont {Jones},
  \citenamefont {Varma},\ and\ \citenamefont
  {Wilkins}}]{jonesLowTemperaturePropertiesTwoImpurity1988}%
  \BibitemOpen
  \bibfield  {author} {\bibinfo {author} {\bibfnamefont {B.~A.}\ \bibnamefont
  {Jones}}, \bibinfo {author} {\bibfnamefont {C.~M.}\ \bibnamefont {Varma}},\
  and\ \bibinfo {author} {\bibfnamefont {J.~W.}\ \bibnamefont {Wilkins}},\
  }\bibfield  {title} {\bibinfo {title} {Low-{{Temperature Properties}} of the
  {{Two-Impurity Kondo Hamiltonian}}},\ }\href
  {https://doi.org/10.1103/PhysRevLett.61.125} {\bibfield  {journal} {\bibinfo
  {journal} {Phys. Rev. Lett.}\ }\textbf {\bibinfo {volume} {61}},\ \bibinfo
  {pages} {125} (\bibinfo {year} {1988})}\BibitemShut {NoStop}%
\bibitem [{\citenamefont {Fye}\ and\ \citenamefont
  {Hirsch}(1989)}]{fyeQuantumMonteCarlo1989}%
  \BibitemOpen
  \bibfield  {author} {\bibinfo {author} {\bibfnamefont {R.~M.}\ \bibnamefont
  {Fye}}\ and\ \bibinfo {author} {\bibfnamefont {J.~E.}\ \bibnamefont
  {Hirsch}},\ }\bibfield  {title} {\bibinfo {title} {Quantum {{Monte Carlo}}
  study of the two-impurity {{Kondo Hamiltonian}}},\ }\href
  {https://doi.org/10.1103/PhysRevB.40.4780} {\bibfield  {journal} {\bibinfo
  {journal} {Phys. Rev. B}\ }\textbf {\bibinfo {volume} {40}},\ \bibinfo
  {pages} {4780} (\bibinfo {year} {1989})}\BibitemShut {NoStop}%
\bibitem [{\citenamefont {Fye}(1994)}]{fyeAnomalousFixedPoint1994}%
  \BibitemOpen
  \bibfield  {author} {\bibinfo {author} {\bibfnamefont {R.~M.}\ \bibnamefont
  {Fye}},\ }\bibfield  {title} {\bibinfo {title} {``{{Anomalous}} fixed point
  behavior'' of two {{Kondo}} impurities: {{A}} reexamination},\ }\href
  {https://doi.org/10.1103/PhysRevLett.72.916} {\bibfield  {journal} {\bibinfo
  {journal} {Phys. Rev. Lett.}\ }\textbf {\bibinfo {volume} {72}},\ \bibinfo
  {pages} {916} (\bibinfo {year} {1994})}\BibitemShut {NoStop}%
\bibitem [{\citenamefont {Affleck}\ \emph {et~al.}(1995)\citenamefont
  {Affleck}, \citenamefont {Ludwig},\ and\ \citenamefont
  {Jones}}]{affleckConformalfieldtheoryApproachTwoimpurity1995}%
  \BibitemOpen
  \bibfield  {author} {\bibinfo {author} {\bibfnamefont {I.}~\bibnamefont
  {Affleck}}, \bibinfo {author} {\bibfnamefont {A.~W.~W.}\ \bibnamefont
  {Ludwig}},\ and\ \bibinfo {author} {\bibfnamefont {B.~A.}\ \bibnamefont
  {Jones}},\ }\bibfield  {title} {\bibinfo {title} {Conformal-field-theory
  approach to the two-impurity {{Kondo}} problem: {{Comparison}} with numerical
  renormalization-group results},\ }\href
  {https://doi.org/10.1103/PhysRevB.52.9528} {\bibfield  {journal} {\bibinfo
  {journal} {Phys. Rev. B}\ }\textbf {\bibinfo {volume} {52}},\ \bibinfo
  {pages} {9528} (\bibinfo {year} {1995})}\BibitemShut {NoStop}%
\bibitem [{\citenamefont {Gan}(1995)}]{ganSolutionTwoimpurityKondo1995}%
  \BibitemOpen
  \bibfield  {author} {\bibinfo {author} {\bibfnamefont {J.}~\bibnamefont
  {Gan}},\ }\bibfield  {title} {\bibinfo {title} {Solution of the two-impurity
  {{Kondo}} model: {{Critical}} point, {{Fermi-liquid}} phase, and crossover},\
  }\href {https://doi.org/10.1103/PhysRevB.51.8287} {\bibfield  {journal}
  {\bibinfo  {journal} {Phys. Rev. B}\ }\textbf {\bibinfo {volume} {51}},\
  \bibinfo {pages} {8287} (\bibinfo {year} {1995})}\BibitemShut {NoStop}%
\bibitem [{\citenamefont {Silva}\ \emph {et~al.}(1996)\citenamefont {Silva},
  \citenamefont {Lima}, \citenamefont {Oliveira}, \citenamefont {Mello},
  \citenamefont {Oliveira},\ and\ \citenamefont
  {Wilkins}}]{silvaParticleHoleAsymmetryTwoImpurity1996}%
  \BibitemOpen
  \bibfield  {author} {\bibinfo {author} {\bibfnamefont {J.~B.}\ \bibnamefont
  {Silva}}, \bibinfo {author} {\bibfnamefont {W.~L.~C.}\ \bibnamefont {Lima}},
  \bibinfo {author} {\bibfnamefont {W.~C.}\ \bibnamefont {Oliveira}}, \bibinfo
  {author} {\bibfnamefont {J.~L.~N.}\ \bibnamefont {Mello}}, \bibinfo {author}
  {\bibfnamefont {L.~N.}\ \bibnamefont {Oliveira}},\ and\ \bibinfo {author}
  {\bibfnamefont {J.~W.}\ \bibnamefont {Wilkins}},\ }\bibfield  {title}
  {\bibinfo {title} {Particle-{{Hole Asymmetry}} in the {{Two-Impurity Kondo
  Model}}},\ }\href {https://doi.org/10.1103/PhysRevLett.76.275} {\bibfield
  {journal} {\bibinfo  {journal} {Phys. Rev. Lett.}\ }\textbf {\bibinfo
  {volume} {76}},\ \bibinfo {pages} {275} (\bibinfo {year} {1996})}\BibitemShut
  {NoStop}%
\bibitem [{\citenamefont {Andrei}\ \emph {et~al.}(1999)\citenamefont {Andrei},
  \citenamefont {Zim{\'a}nyi},\ and\ \citenamefont
  {Sch{\"o}n}}]{andreiQuantumPhaseTransition1999}%
  \BibitemOpen
  \bibfield  {author} {\bibinfo {author} {\bibfnamefont {N.}~\bibnamefont
  {Andrei}}, \bibinfo {author} {\bibfnamefont {G.~T.}\ \bibnamefont
  {Zim{\'a}nyi}},\ and\ \bibinfo {author} {\bibfnamefont {G.}~\bibnamefont
  {Sch{\"o}n}},\ }\bibfield  {title} {\bibinfo {title} {Quantum phase
  transition in coupled quantum dots},\ }\href
  {https://doi.org/10.1103/PhysRevB.60.R5125} {\bibfield  {journal} {\bibinfo
  {journal} {Phys. Rev. B}\ }\textbf {\bibinfo {volume} {60}},\ \bibinfo
  {pages} {R5125} (\bibinfo {year} {1999})}\BibitemShut {NoStop}%
\bibitem [{\citenamefont {Kane}\ and\ \citenamefont
  {Mele}(2005)}]{kaneQuantumSpinHall2005a}%
  \BibitemOpen
  \bibfield  {author} {\bibinfo {author} {\bibfnamefont {C.~L.}\ \bibnamefont
  {Kane}}\ and\ \bibinfo {author} {\bibfnamefont {E.~J.}\ \bibnamefont
  {Mele}},\ }\bibfield  {title} {\bibinfo {title} {Quantum {{Spin Hall Effect}}
  in {{Graphene}}},\ }\href {https://doi.org/10.1103/PhysRevLett.95.226801}
  {\bibfield  {journal} {\bibinfo  {journal} {Phys. Rev. Lett.}\ }\textbf
  {\bibinfo {volume} {95}},\ \bibinfo {pages} {226801} (\bibinfo {year}
  {2005})}\BibitemShut {NoStop}%
\bibitem [{\citenamefont {Bernevig}\ \emph {et~al.}(2006)\citenamefont
  {Bernevig}, \citenamefont {Hughes},\ and\ \citenamefont
  {Zhang}}]{bernevigQuantumSpinHall2006}%
  \BibitemOpen
  \bibfield  {author} {\bibinfo {author} {\bibfnamefont {B.~A.}\ \bibnamefont
  {Bernevig}}, \bibinfo {author} {\bibfnamefont {T.~L.}\ \bibnamefont
  {Hughes}},\ and\ \bibinfo {author} {\bibfnamefont {S.-C.}\ \bibnamefont
  {Zhang}},\ }\bibfield  {title} {\bibinfo {title} {Quantum {{Spin Hall
  Effect}} and {{Topological Phase Transition}} in {{HgTe Quantum Wells}}},\
  }\href {https://doi.org/10.1126/science.1133734} {\bibfield  {journal}
  {\bibinfo  {journal} {Science}\ }\textbf {\bibinfo {volume} {314}},\ \bibinfo
  {pages} {1757} (\bibinfo {year} {2006})}\BibitemShut {NoStop}%
\bibitem [{\citenamefont {Gao}\ \emph {et~al.}(2009)\citenamefont {Gao},
  \citenamefont {Chen}, \citenamefont {Xie},\ and\ \citenamefont
  {Zhang}}]{gaoInplaneNoncollinearExchange2009}%
  \BibitemOpen
  \bibfield  {author} {\bibinfo {author} {\bibfnamefont {J.}~\bibnamefont
  {Gao}}, \bibinfo {author} {\bibfnamefont {W.}~\bibnamefont {Chen}}, \bibinfo
  {author} {\bibfnamefont {X.~C.}\ \bibnamefont {Xie}},\ and\ \bibinfo {author}
  {\bibfnamefont {F.-c.}\ \bibnamefont {Zhang}},\ }\bibfield  {title} {\bibinfo
  {title} {In-plane noncollinear exchange coupling mediated by helical edge
  states in quantum spin {{Hall}} systems},\ }\href
  {https://doi.org/10.1103/PhysRevB.80.241302} {\bibfield  {journal} {\bibinfo
  {journal} {Phys. Rev. B}\ }\textbf {\bibinfo {volume} {80}},\ \bibinfo
  {pages} {241302} (\bibinfo {year} {2009})}\BibitemShut {NoStop}%
\bibitem [{\citenamefont {Lee}\ and\ \citenamefont
  {Lee}(2015)}]{leeElectricalControlInteraction2015}%
  \BibitemOpen
  \bibfield  {author} {\bibinfo {author} {\bibfnamefont {Y.-W.}\ \bibnamefont
  {Lee}}\ and\ \bibinfo {author} {\bibfnamefont {Y.-L.}\ \bibnamefont {Lee}},\
  }\bibfield  {title} {\bibinfo {title} {Electrical control and interaction
  effects of the {{RKKY}} interaction in helical liquids},\ }\href
  {https://doi.org/10.1103/PhysRevB.91.214431} {\bibfield  {journal} {\bibinfo
  {journal} {Phys. Rev. B}\ }\textbf {\bibinfo {volume} {91}},\ \bibinfo
  {pages} {214431} (\bibinfo {year} {2015})}\BibitemShut {NoStop}%
\bibitem [{\citenamefont {Kurilovich}\ \emph {et~al.}(2017)\citenamefont
  {Kurilovich}, \citenamefont {Kurilovich},\ and\ \citenamefont
  {Burmistrov}}]{kurilovichIndirectExchangeInteraction2017}%
  \BibitemOpen
  \bibfield  {author} {\bibinfo {author} {\bibfnamefont {V.~D.}\ \bibnamefont
  {Kurilovich}}, \bibinfo {author} {\bibfnamefont {P.~D.}\ \bibnamefont
  {Kurilovich}},\ and\ \bibinfo {author} {\bibfnamefont {I.~S.}\ \bibnamefont
  {Burmistrov}},\ }\bibfield  {title} {\bibinfo {title} {Indirect exchange
  interaction between magnetic impurities near the helical edge},\ }\href
  {https://doi.org/10.1103/PhysRevB.95.115430} {\bibfield  {journal} {\bibinfo
  {journal} {Phys. Rev. B}\ }\textbf {\bibinfo {volume} {95}},\ \bibinfo
  {pages} {115430} (\bibinfo {year} {2017})}\BibitemShut {NoStop}%
\bibitem [{\citenamefont {Yevtushenko}\ and\ \citenamefont
  {Yudson}(2018)}]{yevtushenkoKondoImpuritiesCoupled2018}%
  \BibitemOpen
  \bibfield  {author} {\bibinfo {author} {\bibfnamefont {O.~M.}\ \bibnamefont
  {Yevtushenko}}\ and\ \bibinfo {author} {\bibfnamefont {V.~I.}\ \bibnamefont
  {Yudson}},\ }\bibfield  {title} {\bibinfo {title} {Kondo {{Impurities
  Coupled}} to a {{Helical Luttinger Liquid}}: {{RKKY-Kondo Physics
  Revisited}}},\ }\href {https://doi.org/10.1103/PhysRevLett.120.147201}
  {\bibfield  {journal} {\bibinfo  {journal} {Phys. Rev. Lett.}\ }\textbf
  {\bibinfo {volume} {120}},\ \bibinfo {pages} {147201} (\bibinfo {year}
  {2018})}\BibitemShut {NoStop}%
\bibitem [{\citenamefont {Cheianov}\ and\ \citenamefont
  {Glazman}(2013)}]{cheianovMesoscopicFluctuationsConductance2013}%
  \BibitemOpen
  \bibfield  {author} {\bibinfo {author} {\bibfnamefont {V.}~\bibnamefont
  {Cheianov}}\ and\ \bibinfo {author} {\bibfnamefont {L.~I.}\ \bibnamefont
  {Glazman}},\ }\bibfield  {title} {\bibinfo {title} {Mesoscopic
  {{Fluctuations}} of {{Conductance}} of a {{Helical Edge Contaminated}} by
  {{Magnetic Impurities}}},\ }\href
  {https://doi.org/10.1103/PhysRevLett.110.206803} {\bibfield  {journal}
  {\bibinfo  {journal} {Phys. Rev. Lett.}\ }\textbf {\bibinfo {volume} {110}},\
  \bibinfo {pages} {206803} (\bibinfo {year} {2013})}\BibitemShut {NoStop}%
\bibitem [{\citenamefont {Altshuler}\ \emph {et~al.}(2013)\citenamefont
  {Altshuler}, \citenamefont {Aleiner},\ and\ \citenamefont
  {Yudson}}]{altshulerLocalizationEdge2D2013}%
  \BibitemOpen
  \bibfield  {author} {\bibinfo {author} {\bibfnamefont {B.~L.}\ \bibnamefont
  {Altshuler}}, \bibinfo {author} {\bibfnamefont {I.~L.}\ \bibnamefont
  {Aleiner}},\ and\ \bibinfo {author} {\bibfnamefont {V.~I.}\ \bibnamefont
  {Yudson}},\ }\bibfield  {title} {\bibinfo {title} {Localization at the
  {{Edge}} of a {{2D Topological Insulator}} by {{Kondo Impurities}} with
  {{Random Anisotropies}}},\ }\href
  {https://doi.org/10.1103/PhysRevLett.111.086401} {\bibfield  {journal}
  {\bibinfo  {journal} {Phys. Rev. Lett.}\ }\textbf {\bibinfo {volume} {111}},\
  \bibinfo {pages} {086401} (\bibinfo {year} {2013})}\BibitemShut {NoStop}%
\bibitem [{\citenamefont {Hsu}\ \emph {et~al.}(2017)\citenamefont {Hsu},
  \citenamefont {Stano}, \citenamefont {Klinovaja},\ and\ \citenamefont
  {Loss}}]{hsuNuclearspininducedLocalizationEdge2017}%
  \BibitemOpen
  \bibfield  {author} {\bibinfo {author} {\bibfnamefont {C.-H.}\ \bibnamefont
  {Hsu}}, \bibinfo {author} {\bibfnamefont {P.}~\bibnamefont {Stano}}, \bibinfo
  {author} {\bibfnamefont {J.}~\bibnamefont {Klinovaja}},\ and\ \bibinfo
  {author} {\bibfnamefont {D.}~\bibnamefont {Loss}},\ }\bibfield  {title}
  {\bibinfo {title} {Nuclear-spin-induced localization of edge states in
  two-dimensional topological insulators},\ }\href
  {https://doi.org/10.1103/PhysRevB.96.081405} {\bibfield  {journal} {\bibinfo
  {journal} {Phys. Rev. B}\ }\textbf {\bibinfo {volume} {96}},\ \bibinfo
  {pages} {081405} (\bibinfo {year} {2017})}\BibitemShut {NoStop}%
\bibitem [{\citenamefont
  {Halperin}(1982)}]{halperinQuantizedHallConductance1982}%
  \BibitemOpen
  \bibfield  {author} {\bibinfo {author} {\bibfnamefont {B.~I.}\ \bibnamefont
  {Halperin}},\ }\bibfield  {title} {\bibinfo {title} {Quantized {{Hall}}
  conductance, current-carrying edge states, and the existence of extended
  states in a two-dimensional disordered potential},\ }\href
  {https://doi.org/10.1103/PhysRevB.25.2185} {\bibfield  {journal} {\bibinfo
  {journal} {Phys. Rev. B}\ }\textbf {\bibinfo {volume} {25}},\ \bibinfo
  {pages} {2185} (\bibinfo {year} {1982})}\BibitemShut {NoStop}%
\bibitem [{\citenamefont
  {Tsvelick}(1985)}]{tsvelickThermodynamicsMultichannelKondo1985}%
  \BibitemOpen
  \bibfield  {author} {\bibinfo {author} {\bibfnamefont {A.~M.}\ \bibnamefont
  {Tsvelick}},\ }\bibfield  {title} {\bibinfo {title} {The thermodynamics of
  multichannel {{Kondo}} problem},\ }\href
  {https://doi.org/10.1088/0022-3719/18/1/020} {\bibfield  {journal} {\bibinfo
  {journal} {J. Phys. C: Solid State Phys.}\ }\textbf {\bibinfo {volume}
  {18}},\ \bibinfo {pages} {159} (\bibinfo {year} {1985})}\BibitemShut
  {NoStop}%
\bibitem [{\citenamefont
  {Kitaev}(2003)}]{kitaevFaulttolerantQuantumComputation2003}%
  \BibitemOpen
  \bibfield  {author} {\bibinfo {author} {\bibfnamefont {{\relax A.
  Yu}.}~\bibnamefont {Kitaev}},\ }\bibfield  {title} {\bibinfo {title}
  {Fault-tolerant quantum computation by anyons},\ }\href
  {https://doi.org/10.1016/S0003-4916(02)00018-0} {\bibfield  {journal}
  {\bibinfo  {journal} {Annals of Physics}\ }\textbf {\bibinfo {volume}
  {303}},\ \bibinfo {pages} {2} (\bibinfo {year} {2003})}\BibitemShut {NoStop}%
\bibitem [{\citenamefont {Nayak}\ \emph {et~al.}(2008)\citenamefont {Nayak},
  \citenamefont {Simon}, \citenamefont {Stern}, \citenamefont {Freedman},\ and\
  \citenamefont {Das~Sarma}}]{nayakNonAbelianAnyonsTopological2008}%
  \BibitemOpen
  \bibfield  {author} {\bibinfo {author} {\bibfnamefont {C.}~\bibnamefont
  {Nayak}}, \bibinfo {author} {\bibfnamefont {S.~H.}\ \bibnamefont {Simon}},
  \bibinfo {author} {\bibfnamefont {A.}~\bibnamefont {Stern}}, \bibinfo
  {author} {\bibfnamefont {M.}~\bibnamefont {Freedman}},\ and\ \bibinfo
  {author} {\bibfnamefont {S.}~\bibnamefont {Das~Sarma}},\ }\bibfield  {title}
  {\bibinfo {title} {Non-{{Abelian}} anyons and topological quantum
  computation},\ }\href {https://doi.org/10.1103/RevModPhys.80.1083} {\bibfield
   {journal} {\bibinfo  {journal} {Rev. Mod. Phys.}\ }\textbf {\bibinfo
  {volume} {80}},\ \bibinfo {pages} {1083} (\bibinfo {year}
  {2008})}\BibitemShut {NoStop}%
\bibitem [{\citenamefont {Komijani}(2020)}]{komijaniIsolatingKondoAnyons2020}%
  \BibitemOpen
  \bibfield  {author} {\bibinfo {author} {\bibfnamefont {Y.}~\bibnamefont
  {Komijani}},\ }\bibfield  {title} {\bibinfo {title} {Isolating {{Kondo}}
  anyons for topological quantum computation},\ }\href
  {https://doi.org/10.1103/PhysRevB.101.235131} {\bibfield  {journal} {\bibinfo
   {journal} {Phys. Rev. B}\ }\textbf {\bibinfo {volume} {101}},\ \bibinfo
  {pages} {235131} (\bibinfo {year} {2020})}\BibitemShut {NoStop}%
\bibitem [{\citenamefont {Lopes}\ \emph {et~al.}(2020)\citenamefont {Lopes},
  \citenamefont {Affleck},\ and\ \citenamefont
  {Sela}}]{lopesAnyonsMultichannelKondo2020}%
  \BibitemOpen
  \bibfield  {author} {\bibinfo {author} {\bibfnamefont {P.~L.~S.}\
  \bibnamefont {Lopes}}, \bibinfo {author} {\bibfnamefont {I.}~\bibnamefont
  {Affleck}},\ and\ \bibinfo {author} {\bibfnamefont {E.}~\bibnamefont
  {Sela}},\ }\bibfield  {title} {\bibinfo {title} {Anyons in multichannel
  {{Kondo}} systems},\ }\href {https://doi.org/10.1103/PhysRevB.101.085141}
  {\bibfield  {journal} {\bibinfo  {journal} {Phys. Rev. B}\ }\textbf {\bibinfo
  {volume} {101}},\ \bibinfo {pages} {085141} (\bibinfo {year}
  {2020})}\BibitemShut {NoStop}%
\bibitem [{\citenamefont {Gabay}\ \emph {et~al.}(2022)\citenamefont {Gabay},
  \citenamefont {Han}, \citenamefont {Lopes}, \citenamefont {Affleck},\ and\
  \citenamefont {Sela}}]{gabayMultiimpurityChiralKondo2022}%
  \BibitemOpen
  \bibfield  {author} {\bibinfo {author} {\bibfnamefont {D.}~\bibnamefont
  {Gabay}}, \bibinfo {author} {\bibfnamefont {C.}~\bibnamefont {Han}}, \bibinfo
  {author} {\bibfnamefont {P.~L.~S.}\ \bibnamefont {Lopes}}, \bibinfo {author}
  {\bibfnamefont {I.}~\bibnamefont {Affleck}},\ and\ \bibinfo {author}
  {\bibfnamefont {E.}~\bibnamefont {Sela}},\ }\bibfield  {title} {\bibinfo
  {title} {Multi-impurity chiral {{Kondo}} model: {{Correlation}} functions and
  anyon fusion rules},\ }\href {https://doi.org/10.1103/PhysRevB.105.035151}
  {\bibfield  {journal} {\bibinfo  {journal} {Phys. Rev. B}\ }\textbf {\bibinfo
  {volume} {105}},\ \bibinfo {pages} {035151} (\bibinfo {year}
  {2022})}\BibitemShut {NoStop}%
\bibitem [{\citenamefont {Lotem}\ \emph {et~al.}(2022)\citenamefont {Lotem},
  \citenamefont {Sela},\ and\ \citenamefont
  {Goldstein}}]{lotemManipulatingNonAbelianAnyons2022}%
  \BibitemOpen
  \bibfield  {author} {\bibinfo {author} {\bibfnamefont {M.}~\bibnamefont
  {Lotem}}, \bibinfo {author} {\bibfnamefont {E.}~\bibnamefont {Sela}},\ and\
  \bibinfo {author} {\bibfnamefont {M.}~\bibnamefont {Goldstein}},\ }\bibfield
  {title} {\bibinfo {title} {Manipulating {{Non-Abelian Anyons}} in a {{Chiral
  Multichannel Kondo Model}}},\ }\href
  {https://doi.org/10.1103/PhysRevLett.129.227703} {\bibfield  {journal}
  {\bibinfo  {journal} {Phys. Rev. Lett.}\ }\textbf {\bibinfo {volume} {129}},\
  \bibinfo {pages} {227703} (\bibinfo {year} {2022})}\BibitemShut {NoStop}%
\bibitem [{\citenamefont {Iftikhar}\ \emph {et~al.}(2015)\citenamefont
  {Iftikhar}, \citenamefont {Jezouin}, \citenamefont {Anthore}, \citenamefont
  {Gennser}, \citenamefont {Parmentier}, \citenamefont {Cavanna},\ and\
  \citenamefont {Pierre}}]{iftikharTwochannelKondoEffect2015}%
  \BibitemOpen
  \bibfield  {author} {\bibinfo {author} {\bibfnamefont {Z.}~\bibnamefont
  {Iftikhar}}, \bibinfo {author} {\bibfnamefont {S.}~\bibnamefont {Jezouin}},
  \bibinfo {author} {\bibfnamefont {A.}~\bibnamefont {Anthore}}, \bibinfo
  {author} {\bibfnamefont {U.}~\bibnamefont {Gennser}}, \bibinfo {author}
  {\bibfnamefont {F.~D.}\ \bibnamefont {Parmentier}}, \bibinfo {author}
  {\bibfnamefont {A.}~\bibnamefont {Cavanna}},\ and\ \bibinfo {author}
  {\bibfnamefont {F.}~\bibnamefont {Pierre}},\ }\bibfield  {title} {\bibinfo
  {title} {Two-channel {{Kondo}} effect and renormalization flow with
  macroscopic quantum charge states},\ }\href
  {https://doi.org/10.1038/nature15384} {\bibfield  {journal} {\bibinfo
  {journal} {Nature}\ }\textbf {\bibinfo {volume} {526}},\ \bibinfo {pages}
  {233} (\bibinfo {year} {2015})}\BibitemShut {NoStop}%
\bibitem [{\citenamefont {Iftikhar}\ \emph {et~al.}(2018)\citenamefont
  {Iftikhar}, \citenamefont {Anthore}, \citenamefont {Mitchell}, \citenamefont
  {Parmentier}, \citenamefont {Gennser}, \citenamefont {Ouerghi}, \citenamefont
  {Cavanna}, \citenamefont {Mora}, \citenamefont {Simon},\ and\ \citenamefont
  {Pierre}}]{iftikharTunableQuantumCriticality2018}%
  \BibitemOpen
  \bibfield  {author} {\bibinfo {author} {\bibfnamefont {Z.}~\bibnamefont
  {Iftikhar}}, \bibinfo {author} {\bibfnamefont {A.}~\bibnamefont {Anthore}},
  \bibinfo {author} {\bibfnamefont {A.~K.}\ \bibnamefont {Mitchell}}, \bibinfo
  {author} {\bibfnamefont {F.~D.}\ \bibnamefont {Parmentier}}, \bibinfo
  {author} {\bibfnamefont {U.}~\bibnamefont {Gennser}}, \bibinfo {author}
  {\bibfnamefont {A.}~\bibnamefont {Ouerghi}}, \bibinfo {author} {\bibfnamefont
  {A.}~\bibnamefont {Cavanna}}, \bibinfo {author} {\bibfnamefont
  {C.}~\bibnamefont {Mora}}, \bibinfo {author} {\bibfnamefont {P.}~\bibnamefont
  {Simon}},\ and\ \bibinfo {author} {\bibfnamefont {F.}~\bibnamefont
  {Pierre}},\ }\bibfield  {title} {\bibinfo {title} {Tunable quantum
  criticality and super-ballistic transport in a ``charge'' {{Kondo}}
  circuit},\ }\href {https://doi.org/10.1126/science.aan5592} {\bibfield
  {journal} {\bibinfo  {journal} {Science}\ }\textbf {\bibinfo {volume}
  {360}},\ \bibinfo {pages} {1315} (\bibinfo {year} {2018})}\BibitemShut
  {NoStop}%
\bibitem [{\citenamefont {Landau}\ \emph {et~al.}(2018)\citenamefont {Landau},
  \citenamefont {Cornfeld},\ and\ \citenamefont
  {Sela}}]{landauChargeFractionalizationTwoChannel2018}%
  \BibitemOpen
  \bibfield  {author} {\bibinfo {author} {\bibfnamefont {L.~A.}\ \bibnamefont
  {Landau}}, \bibinfo {author} {\bibfnamefont {E.}~\bibnamefont {Cornfeld}},\
  and\ \bibinfo {author} {\bibfnamefont {E.}~\bibnamefont {Sela}},\ }\bibfield
  {title} {\bibinfo {title} {Charge {{Fractionalization}} in the {{Two-Channel
  Kondo Effect}}},\ }\href {https://doi.org/10.1103/PhysRevLett.120.186801}
  {\bibfield  {journal} {\bibinfo  {journal} {Phys. Rev. Lett.}\ }\textbf
  {\bibinfo {volume} {120}},\ \bibinfo {pages} {186801} (\bibinfo {year}
  {2018})}\BibitemShut {NoStop}%
\bibitem [{\citenamefont {{van Dalum}}\ \emph {et~al.}(2020)\citenamefont {{van
  Dalum}}, \citenamefont {Mitchell},\ and\ \citenamefont
  {Fritz}}]{vandalumWiedemannFranzLawNonFermi2020}%
  \BibitemOpen
  \bibfield  {author} {\bibinfo {author} {\bibfnamefont {G.~A.~R.}\
  \bibnamefont {{van Dalum}}}, \bibinfo {author} {\bibfnamefont {A.~K.}\
  \bibnamefont {Mitchell}},\ and\ \bibinfo {author} {\bibfnamefont
  {L.}~\bibnamefont {Fritz}},\ }\bibfield  {title} {\bibinfo {title}
  {Wiedemann-{{Franz}} law in a non-{{Fermi}} liquid and {{Majorana}} central
  charge: {{Thermoelectric}} transport in a two-channel {{Kondo}} system},\
  }\href {https://doi.org/10.1103/PhysRevB.102.041111} {\bibfield  {journal}
  {\bibinfo  {journal} {Phys. Rev. B}\ }\textbf {\bibinfo {volume} {102}},\
  \bibinfo {pages} {041111} (\bibinfo {year} {2020})}\BibitemShut {NoStop}%
\bibitem [{\citenamefont {Nguyen}\ and\ \citenamefont
  {Kiselev}(2020)}]{nguyenThermoelectricTransportThreeChannel2020}%
  \BibitemOpen
  \bibfield  {author} {\bibinfo {author} {\bibfnamefont {T.~K.~T.}\
  \bibnamefont {Nguyen}}\ and\ \bibinfo {author} {\bibfnamefont {M.~N.}\
  \bibnamefont {Kiselev}},\ }\bibfield  {title} {\bibinfo {title}
  {Thermoelectric {{Transport}} in a {{Three-Channel Charge Kondo Circuit}}},\
  }\href {https://doi.org/10.1103/PhysRevLett.125.026801} {\bibfield  {journal}
  {\bibinfo  {journal} {Phys. Rev. Lett.}\ }\textbf {\bibinfo {volume} {125}},\
  \bibinfo {pages} {026801} (\bibinfo {year} {2020})}\BibitemShut {NoStop}%
\bibitem [{\citenamefont {Han}\ \emph {et~al.}(2022)\citenamefont {Han},
  \citenamefont {Iftikhar}, \citenamefont {Kleeorin}, \citenamefont {Anthore},
  \citenamefont {Pierre}, \citenamefont {Meir}, \citenamefont {Mitchell},\ and\
  \citenamefont {Sela}}]{hanFractionalEntropyMultichannel2022}%
  \BibitemOpen
  \bibfield  {author} {\bibinfo {author} {\bibfnamefont {C.}~\bibnamefont
  {Han}}, \bibinfo {author} {\bibfnamefont {Z.}~\bibnamefont {Iftikhar}},
  \bibinfo {author} {\bibfnamefont {Y.}~\bibnamefont {Kleeorin}}, \bibinfo
  {author} {\bibfnamefont {A.}~\bibnamefont {Anthore}}, \bibinfo {author}
  {\bibfnamefont {F.}~\bibnamefont {Pierre}}, \bibinfo {author} {\bibfnamefont
  {Y.}~\bibnamefont {Meir}}, \bibinfo {author} {\bibfnamefont {A.~K.}\
  \bibnamefont {Mitchell}},\ and\ \bibinfo {author} {\bibfnamefont
  {E.}~\bibnamefont {Sela}},\ }\bibfield  {title} {\bibinfo {title} {Fractional
  {{Entropy}} of {{Multichannel Kondo Systems}} from {{Conductance-Charge
  Relations}}},\ }\href {https://doi.org/10.1103/PhysRevLett.128.146803}
  {\bibfield  {journal} {\bibinfo  {journal} {Phys. Rev. Lett.}\ }\textbf
  {\bibinfo {volume} {128}},\ \bibinfo {pages} {146803} (\bibinfo {year}
  {2022})}\BibitemShut {NoStop}%
\bibitem [{\citenamefont {Pouse}\ \emph {et~al.}(2023)\citenamefont {Pouse},
  \citenamefont {Peeters}, \citenamefont {Hsueh}, \citenamefont {Gennser},
  \citenamefont {Cavanna}, \citenamefont {Kastner}, \citenamefont {Mitchell},\
  and\ \citenamefont {{Goldhaber-Gordon}}}]{pouseQuantumSimulationExotic2023}%
  \BibitemOpen
  \bibfield  {author} {\bibinfo {author} {\bibfnamefont {W.}~\bibnamefont
  {Pouse}}, \bibinfo {author} {\bibfnamefont {L.}~\bibnamefont {Peeters}},
  \bibinfo {author} {\bibfnamefont {C.~L.}\ \bibnamefont {Hsueh}}, \bibinfo
  {author} {\bibfnamefont {U.}~\bibnamefont {Gennser}}, \bibinfo {author}
  {\bibfnamefont {A.}~\bibnamefont {Cavanna}}, \bibinfo {author} {\bibfnamefont
  {M.~A.}\ \bibnamefont {Kastner}}, \bibinfo {author} {\bibfnamefont {A.~K.}\
  \bibnamefont {Mitchell}},\ and\ \bibinfo {author} {\bibfnamefont
  {D.}~\bibnamefont {{Goldhaber-Gordon}}},\ }\bibfield  {title} {\bibinfo
  {title} {Quantum simulation of an exotic quantum critical point in a two-site
  charge {{Kondo}} circuit},\ }\href
  {https://doi.org/10.1038/s41567-022-01905-4} {\bibfield  {journal} {\bibinfo
  {journal} {Nat. Phys.}\ ,\ \bibinfo {pages} {1}} (\bibinfo {year}
  {2023})}\BibitemShut {NoStop}%
\bibitem [{\citenamefont
  {Wilson}(1975)}]{wilsonRenormalizationGroupCritical1975}%
  \BibitemOpen
  \bibfield  {author} {\bibinfo {author} {\bibfnamefont {K.~G.}\ \bibnamefont
  {Wilson}},\ }\bibfield  {title} {\bibinfo {title} {The renormalization group:
  {{Critical}} phenomena and the {{Kondo}} problem},\ }\href
  {https://doi.org/10.1103/RevModPhys.47.773} {\bibfield  {journal} {\bibinfo
  {journal} {Rev. Mod. Phys.}\ }\textbf {\bibinfo {volume} {47}},\ \bibinfo
  {pages} {773} (\bibinfo {year} {1975})}\BibitemShut {NoStop}%
\bibitem [{\citenamefont {Bulla}\ \emph {et~al.}(2008)\citenamefont {Bulla},
  \citenamefont {Costi},\ and\ \citenamefont
  {Pruschke}}]{bullaNumericalRenormalizationGroup2008}%
  \BibitemOpen
  \bibfield  {author} {\bibinfo {author} {\bibfnamefont {R.}~\bibnamefont
  {Bulla}}, \bibinfo {author} {\bibfnamefont {T.~A.}\ \bibnamefont {Costi}},\
  and\ \bibinfo {author} {\bibfnamefont {T.}~\bibnamefont {Pruschke}},\
  }\bibfield  {title} {\bibinfo {title} {Numerical renormalization group method
  for quantum impurity systems},\ }\href
  {https://doi.org/10.1103/RevModPhys.80.395} {\bibfield  {journal} {\bibinfo
  {journal} {Rev. Mod. Phys.}\ }\textbf {\bibinfo {volume} {80}},\ \bibinfo
  {pages} {395} (\bibinfo {year} {2008})}\BibitemShut {NoStop}%
\bibitem [{\citenamefont {Ingersent}\ \emph {et~al.}(1992)\citenamefont
  {Ingersent}, \citenamefont {Jones},\ and\ \citenamefont
  {Wilkins}}]{ingersentStudyTwoimpurityTwochannel1992}%
  \BibitemOpen
  \bibfield  {author} {\bibinfo {author} {\bibfnamefont {K.}~\bibnamefont
  {Ingersent}}, \bibinfo {author} {\bibfnamefont {B.~A.}\ \bibnamefont
  {Jones}},\ and\ \bibinfo {author} {\bibfnamefont {J.~W.}\ \bibnamefont
  {Wilkins}},\ }\bibfield  {title} {\bibinfo {title} {Study of the
  two-impurity, two-channel {{Kondo Hamiltonian}}},\ }\href
  {https://doi.org/10.1103/PhysRevLett.69.2594} {\bibfield  {journal} {\bibinfo
   {journal} {Phys. Rev. Lett.}\ }\textbf {\bibinfo {volume} {69}},\ \bibinfo
  {pages} {2594} (\bibinfo {year} {1992})}\BibitemShut {NoStop}%
\bibitem [{\citenamefont {Campo}\ and\ \citenamefont
  {Oliveira}(2005)}]{campoAlternativeDiscretizationNumerical2005}%
  \BibitemOpen
  \bibfield  {author} {\bibinfo {author} {\bibfnamefont {V.~L.}\ \bibnamefont
  {Campo}}\ and\ \bibinfo {author} {\bibfnamefont {L.~N.}\ \bibnamefont
  {Oliveira}},\ }\bibfield  {title} {\bibinfo {title} {Alternative
  discretization in the numerical renormalization-group method},\ }\href
  {https://doi.org/10.1103/PhysRevB.72.104432} {\bibfield  {journal} {\bibinfo
  {journal} {Phys. Rev. B}\ }\textbf {\bibinfo {volume} {72}},\ \bibinfo
  {pages} {104432} (\bibinfo {year} {2005})}\BibitemShut {NoStop}%
\bibitem [{\citenamefont {Mitchell}\ \emph {et~al.}(2012)\citenamefont
  {Mitchell}, \citenamefont {Sela},\ and\ \citenamefont
  {Logan}}]{mitchellTwoChannelKondoPhysics2012}%
  \BibitemOpen
  \bibfield  {author} {\bibinfo {author} {\bibfnamefont {A.~K.}\ \bibnamefont
  {Mitchell}}, \bibinfo {author} {\bibfnamefont {E.}~\bibnamefont {Sela}},\
  and\ \bibinfo {author} {\bibfnamefont {D.~E.}\ \bibnamefont {Logan}},\
  }\bibfield  {title} {\bibinfo {title} {Two-{{Channel Kondo Physics}} in
  {{Two-Impurity Kondo Models}}},\ }\href
  {https://doi.org/10.1103/PhysRevLett.108.086405} {\bibfield  {journal}
  {\bibinfo  {journal} {Phys. Rev. Lett.}\ }\textbf {\bibinfo {volume} {108}},\
  \bibinfo {pages} {086405} (\bibinfo {year} {2012})}\BibitemShut {NoStop}%
\bibitem [{\citenamefont {Mitchell}\ \emph {et~al.}(2015)\citenamefont
  {Mitchell}, \citenamefont {Derry},\ and\ \citenamefont
  {Logan}}]{mitchellMultipleMagneticImpurities2015}%
  \BibitemOpen
  \bibfield  {author} {\bibinfo {author} {\bibfnamefont {A.~K.}\ \bibnamefont
  {Mitchell}}, \bibinfo {author} {\bibfnamefont {P.~G.}\ \bibnamefont
  {Derry}},\ and\ \bibinfo {author} {\bibfnamefont {D.~E.}\ \bibnamefont
  {Logan}},\ }\bibfield  {title} {\bibinfo {title} {Multiple magnetic
  impurities on surfaces: {{Scattering}} and quasiparticle interference},\
  }\href {https://doi.org/10.1103/PhysRevB.91.235127} {\bibfield  {journal}
  {\bibinfo  {journal} {Phys. Rev. B}\ }\textbf {\bibinfo {volume} {91}},\
  \bibinfo {pages} {235127} (\bibinfo {year} {2015})}\BibitemShut {NoStop}%
\bibitem [{\citenamefont {Lechtenberg}\ \emph {et~al.}(2017)\citenamefont
  {Lechtenberg}, \citenamefont {Eickhoff},\ and\ \citenamefont
  {Anders}}]{lechtenbergRealisticQuantumCritical2017}%
  \BibitemOpen
  \bibfield  {author} {\bibinfo {author} {\bibfnamefont {B.}~\bibnamefont
  {Lechtenberg}}, \bibinfo {author} {\bibfnamefont {F.}~\bibnamefont
  {Eickhoff}},\ and\ \bibinfo {author} {\bibfnamefont {F.~B.}\ \bibnamefont
  {Anders}},\ }\bibfield  {title} {\bibinfo {title} {Realistic quantum critical
  point in one-dimensional two-impurity models},\ }\href
  {https://doi.org/10.1103/PhysRevB.96.041109} {\bibfield  {journal} {\bibinfo
  {journal} {Phys. Rev. B}\ }\textbf {\bibinfo {volume} {96}},\ \bibinfo
  {pages} {041109} (\bibinfo {year} {2017})}\BibitemShut {NoStop}%
\bibitem [{\citenamefont {Eickhoff}\ \emph {et~al.}(2018)\citenamefont
  {Eickhoff}, \citenamefont {Lechtenberg},\ and\ \citenamefont
  {Anders}}]{eickhoffEffectiveLowenergyDescription2018}%
  \BibitemOpen
  \bibfield  {author} {\bibinfo {author} {\bibfnamefont {F.}~\bibnamefont
  {Eickhoff}}, \bibinfo {author} {\bibfnamefont {B.}~\bibnamefont
  {Lechtenberg}},\ and\ \bibinfo {author} {\bibfnamefont {F.~B.}\ \bibnamefont
  {Anders}},\ }\bibfield  {title} {\bibinfo {title} {Effective low-energy
  description of the two-impurity {{Anderson}} model: {{RKKY}} interaction and
  quantum criticality},\ }\href {https://doi.org/10.1103/PhysRevB.98.115103}
  {\bibfield  {journal} {\bibinfo  {journal} {Phys. Rev. B}\ }\textbf {\bibinfo
  {volume} {98}},\ \bibinfo {pages} {115103} (\bibinfo {year}
  {2018})}\BibitemShut {NoStop}%
\bibitem [{\citenamefont {Eickhoff}\ and\ \citenamefont
  {Anders}(2020)}]{eickhoffStronglyCorrelatedMultiimpurity2020}%
  \BibitemOpen
  \bibfield  {author} {\bibinfo {author} {\bibfnamefont {F.}~\bibnamefont
  {Eickhoff}}\ and\ \bibinfo {author} {\bibfnamefont {F.~B.}\ \bibnamefont
  {Anders}},\ }\bibfield  {title} {\bibinfo {title} {Strongly correlated
  multi-impurity models: {{The}} crossover from a single-impurity problem to
  lattice models},\ }\href {https://doi.org/10.1103/PhysRevB.102.205132}
  {\bibfield  {journal} {\bibinfo  {journal} {Phys. Rev. B}\ }\textbf {\bibinfo
  {volume} {102}},\ \bibinfo {pages} {205132} (\bibinfo {year}
  {2020})}\BibitemShut {NoStop}%
\bibitem [{\citenamefont {Liu}\ \emph {et~al.}(2016)\citenamefont {Liu},
  \citenamefont {Wang},\ and\ \citenamefont
  {Wang}}]{liuQuantumImpuritiesChannel2016}%
  \BibitemOpen
  \bibfield  {author} {\bibinfo {author} {\bibfnamefont {J.-G.}\ \bibnamefont
  {Liu}}, \bibinfo {author} {\bibfnamefont {D.}~\bibnamefont {Wang}},\ and\
  \bibinfo {author} {\bibfnamefont {Q.-H.}\ \bibnamefont {Wang}},\ }\bibfield
  {title} {\bibinfo {title} {Quantum impurities in channel mixing baths},\
  }\href {https://doi.org/10.1103/PhysRevB.93.035102} {\bibfield  {journal}
  {\bibinfo  {journal} {Phys. Rev. B}\ }\textbf {\bibinfo {volume} {93}},\
  \bibinfo {pages} {035102} (\bibinfo {year} {2016})}\BibitemShut {NoStop}%
\bibitem [{\citenamefont {Bruognolo}\ \emph {et~al.}(2017)\citenamefont
  {Bruognolo}, \citenamefont {Linden}, \citenamefont {Schwarz}, \citenamefont
  {Lee}, \citenamefont {Stadler}, \citenamefont {Weichselbaum}, \citenamefont
  {Vojta}, \citenamefont {Anders},\ and\ \citenamefont {{von
  Delft}}}]{bruognoloOpenWilsonChains2017}%
  \BibitemOpen
  \bibfield  {author} {\bibinfo {author} {\bibfnamefont {B.}~\bibnamefont
  {Bruognolo}}, \bibinfo {author} {\bibfnamefont {N.-O.}\ \bibnamefont
  {Linden}}, \bibinfo {author} {\bibfnamefont {F.}~\bibnamefont {Schwarz}},
  \bibinfo {author} {\bibfnamefont {S.-S.~B.}\ \bibnamefont {Lee}}, \bibinfo
  {author} {\bibfnamefont {K.}~\bibnamefont {Stadler}}, \bibinfo {author}
  {\bibfnamefont {A.}~\bibnamefont {Weichselbaum}}, \bibinfo {author}
  {\bibfnamefont {M.}~\bibnamefont {Vojta}}, \bibinfo {author} {\bibfnamefont
  {F.~B.}\ \bibnamefont {Anders}},\ and\ \bibinfo {author} {\bibfnamefont
  {J.}~\bibnamefont {{von Delft}}},\ }\bibfield  {title} {\bibinfo {title}
  {Open {{Wilson}} chains for quantum impurity models: {{Keeping}} track of all
  bath modes},\ }\href {https://doi.org/10.1103/PhysRevB.95.121115} {\bibfield
  {journal} {\bibinfo  {journal} {Phys. Rev. B}\ }\textbf {\bibinfo {volume}
  {95}},\ \bibinfo {pages} {121115} (\bibinfo {year} {2017})}\BibitemShut
  {NoStop}%
\bibitem [{\citenamefont {{\v
  Z}itko}(2009)}]{zitkoAdaptiveLogarithmicDiscretization2009}%
  \BibitemOpen
  \bibfield  {author} {\bibinfo {author} {\bibfnamefont {R.}~\bibnamefont {{\v
  Z}itko}},\ }\bibfield  {title} {\bibinfo {title} {Adaptive logarithmic
  discretization for numerical renormalization group methods},\ }\href
  {https://doi.org/10.1016/j.cpc.2009.02.007} {\bibfield  {journal} {\bibinfo
  {journal} {Computer Physics Communications}\ }\textbf {\bibinfo {volume}
  {180}},\ \bibinfo {pages} {1271} (\bibinfo {year} {2009})}\BibitemShut
  {NoStop}%
\bibitem [{\citenamefont {{\v Z}itko}\ and\ \citenamefont
  {Pruschke}(2009)}]{zitkoEnergyResolutionDiscretization2009}%
  \BibitemOpen
  \bibfield  {author} {\bibinfo {author} {\bibfnamefont {R.}~\bibnamefont {{\v
  Z}itko}}\ and\ \bibinfo {author} {\bibfnamefont {T.}~\bibnamefont
  {Pruschke}},\ }\bibfield  {title} {\bibinfo {title} {Energy resolution and
  discretization artifacts in the numerical renormalization group},\ }\href
  {https://doi.org/10.1103/PhysRevB.79.085106} {\bibfield  {journal} {\bibinfo
  {journal} {Phys. Rev. B}\ }\textbf {\bibinfo {volume} {79}},\ \bibinfo
  {pages} {085106} (\bibinfo {year} {2009})}\BibitemShut {NoStop}%
\bibitem [{\citenamefont
  {Weichselbaum}(2012{\natexlab{a}})}]{weichselbaumTensorNetworksNumerical2012}%
  \BibitemOpen
  \bibfield  {author} {\bibinfo {author} {\bibfnamefont {A.}~\bibnamefont
  {Weichselbaum}},\ }\bibfield  {title} {\bibinfo {title} {Tensor networks and
  the numerical renormalization group},\ }\href
  {https://doi.org/10.1103/PhysRevB.86.245124} {\bibfield  {journal} {\bibinfo
  {journal} {Phys. Rev. B}\ }\textbf {\bibinfo {volume} {86}},\ \bibinfo
  {pages} {245124} (\bibinfo {year} {2012}{\natexlab{a}})}\BibitemShut
  {NoStop}%
\bibitem [{\citenamefont {Mitchell}\ \emph {et~al.}(2014)\citenamefont
  {Mitchell}, \citenamefont {Galpin}, \citenamefont {{Wilson-Fletcher}},
  \citenamefont {Logan},\ and\ \citenamefont
  {Bulla}}]{mitchellGeneralizedWilsonChain2014}%
  \BibitemOpen
  \bibfield  {author} {\bibinfo {author} {\bibfnamefont {A.~K.}\ \bibnamefont
  {Mitchell}}, \bibinfo {author} {\bibfnamefont {M.~R.}\ \bibnamefont
  {Galpin}}, \bibinfo {author} {\bibfnamefont {S.}~\bibnamefont
  {{Wilson-Fletcher}}}, \bibinfo {author} {\bibfnamefont {D.~E.}\ \bibnamefont
  {Logan}},\ and\ \bibinfo {author} {\bibfnamefont {R.}~\bibnamefont {Bulla}},\
  }\bibfield  {title} {\bibinfo {title} {Generalized {{Wilson}} chain for
  solving multichannel quantum impurity problems},\ }\href
  {https://doi.org/10.1103/PhysRevB.89.121105} {\bibfield  {journal} {\bibinfo
  {journal} {Phys. Rev. B}\ }\textbf {\bibinfo {volume} {89}},\ \bibinfo
  {pages} {121105} (\bibinfo {year} {2014})}\BibitemShut {NoStop}%
\bibitem [{\citenamefont {Stadler}\ \emph {et~al.}(2016)\citenamefont
  {Stadler}, \citenamefont {Mitchell}, \citenamefont {{von Delft}},\ and\
  \citenamefont
  {Weichselbaum}}]{stadlerInterleavedNumericalRenormalization2016}%
  \BibitemOpen
  \bibfield  {author} {\bibinfo {author} {\bibfnamefont {K.~M.}\ \bibnamefont
  {Stadler}}, \bibinfo {author} {\bibfnamefont {A.~K.}\ \bibnamefont
  {Mitchell}}, \bibinfo {author} {\bibfnamefont {J.}~\bibnamefont {{von
  Delft}}},\ and\ \bibinfo {author} {\bibfnamefont {A.}~\bibnamefont
  {Weichselbaum}},\ }\bibfield  {title} {\bibinfo {title} {Interleaved
  numerical renormalization group as an efficient multiband impurity solver},\
  }\href {https://doi.org/10.1103/PhysRevB.93.235101} {\bibfield  {journal}
  {\bibinfo  {journal} {Phys. Rev. B}\ }\textbf {\bibinfo {volume} {93}},\
  \bibinfo {pages} {235101} (\bibinfo {year} {2016})}\BibitemShut {NoStop}%
\bibitem [{\citenamefont
  {Weichselbaum}(2012{\natexlab{b}})}]{weichselbaumNonabelianSymmetriesTensor2012}%
  \BibitemOpen
  \bibfield  {author} {\bibinfo {author} {\bibfnamefont {A.}~\bibnamefont
  {Weichselbaum}},\ }\bibfield  {title} {\bibinfo {title} {Non-abelian
  symmetries in tensor networks: {{A}} quantum symmetry space approach},\
  }\href {https://doi.org/10.1016/j.aop.2012.07.009} {\bibfield  {journal}
  {\bibinfo  {journal} {Annals of Physics}\ }\textbf {\bibinfo {volume}
  {327}},\ \bibinfo {pages} {2972} (\bibinfo {year}
  {2012}{\natexlab{b}})}\BibitemShut {NoStop}%
\bibitem [{\citenamefont
  {Weichselbaum}(2020)}]{weichselbaumXsymbolsNonAbelianSymmetries2020}%
  \BibitemOpen
  \bibfield  {author} {\bibinfo {author} {\bibfnamefont {A.}~\bibnamefont
  {Weichselbaum}},\ }\bibfield  {title} {\bibinfo {title} {X-symbols for
  non-{{Abelian}} symmetries in tensor networks},\ }\href
  {https://doi.org/10.1103/PhysRevResearch.2.023385} {\bibfield  {journal}
  {\bibinfo  {journal} {Phys. Rev. Research}\ }\textbf {\bibinfo {volume}
  {2}},\ \bibinfo {pages} {023385} (\bibinfo {year} {2020})}\BibitemShut
  {NoStop}%
\bibitem [{\citenamefont {Schrieffer}\ and\ \citenamefont
  {Wolff}(1966)}]{schriefferRelationAndersonKondo1966a}%
  \BibitemOpen
  \bibfield  {author} {\bibinfo {author} {\bibfnamefont {J.~R.}\ \bibnamefont
  {Schrieffer}}\ and\ \bibinfo {author} {\bibfnamefont {P.~A.}\ \bibnamefont
  {Wolff}},\ }\bibfield  {title} {\bibinfo {title} {Relation between the
  {{Anderson}} and {{Kondo Hamiltonians}}},\ }\href
  {https://doi.org/10.1103/PhysRev.149.491} {\bibfield  {journal} {\bibinfo
  {journal} {Phys. Rev.}\ }\textbf {\bibinfo {volume} {149}},\ \bibinfo {pages}
  {491} (\bibinfo {year} {1966})}\BibitemShut {NoStop}%
\bibitem [{\citenamefont {Horodecki}\ \emph {et~al.}(2009)\citenamefont
  {Horodecki}, \citenamefont {Horodecki}, \citenamefont {Horodecki},\ and\
  \citenamefont {Horodecki}}]{horodeckiQuantumEntanglement2009}%
  \BibitemOpen
  \bibfield  {author} {\bibinfo {author} {\bibfnamefont {R.}~\bibnamefont
  {Horodecki}}, \bibinfo {author} {\bibfnamefont {P.}~\bibnamefont
  {Horodecki}}, \bibinfo {author} {\bibfnamefont {M.}~\bibnamefont
  {Horodecki}},\ and\ \bibinfo {author} {\bibfnamefont {K.}~\bibnamefont
  {Horodecki}},\ }\bibfield  {title} {\bibinfo {title} {Quantum entanglement},\
  }\href {https://doi.org/10.1103/RevModPhys.81.865} {\bibfield  {journal}
  {\bibinfo  {journal} {Rev. Mod. Phys.}\ }\textbf {\bibinfo {volume} {81}},\
  \bibinfo {pages} {865} (\bibinfo {year} {2009})}\BibitemShut {NoStop}%
\bibitem [{\citenamefont
  {Kamenev}(2011)}]{kamenevFieldTheoryNonEquilibrium2011}%
  \BibitemOpen
  \bibfield  {author} {\bibinfo {author} {\bibfnamefont {A.}~\bibnamefont
  {Kamenev}},\ }\href {https://doi.org/10.1017/CBO9781139003667} {\emph
  {\bibinfo {title} {Field {{Theory}} of {{Non-Equilibrium Systems}}}}}\
  (\bibinfo  {publisher} {{Cambridge University Press}},\ \bibinfo {address}
  {{Cambridge}},\ \bibinfo {year} {2011})\BibitemShut {NoStop}%
\bibitem [{\citenamefont {Ferrer}\ \emph {et~al.}(2022)\citenamefont {Ferrer},
  \citenamefont {Yevtushenko},\ and\ \citenamefont
  {Weichselbaum}}]{ferrerRKKYKondoCrossover2022}%
  \BibitemOpen
  \bibfield  {author} {\bibinfo {author} {\bibfnamefont {P.~A.-C.}\
  \bibnamefont {Ferrer}}, \bibinfo {author} {\bibfnamefont {O.~M.}\
  \bibnamefont {Yevtushenko}},\ and\ \bibinfo {author} {\bibfnamefont
  {A.}~\bibnamefont {Weichselbaum}},\ }\href
  {https://doi.org/10.48550/arXiv.2208.02275} {\bibinfo {title} {{{RKKY}} to
  {{Kondo}} crossover in {{Helical Edge}} of a {{Topological Insulator}}}}
  (\bibinfo {year} {2022}),\ \Eprint {https://arxiv.org/abs/arXiv:2208.02275}
  {arxiv:arXiv:2208.02275} \BibitemShut {NoStop}%
\bibitem [{\citenamefont {Lee}\ and\ \citenamefont
  {Weichselbaum}(2016)}]{leeAdaptiveBroadeningImprove2016}%
  \BibitemOpen
  \bibfield  {author} {\bibinfo {author} {\bibfnamefont {S.-S.~B.}\
  \bibnamefont {Lee}}\ and\ \bibinfo {author} {\bibfnamefont {A.}~\bibnamefont
  {Weichselbaum}},\ }\bibfield  {title} {\bibinfo {title} {Adaptive broadening
  to improve spectral resolution in the numerical renormalization group},\
  }\href {https://doi.org/10.1103/PhysRevB.94.235127} {\bibfield  {journal}
  {\bibinfo  {journal} {Phys. Rev. B}\ }\textbf {\bibinfo {volume} {94}},\
  \bibinfo {pages} {235127} (\bibinfo {year} {2016})}\BibitemShut {NoStop}%
\bibitem [{\citenamefont {Sakurai}\ and\ \citenamefont
  {Napolitano}(2017)}]{sakuraiModernQuantumMechanics2017b}%
  \BibitemOpen
  \bibfield  {author} {\bibinfo {author} {\bibfnamefont {J.~J.}\ \bibnamefont
  {Sakurai}}\ and\ \bibinfo {author} {\bibfnamefont {J.}~\bibnamefont
  {Napolitano}},\ }\href {https://doi.org/10.1017/9781108499996} {\emph
  {\bibinfo {title} {Modern {{Quantum Mechanics}}}}},\ \bibinfo {edition}
  {2nd}\ ed.\ (\bibinfo  {publisher} {{Cambridge University Press}},\ \bibinfo
  {address} {{Cambridge}},\ \bibinfo {year} {2017})\BibitemShut {NoStop}%
\end{thebibliography}%

\end{document}